\documentclass[a4paper,fleqn,usenatbib]{mnras}

\usepackage{mathptmx}

\usepackage[T1]{fontenc}
\usepackage{ae,aecompl}

\usepackage{graphicx}	
\usepackage{amsmath}	
\usepackage{amssymb}	

\usepackage{url}
\usepackage{verbatim}
\usepackage{hyperref}
\usepackage{bm}

\def\IRACcolour{[3.6$\mu$m]-[4.5$\mu$m]~}
\def\chone{[3.6$\mu$m]}
\def\chtwo{[4.5$\mu$m]}
\def\sfrsdunits{${\rm M}_{\odot}\,{\rm yr}^{-1}\,{\rm kpc}^{-2}$}
\def\sfrsd{$\Sigma_{\rm SFR}$}
\def\oiii{[O{\scshape III}]}

\title[HST/WFC3 imaging of bright $z \simeq 7$ LBGs]{Unveiling the nature of bright $\bm{z \simeq 7}$ galaxies with the \emph{Hubble Space Telescope}}  
\author[R. A. A. Bowler et al.]{R. A. A. Bowler$^{1,2}$\thanks{E-mail:
rebecca.bowler@physics.ox.ac.uk}, J. S. Dunlop$^{2}$, R. J. McLure$^{2}$, D. J. McLeod$^{2}$ \\
$^{1}$Astrophysics, The Denys Wilkinson Building, University of Oxford, Keble Road, Oxford, OX1 3RH \\
$^{2}$Institute for Astronomy, University of Edinburgh, Royal Observatory, Edinburgh, EH9 3HJ}

\pubyear{2016}

\begin{document}
\label{firstpage}
\pagerange{\pageref{firstpage}--\pageref{lastpage}}
\maketitle

\begin{abstract}
We present new~\emph{Hubble Space Telescope}/Wide Field Camera 3 imaging of $25$ extremely luminous ($-23.2 \le M_{\rm UV} \lesssim -21.2$) Lyman-break galaxies (LBGs) at $z \simeq 7$.  
The sample was initially selected from $1.65$~deg$^2$ of ground-based imaging in the UltraVISTA/COSMOS and UDS/SXDS fields, and includes the extreme Lyman-$\alpha$ emitters, `Himiko' and `CR7'. 
A deconfusion analysis of the deep~\emph{Spitzer} photometry available suggests that these galaxies exhibit strong rest-frame optical nebular emission lines ($EW_{0}$(H$\beta$ + \oiii) $ > 600$\AA).
We find that irregular, multiple-component morphologies suggestive of clumpy or merging systems are common ($f_{\rm multi} > 0.4$) in bright $z \simeq 7$ galaxies, and ubiquitous at the very bright end ($M_{\rm UV} < -22.5$).
The galaxies have half-light radii in the range $r_{1/2} \sim 0.5$--$3\,{\rm kpc}$.
The size measurements provide the first determination of the size-luminosity relation at $z \simeq 7$ that extends to $M_{\rm UV} \sim -23$.
We find the relation to be steep with $r_{1/2} \propto L^{1/2}$.
Excluding clumpy, multi-component galaxies however, we find a shallower relation that implies an increased star-formation rate surface density in bright LBGs.
Using the new, independent,~\emph{HST}/WFC3 data we confirm that the rest-frame UV luminosity function at $z \simeq 7$ favours a power-law decline at the bright-end, compared to an exponential Schechter function drop-off. 
Finally, these results have important implications for the~\emph{Euclid} mission, which we predict will detect $> 1000$ similarly bright galaxies at $z \simeq 7$.
Our new~\emph{HST} imaging suggests that the vast majority of these galaxies will be spatially resolved by~\emph{Euclid}, mitigating concerns over dwarf star contamination.
\end{abstract}


\section{Introduction}

The study of extremely high-redshift galaxies provides key insight into the earliest stages of galaxy evolution. 
Over the last decade, the observational frontier has extended well into the first billion years of the history of the Universe, with hundreds of Lyman-break galaxies (LBGs) now known at $z > 6$ (e.g.~\citealp{Bouwens2015, Finkelstein2015, Bowler2014, McLure2013}), and samples extending to $z \simeq 9$ (e.g.~\citealp{Oesch2014, McLeod2016}).
The selection of $z > 6.5$ LBGs requires deep near-infrared imaging to detect the rest-frame UV emission as it is redshifted beyond $\lambda_{\rm obs} \sim 1 \mu {\rm m}$.
As a consequence the field has expanded rapidly since the installation of the Wide Field Camera 3 (WFC3) on the~\emph{Hubble Space Telescope (HST)} in 2009, through deep optical and near-infrared surveys from~\emph{HST} such as the Ultra Deep Field (UDF;~\citealp{Beckwith2006, Ellis2013, Illingworth2013}), the Cosmic Origins Deep Extragalactic Legacy Survey (CANDELS;~\citealp{Grogin2011, Koekemoer2011}), the Cluster Lensing and Supernova Survey with~\emph{Hubble} (CLASH;~\citealp{Postman2012}), and the Frontier Fields program (HFF; PI Lotz).
However, despite their success, these~\emph{HST} surveys have only sampled relatively small cosmological volumes (covering at most $\sim 0.2\,{\rm deg}^2$ on the sky) and therefore provide samples dominated by sub-$L^*$ galaxies.
Instead, wide-area ground-based imaging, such as the UK Infrared Telescope (UKIRT) Ultra Deep Survey~\citep{Lawrence2007} and the UltraVISTA survey~\citep{McCracken2012}, which cover several square degrees on the sky, have led the way in the selection of the brightest known $z > 6$ galaxies~\citep{Bowler2012, Bowler2014, Bowler2015, Willott2013}.
Combined, these imaging campaigns have revolutionised our understanding of the luminosity functions (e.g.~\citealp{Bouwens2015, Finkelstein2015, Bowler2015, McLure2013}) and stellar populations (e.g.~\citealt{Bouwens2014beta,Rogers2014, Dunlop2013}) of LBGs from $z = 6$ to $z = 8$.
A detailed size and morphological analysis of high-redshift galaxies however, has been confined to samples of fainter LBGs detected by~\emph{HST}/WFC3, simply due to the compact nature of these galaxies when compared to the typical seeing of ground-based near-infrared observations (typical full-width at half-max, FWHM $\simeq 0.8$-arcsec).
Even with the superior resolution of~\emph{HST} (FWHM $\simeq 0.2$ arcsec), size measurements of high-redshift galaxies are challenging, with typical LBGs at $z > 6$ showing half-light radii of $r_{1/2} < 0.5$ kpc ($< 0.1\, {\rm arcsec}$;~\citealp{Kawamata2015};~\citealp{Ono2013};~\citealp{Oesch2010}).
Nevertheless, the evolution in galaxy size from $z \simeq 4$ to $z \simeq 1$ has now been reasonably well constrained by several studies (e.g.~\citealp{Bouwens2006, Mosleh2011}), which have shown a strong evolution in the typical sizes of faint ($L < L^*$) LBGs during this period.
From $z = 4$ -- $8$ the size evolution is less certain and can depend strongly on the selection procedure (e.g.~\citealp{Grazian2012};~\citealp{Huang2013}, and see the discussion of selection effects in the work of~\citealp{Shibuya2015size} in~\citealt{Curtis-Lake2016}) and the definition of the `typical' galaxy size.

In addition to charting the evolution in the galaxy size with redshift, another key constraint on the formation mechanisms of galaxies comes from the observed size-luminosity or size-mass relation~\citep{Shen2003}.
A size-mass relation has been claimed to exist up to $z \simeq 4$~\citep{Law2012}, suggesting an early onset to the key astrophysical processes thought to impact the relation, such as feedback and/or mergers (e.g.~\citealp{Lacey2016}).
At $z > 4$ however, the masses derived from SED fitting become significantly more uncertain and the relation has yet to be confirmed~\citep{Mosleh2011}.
Instead, at the highest redshifts, the correlation between rest-frame UV luminosity and size is frequently measured. 
At $z\simeq 7$, studies of the sizes of galaxies in the UDF~\citep{Ono2013,Oesch2010} and CANDELS~\citep{Curtis-Lake2016, Shibuya2015size, Grazian2012} have shown evidence for a size-luminosity relation, however there are still large uncertainties in the derived slope, due in-part to the limited dynamic range available from the small area surveys studied.
For the brightest galaxies at $z \simeq 6$--$7$, measurements from the ground-based imaging used to select them has hinted at larger sizes~\citep{Willott2013, Bowler2014}, however the uncertainties have been too large to place strong constraints on the size-luminosity relation.

While measurements of size as a function of redshift and luminosity can provide a broad overview of the build-up of galaxies with time, a morphological analysis can reveal further details about the formation mechanisms.
A disturbed or irregular shape is often attributed to a merger or interaction, however simulations of early galaxies formation predict that star-forming galaxies at high-redshift should also contain large star-forming clumps due to the increased gas content and density (e.g.~\citealp{Dekel2014}; see discussion in~\citealp{Guo2015}).
At lower redshift, the fraction of star-forming galaxies with disturbed morphologies is observed to rise~\citep{Talia2014, Tasca2009}, with $30-60$ percent showing irregular or multiple components at $z \sim 2$--$3$~\citep{Law2012, Ravindranath2006}.
Work at $z \simeq 6$ suggests that this trend continues to the highest redshifts, with $40\,$-$\,50\,$\% of the brightest galaxies ($M_{\rm UV} < -20.5$) showing a disturbed, clumpy morphology~\citep{Jiang2013b, Willott2013}.
At $z \simeq 7$ the morphology of the brightest LBGs is uncertain due to the lack of high-resolution near-infrared imaging of these rare galaxies, however there have been hints of highly clumpy systems from the detailed study of two extremely bright Lyman-$\alpha$ emitters (LAEs) nicknamed `Himiko'~\citep{Ouchi2009} and `CR7'~\citep{Sobral2015}.
Both of these LAEs are also bright in the rest-frame UV continuum ($m_{\rm AB} \sim 25$), where they appear to consist of multiple components in high-resolution~\emph{HST}/WFC3 imaging, suggestive of a merging system~\citep{Ouchi2013}.
Despite the intense study of these galaxies, with only two objects it is unclear whether the extended clumpy morphology observed is typical for all bright star-forming galaxies at $z \simeq 7$ or is exclusive to strong Lyman-$\alpha$ emitters.

Our recent detection of a sample of extremely bright LBGs at $z \simeq 7$~\citep{Bowler2012, Bowler2014} provides an ideal sample with which to investigate the sizes and morphologies of the brightest galaxies at high-redshift.
Selected from the $1.65\,{\rm deg}^2$ of optical/near-infrared imaging provided by the UltraVISTA/Cosmological Origins Survey (COSMOS) and UKIDSS UDS/ Subaru XMM-Newton Deep Survey (SXDS) fields, the sample of 34 galaxies at $z \simeq 7$ contains the brightest ($M_{\rm UV} < -22$) known galaxies at this epoch. 
In contrast to the narrow-band selected Lyman-$\alpha$ emitters `Himiko' and `CR7', our sample was selected based on the rest-frame UV continuum luminosity via the Lyman-break technique and hence provides the first magnitude limited sample of $z \simeq 7$ star-forming galaxies.
The sample allowed the very bright end of the rest-frame UV luminosity function (LF) be determined robustly for the first time, showing that the number counts of galaxies do not decline as rapidly as predicted by the commonly assumed Schechter function, and rather a double-power law (DPL) provides a good description of the bright-end at $z \simeq 7$.
When compared to the LF at $z = 5$ and $z = 6$, the derived LF using the ground-based data at the bright end showed evidence for evolution around the knee of the function~\citep{Bowler2015}, in contrast to other works based exclusively on~\emph{HST} fields~\citep{Bouwens2015, Finkelstein2015}.
To unveil the sizes and morphological properties of the brightest LBGs at $z\sim7$, targeted follow-up of ground-based samples with~\emph{HST} is essential, and in this paper we present the results of our~\emph{HST}/WFC3 Cycle 22 campaign to provide the first high-resolution data for the~\citet{Bowler2012, Bowler2014} objects.
The results of this analysis provide the first magnitude limited study of a sample of $z \simeq 7$ LBGs extending from $L \simeq 2.5\,L^*$ to $L > 10\,L^*$ with~\emph{HST}/WFC3.

The structure of this paper is as follows.
In Section~\ref{sect:data} we describe the ground based and~\emph{HST}/WFC3 data utilized in this study.
The initial results of the~\emph{HST} follow-up are described in Section~\ref{sect:initialresults}, where we discuss a newly identified cross-talk artefact identified in the VISTA/VIRCAM data.
The final sample properties, visual morphologies and the derived merger fraction are discussed in Section~\ref{sect:sample}, where we also present the results of a deconfusion analysis of the available~\emph{Spitzer} data.
In Section~\ref{sect:lf} we calculate an updated luminosity function derived from the~\citet{Bowler2014} sample.
The galaxy sizes and size-luminosity relation are presented in Section~\ref{sect:sizes}, and in Section~\ref{sect:discussion} we discuss the nature of the bright LBGs in our sample.
We end with our conclusions in Section~\ref{sect:conclusions}.
We assume a cosmology with $H_{0} = 70\, {\rm km}\,{\rm s}^{-1}\,{\rm Mpc}^{-1}$, $\Omega_{m} = 0.3$ and $\Omega_{\Lambda} = 0.7$.
In this cosmology, one arcsec corresponds to a physical size of $5.5\,{\rm kpc}$ and $5.2\,{\rm kpc}$ at $z = 6.5$ and $z = 7.0$ respectively.
All magnitudes are quoted in the AB system~\citep{Oke1974, Oke1983}.

\section{Observations and data reduction}\label{sect:data}

In Table~\ref{table:data} we show the coordinates of the galaxy candidates analysed in this work and the origin of the~\emph{HST}/WFC3 data for each object.
The sample of bright $z \simeq 7$ LBGs was selected initially in~\citet{Bowler2014}, and the majority of the~\emph{HST} data was obtained as part of our Cycle 22 WFC3 imaging program.
For reference, we present the key ground- and space-based photometric filters used in this work in Fig.~\ref{fig:filters}.

\subsection{Initial galaxy sample}\label{sect:initialsample}

In~\citet{Bowler2014} we utilized the ground-based datasets in the UltraVISTA/COSMOS and UDS/SXDS fields.
The available datasets for the field are described in detail in Section~\ref{sect:grounddata} below, and consist of multi-band photometry from the optical to the near-infrared, including crucially deep $z$, $Y$, $J$, $H$ and $K$ band observations which allow a robust selection of $z \simeq 7$ LBGs.
The galaxy samples were selected in a stacked $Y+J$ image in the UltraVISTA/COSMOS dataset, with the requirement that $m_{Y} < 25.8$ or $m_{J} < 25.4$ (in a 1.8 arcsec diameter circular aperture).
For the UDS/SXDS field the $J$-band imaging was used for selection, with the requirement that $m_{J} < 25.7$~\emph{and} that the $Y$-band flux exceeded the two-sigma limit ($m_{Y} < 25.8$).
The different selection conditions in the UDS/SXDS field were driven by the shallower $Y$-band imaging available, which is necessary for the robust identification and removal of contaminant low-redshift galaxies or galactic brown dwarfs.
This condition ultimately resulted in significantly fewer LBGs selected in this field (4 in the UDS/SXDS, 30 in UltraVISTA/COSMOS).
After the initial magnitude cuts, we also required a non-detection at the $2$--$\sigma$ level in the optical bands blue-ward of the $z$-band.
The final sample of high-redshift candidates was then selected using a spectral-energy distribution (SED) fitting analysis, using the photometric redshift fitting code {\sc Le Phare}~\citep{Arnouts1999, Ilbert2006}.
Full details of the photometric redshift fitting procedure are available in~\citet{Bowler2014}, however in brief we fitted~\citet{Bruzual2003} models with exponentially rising $\tau$ star-formation histories (SFHs), assuming the~\citet{Calzetti2000} dust law with a range of rest-frame V-band attenuation ($0.0 < A_{\rm V} \le 4.0$).
Two values of metallicities were assumed ($Z = 1/5\,Z_{\sun}$ and $Z = 1\,Z_{\sun}$) and absorption by intergalactic neutral Hydrogen along the line-of-sight was applied using the~\citet{Madau1995} prescription.
In addition, we also fitted models that included a Lyman-$\alpha$ emission line in the galaxy template (with rest-frame equivalent width, $EW_{0}$, of up to $240$\AA), and included galaxies in our final sample that could be at $z > 6.5$ with the addition of Lyman-$\alpha$ in the SED.
In addition to galaxy templates, we also fit with brown dwarf standard spectra to exclude cool galactic sub-stellar objects as contaminants in our sample.

The result of this analysis was a sample of 34 candidate LBGs with photometric redshifts in the range $6.0 < z < 7.5$.
The sample included the $z = 6.6$ LAE `Himiko'~\citep{Ouchi2009}, which was reselected as a lower redshift LBG ($z \sim 6.4$) using our broad-band data.
The galaxies were in the luminosity range $-23.0 < M_{\rm UV} < -21.2$ in the rest-frame UV, with star-formation rates of ${\rm SFR} \sim 10$--$50\,{\rm M}_{\sun}/{\rm yr}$ and stellar masses of $M_{\star} \sim 10^{10}\,{\rm M}_{\sun}$.
An analysis of the sizes of the galaxies in the ground-based data showed that over half appeared resolved under ground-based seeing (FWHM $\simeq 0.8$ arcsec), suggesting half-light radii in excess of $1.5\,{\rm kpc}$ for the brightest galaxies when assuming a single S{\'e}rsic profile~\citep{Bowler2014}.

\begin{table*}
\caption{The central coordinates for the full sample of 25 LBGs from~\citet{Bowler2014} targeted in this work, with the corresponding~\emph{HST}/WFC3 imaging data available for each object.
The majority of the data was obtained from program ID $=13793$ (PI: Bowler), with additional imaging from the CANDELS survey and program ID $=12578$ (PI F{\"o}rster Schrieber; centred on `ZC401925'). 
The second to last column details the sub-sample of each object, with `P' denoting the primary sample consisting of LBGs from the~\citet{Bowler2014} sample with $M_{\rm UV} < -21.5$ and $6.5 < z_{\rm phot} < 7.5$ without Lyman-$\alpha$ emission.
Additional galaxies with~\emph{HST}/WFC3 data that were not in the primary sample are denoted `F' if they were excluded for being too faint, and/or `L' if the photometric redshift only exceeds $z = 6.5$ when Lyman-$\alpha$ emission is included in the SED fitting.
}
\begin{tabular}{l l l l c l }

\hline
Object ID & R.A.(J2000) & Dec.(J2000) & Dataset & Sample & Notes. \\
\hline
UVISTA-136380 & 09:59:15.89 & +02:07:32.0 & Bowler (O1)& P & \\
UVISTA-28495 & 10:00:28.13 & +01:47:54.4 & Bowler (O2)& P & \\
UVISTA-268511 & 10:00:02.35 & +02:35:52.4 & Bowler (O3)& P & Likely cross-talk artefact (see Section~\ref{sect:ct}) \\
UVISTA-268037 & 09:59:20.69 & +02:31:12.4 & Bowler (O4)& P & Likely cross-talk artefact (see Section~\ref{sect:ct})\\
UVISTA-65666 & 10:01:40.69 & +01:54:52.5 & Bowler (O5)& P & \\
UVISTA-211127 & 10:00:23.77 & +02:20:37.0 & CANDELS& P & $z_{\rm spec} = 7.154$~\citep{Stark2016} \\
UVISTA-137559 & 10:02:02.55 & +02:07:42.0 & Bowler (O6)& P & Likely cross-talk artefact (see Section~\ref{sect:ct})\\
UVISTA-282894 & 10:00:30.49 & +02:33:46.3 & Bowler (O7)& P & \\
UVISTA-238225 & 10:01:52.31 & +02:25:42.3 & Bowler (O8)& P & \\
UVISTA-305036 & 10:00:46.79 & +02:35:52.9 & Bowler (O9)& P & \\
UVISTA-35327 & 10:01:46.18 & +01:49:07.7 & Bowler (O10)& P & \\
UVISTA-304416 & 10:00:43.37 & +02:37:51.6 & Bowler (O11)& P & \\
UVISTA-185070 & 10:00:30.19 & +02:15:59.8 & CANDELS& P & \\
UVISTA-169850 & 10:02:06.48 & +02:13:24.2 & Bowler (O12)& P & Primary target for orbit 12\\
UVISTA-170216 & 10:02:03.82 & +02:13:25.1 & Bowler (O12)& F & Additional LBG imaged in orbit 12\\
UVISTA-304384 & 10:01:36.86 & +02:37:49.2 & Bowler (O13)& P & Primary target for orbit 13 \\
UVISTA-328993 & 10:01:35.33 & +02:38:46.3 & Bowler (O13)& LF & Additional LBG imaged in orbit 13\\
UVISTA-279127 & 10:01:58.50 & +02:33:08.5 & Bowler (O14)& P & \\
UVISTA-104600 & 10:00:42.13 & +02:01:57.1 & Bowler (O15)& P & \\
UVISTA-268576 & 10:00:23.39 & +02:31:14.8 & CANDELS& P & \\
UVISTA-271028 & 10:00:45.17 & +02:31:40.2 & CANDELS& L & \\
UVISTA-30425 & 10:00:58.01 & +01:48:15.3 & ZC401925& L & LAE `CR7'~\citep{Sobral2015}\\
\hline
UDS-35314 & 02:19:09.49 & --05:23:20.6 & Bowler (O16)& P & \\
UDS-118717 & 02:18:11.50 & --05:00:59.4 & Bowler (O17)& P & \\
UDS-88759 & 02:17:57.58 & --05:08:44.8 & CANDELS& L & LAE `Himiko'~\citep{Ouchi2009}\\
\hline

\end{tabular}\label{table:data}
\end{table*}

\subsection{Follow-up observations with~\emph{HST}/WFC3}

Observations with~\emph{HST}/WFC3 of the 17 brightest $z \simeq 7$ galaxies selected in~\citet{Bowler2014} were awarded in Cycle 22 for the General Observer program ID 13793 (PI Bowler).
The sub-sample chosen for~\emph{HST} follow-up was required to have $M_{\rm UV} \lesssim -21.5$ and $6.5 < z_{\rm phot} < 7.5$ (assuming no Lyman-$\alpha$ emission).
These conditions resulted in an initial sample of 20 of the 34 galaxies from~\citet{Bowler2014} proposed for observation.
Of this sample,~\emph{HST}/WFC3 imaging already existed for three objects taken as part of the CANDELS survey, and hence 17 objects were targeted for imaging.
As the galaxies are widely separated on the sky, 17 individual pointings were required.
In fact, two additional~\citet{Bowler2014} galaxies were serendipitously covered by our~\emph{HST}/WFC3 pointings, ID170216 (primary target for that orbit was ID169850), and ID 328993 (primary target ID304384).
Hence, our~\emph{HST}/WFC3 follow-up provides high-resolution imaging of 19 LBGs from~\citet{Bowler2014} as presented in Table~\ref{table:data}.

\begin{figure}
\includegraphics[width = 0.49\textwidth, trim = 0.6cm 1.5cm 0 0.5]{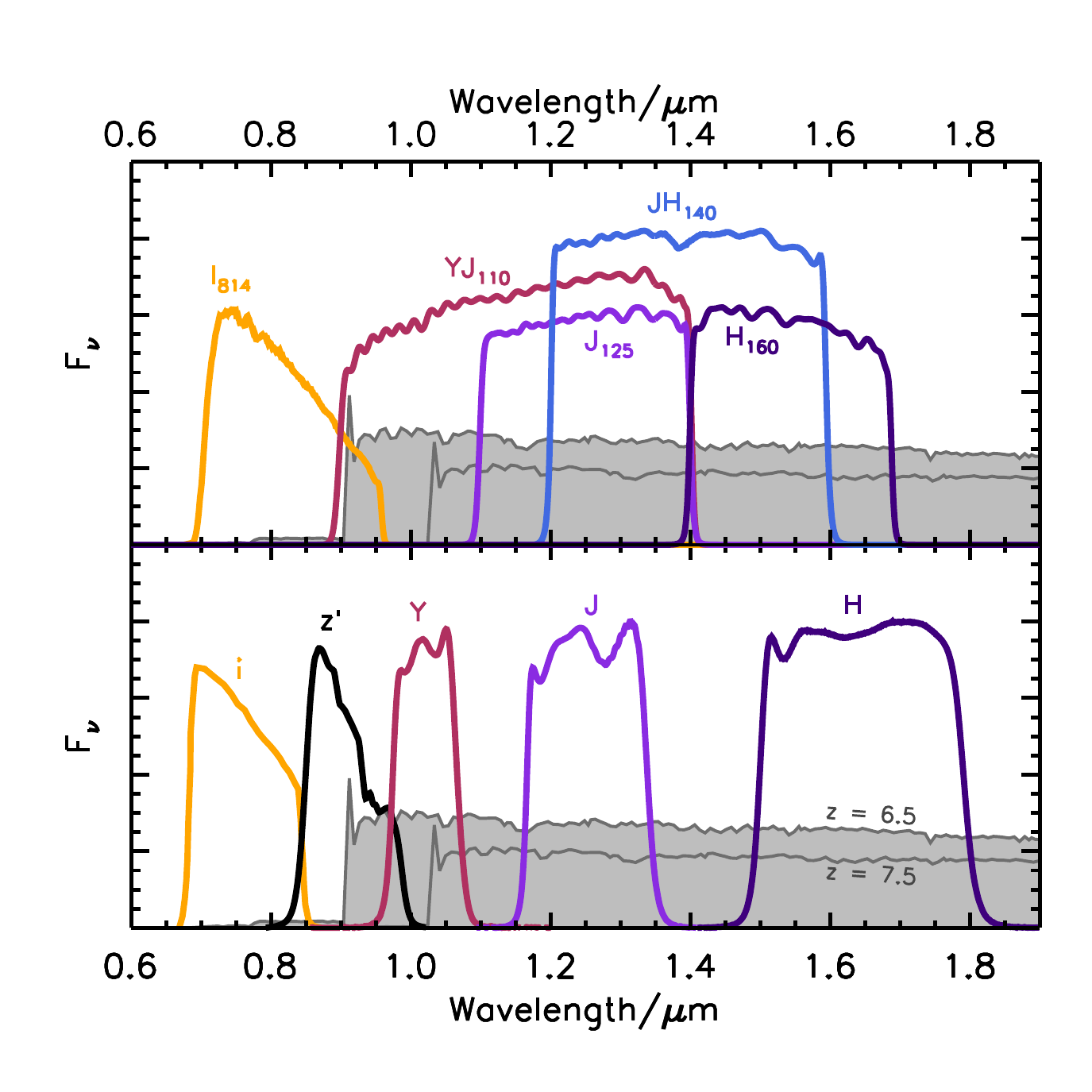}
\caption{A comparison of the key filter transmission curves from~\emph{HST} (top) and the ground-based datasets (bottom) used in this work, restricted to the wavelength range $\lambda = 0.6$--$1.9\mu{\rm m}$.
The upper panel shows the~\emph{HST}/WFC3 filters used primarily in this work to determine the rest-frame UV sizes ($JH_{140}$), compared with those available in the CANDELS survey ($J_{125}$ and $H_{160}$) and in archival imaging of `CR7' ($YJ_{110}$ and $H_{160}$).
The lower panel shows a selection of the ground-based filters used (in this wavelength range) for comparison.
Here the near-infrared filters are from the VISTA telescope, the $i$-band is from CFHT and the $z'$-band is from Subaru.
Note that the UKIRT $J$ and $H$ band transmission curves are very similar to the VISTA profiles (as shown in figure 1 of~\citealp{Bowler2015}).
Two example model SEDs are plotted at $z = 6.5$ and $z = 7.5$.
The models were taken from the~\citet{Bruzual2003} library, with an age of $100\,{\rm Myr}$ a $\tau = 50\,{\rm Myr}$, $A_{\rm V} = 0.1$ and a Lyman-$\alpha$ emission line with $EW_{0} = 10$\AA.
IGM absorption has been applied assuming the~\citet{Madau1995} prescription.
}\label{fig:filters}
\end{figure}

The observational requirements were to obtain a signal-to-noise $S/N > 15$ for the faintest object in our sample, to ensure a reliable size estimate (\citealp{Ono2012, Mosleh2012}).
In addition, the high signal-to-noise threshold substantially improves our sensitivity to individual, fainter, components in the case that the galaxy fragments under~\emph{HST} resolution (e.g. for objects similar to `Himiko' that appear as three separate components in~\emph{HST} imaging;~\citealp{Ouchi2013}).
We obtained imaging to a single orbit depth for each object, using the wide F140W (hereafter $JH_{140}$; see Fig.~\ref{fig:filters}) filter to maximise the $S/N$ in the rest-frame UV.
The observations were spread through the cycle and taken from November 2014 to December 2015.
Each orbit was split into four separate exposures of $\simeq 650$ seconds, using a simple box dither pattern for improved point-spread function (PSF) reconstruction.
The SPARS50 sample rate was used with NSAMP $= 14$ to ensure a high number of non-destructive reads for cosmic ray rejection.
The total integration time for each object was $2612$ seconds.
The median $5\sigma$ depth for the 17 separate pointings of the $JH_{140}$ filter was $m_{\rm AB} = 26.9 \pm 0.1$ in a $0.6$~arcsec diameter circular aperture.

\subsection{Data reduction}\label{sect:dr}

We reduced the~\emph{HST}/WFC3 data using the standard {\sc Astrodrizzle} software to combine the calibrated individual exposures (the {\sc flt} files) into a single image.
The final pixel scale was set to $0.06$~arcsec/pixel (as in the CANDELS imaging;~\citealp{Koekemoer2011}), and a {\sc pix\_frac} of 0.8 was used to provide optimal reconstruction of the PSF.
The astrometry of the images was matched to the ground-based imaging (using a $J$-band selected catalogue) using the {\sc IRAF} package {\sc CCMAP} to account for the astrometric offsets due to errors in the guide star positioning (typically $0.1$--$0.5\,$arcsec for our data).
A zeropoint of $26.452$ was assumed for the $JH_{140}$ data.\footnote{\url{http://www.stsci.edu/hst/wfc3/phot_zp_lbn}}

\subsection{Pre-existing~\emph{HST} imaging}\label{sect:hstdata}

A subset of the initial~\citet{Bowler2014} sample lies within~\emph{HST} data obtained as part of the CANDELS survey, and hence no pointed observations were made of these objects as part of our proposal.
We used the most recent data reductions of the CANDELS data described in~\citet{Koekemoer2011}, with a pixel scale of $0.06$~arcsec/pixel.
The CANDELS imaging in the UDS and COSMOS fields was taken as part of the `wide' component of the survey, which provided WFC3 data over $4 \times 11$ tiles (one pointing of WFC3, area $\simeq 4.5\,{\rm arcmin}^2$) in the centre of both fields.
The total area of each CANDELS field in the UDS and COSMOS was approximately $200\,{\rm arcmin}^2$, with observations taken for $2/3$ of an orbit in the $J_{125}$ filter and $4/3$ of an orbit in the $H_{160}$ filter.
To facilitate a more straightforward comparison to the $JH_{140}$ data available for the majority of the sample, we created an inverse variance weighted stack of the $J_{125}$ and $H_{160}$ data which was used in all subsequent analysis. 
Zeropoints were taken as $m_{\rm AB} = 26.25$ and $m_{\rm AB} = 25.96$ from~\citet{Koekemoer2011}.
The $5\sigma$ depths in the two filters were $m_{\rm AB} = 26.9$ ($0.6$ arcsec diameter aperture) consistent with previous measurements~\citep{McLure2013}, with the depth in the JH stack reaching $m_{\rm AB} = 27.2$.

To provide a complete analysis of the available~\emph{HST} data for the~\citet{Bowler2014} sample, we also include the~\emph{HST}/WFC3 imaging of the Lyman-$\alpha$ emitted galaxy nicknamed `CR7' studied in detail by~\citet{Sobral2015}.
 The LAE was initially selected by~\citet{Bowler2012,Bowler2014}, however was excluded from our~\emph{HST}/WFC3 imaging program as the photometric redshift (without Lyman-$\alpha$) was $z < 6.5$.
 The object was followed-up spectroscopically by~\citet{Sobral2015}, who selected it based on narrow-band photometry, revealing the object to be a strong Lyman-$\alpha$ emitter at $z_{\rm spec} = 6.604$.
As presented in~\citet{Sobral2015}, `CR7' has been imaged by~\emph{HST}/WFC3 serendipitously, in a program designed to target massive star-forming galaxies at $z = 2$ (PI: F{\"o}rster Schreiber, ID 12578).
The imaging consisted of one orbit in the $YJ_{110}$ band, and two in the $H_{160}$ band, centred on object ZC401925.
We performed our own reduction of the data, following the methodology described above in Section~\ref{sect:dr}.
We excluding one of the $H_{160}$ orbits available which poorly overlaps with `CR7'.
The images were registered onto the same pixel scale using the {\sc Astrodrizzle} program {\sc tweakreg}.
The data reached a $5\sigma$ depth of $m_{\rm AB} = 27.4$ in the $YJ_{110}$ filter, and $m_{\rm AB} = 26.6$ in the $H_{160}$ imaging (0.6 arcsec diameter circular apertures).

Finally, we utilize the high-resolution~\emph{HST}/Advanced Camera for Surveys (ACS) $I_{814}$-band imaging available for the sample as a reference high-resolution optical image.
Single orbit depth imaging in the $I_{814}$ band is available in the full $2\,{\rm deg}^2$ COSMOS field~\citep{Koekemoer2007, Scoville2007a, Massey2010} and within the CANDELS COSMOS and UDS `wide' fields to a $5\sigma$ depth of $m_{\rm AB} = 27.2 (27.6)$ in a $0.6 (0.4)$ arcsec diameter circular aperture.
 
\subsection{Ground-based optical and near-infrared data}\label{sect:grounddata}

The UltraVISTA/COSMOS and UDS/SXDS fields contain a wealth of data from the optical to near- and mid-infrared, which makes them ideal fields for the study of high-redshift galaxies.
In the UltraVISTA/COSMOS field the optical imaging was obtained from the T0007 release of the Canada-France-Hawaii Legacy Survey (CFHTLS field D1, $u^*griz$ filters), and in the UDS/SXDS field from the SXDS survey with the Subaru/Suprime-Cam~\citep{Furusawa2008}.
Additional deeper $z'$-band imaging than these surveys was obtained in both fields from Subaru/Suprime-Cam (~\citealp{Furusawa2016}; see~\citealp{Bowler2014} for details).
In the near-infrared, $YJHK_{s}$ photometry is provided in the COSMOS field as part of the third data release of the UltraVISTA survey~\citep{McCracken2012}.
The UltraVISTA data consists of a shallower component that covers the $1.5\,{\rm deg}^2$, and four strips that form the `ultra-deep' part of the survey and extend $\sim 1\,$mag deeper.
The resulting $5\sigma$ depths from the DR3 `ultra-deep' data are as follows $Y = 26.1$, $J = 25.9$, $H = 25.5$, $K_{s} = 25.4$ for the deepest strip, with the other strips being shallower by $\delta m = 0.1$--$0.2\,$mag.
In the gaps between these `ultra-deep' strips, the data reaches $Y = 25.2$, $J = 24.9$, $H = 24.5$, $K_{s} =24.2$ uniformly over the image.
In the UDS/SXDS field we utilise the DR10 of the UKIRT Infrared Deep Sky Survey UDS program, which provides $JHK$ imaging.
The $Y$-band data is provided by the VISTA Deep Extragalactic Observations Survey (VIDEO;~\citealp{Jarvis2013}).
In this paper we exploit the most recent data release of VIDEO, which extends $0.5\,{\rm mag}$ deeper than that analysed previously in~\citet{Bowler2014}, reaching a $5\sigma$ depth of $m_{\rm AB} = 25.3$ (1.8 arcsec diameter circular aperture).

\subsection{\emph{Spitzer}/IRAC data}\label{sect:spitzer}

We utilize deep imaging in the mid-infrared (or rest-frame optical at $z \simeq 7$) from the~\emph{Spitzer Space Telescope} Infrared Array Camera (IRAC) taken as part of the~\emph{Spitzer} Extended Deep Survey (SEDS;~\citealp{Ashby2013}) and the~\emph{Spitzer} Large Area Survey with Hyper-Suprime Cam survey (SPLASH;~\citealp{Steinhardt2014}).
The SPLASH survey provides $\sim 1200$ hours of IRAC data in the \chone~and \chtwo~bands, covering the full COSMOS/UltraVISTA and UDS/SXDS ground-based fields used in this work.
We created a mosaic from the calibrated Level 2 files, which were downloaded from the~\emph{Spitzer} Heritage Archive.
Each image was astrometrically matched to the ground-based $K$-band image in each field, using the {\sc IRAF} package {\sc CCMAP} and background-subtracted using {\sc SExtractor}.
The processed exposures from SPLASH where then combined into a single mosaic including the SEDS data using {\sc Swarp}.
The final images had $5\sigma$ depths of \chone $= 25.3$ and \chtwo $= 25.1$ in a $2.8$ arcsec diameter circular aperture.

To fully exploit the SPLASH data, in particular in cases where nearby low-redshift galaxies contaminate the IRAC imaging of the high-redshift galaxy of interest, we performed a deconfusion analysis.
A full description of the methodology can be found in~\citet{McLeod2015}, however we briefly describe the steps here.
We used {\sc TPHOT}~\citep{Merlin2015} to perform the deconfusion, using the ground-based $J$ or $K$ band images as the high-resolution input data.
Models of the galaxy surface brightness distribution were created from this input image, and then convolved with a transfer kernel to match the resolution of the~\emph{Spitzer}/IRAC data (FWHM $\simeq 2$ arcsec). 
The model galaxies are then fitted to the observed~\emph{Spitzer}/IRAC data by simultaneously varying the flux of each galaxy to provide the best fit.
Errors on the resulting photometry were calculated from the RMS map produced in {\sc TPHOT}.

\section{Initial Results}\label{sect:initialresults}

In~\citet{Bowler2014}, we presented a sample of 34 LBGs at $z \simeq 7$, selected from $1.65\,{\rm deg}^2$ of deep ground-based optical and near-infrared data in the UltraVISTA/COSMOS and UDS/SXDS fields.
Combining our new Cycle 22~\emph{HST}/WFC3 data, the data available from CANDELS and archival imaging (PI F{\"o}rster Schreiber), we present in this paper~\emph{HST}/WFC3 imaging for 25 galaxies.
Of these 25 LBGs, 17 were targeted in our Cycle 22~\emph{HST}/WFC3 proposal, which also serendipitously covers two fainter~\citet{Bowler2014} galaxies.
In addition we include~\emph{HST}/WFC3 imaging for 5 LBGs that lie within the CANDELS imaging and archival data (PI F{\"o}rster Schreiber) that exists for a further object in the~\citet{Bowler2012, Bowler2014} sample (`CR7').
The LAE `CR7' was initially selected in~\citet{Bowler2012} and again in the deeper UltraVISTA DR2 data in~\citet{Bowler2014}, and has been confirmed to be a strong narrow-band emitter by~\citet{Sobral2015}.
We checked the MAST archive to ensure no other serendipitous imaging existed for our sample.\footnote{For object ID583226 there exists $H_{160}$ from PID 12990 (PI Muzzin), with an exposure time of 1062 seconds.  The object is weakly detected in this imaging.  However, to ensure a homogeneous dataset in terms of wavelength coverage and depth, we exclude this imaging from our analysis.}
In total therefore, we present a complete analysis of the~\emph{HST}/WFC3 imaging available for the sample presented in~\citet{Bowler2014}, which includes the LAEs `Himiko' and `CR7', comprising data for 25 of the 34 ground-based selected $z \simeq 7$ galaxies.
As shown in Table~\ref{table:data}, the majority of the sample (20/25 objects) have $M_{\rm UV} < -21.5$ and $z_{\rm phot} > 6.5$ without Lyman-$\alpha$ emission (according to the~\citealp{Bowler2014} values).
The final five objects (two of which are spectroscopically confirmed at $z = 6.6$) are included to provide a complete analysis of the available~\emph{HST}/WFC3 data for the~\citet{Bowler2014} sample, and also to provide a unique comparison between bright LAEs and LBGs.
These five objects were either slightly too faint to be included in the primary sample above (ID170216, ID328993), and/or have best-fitting photometric redshifts in the range $6.0 < z < 6.5$ without the inclusion of Lyman-$\alpha$ emission in the SEDs (ID271028, `CR7' and `Himiko').
Finally, ID211127 has recently been spectroscopically confirmed to be at $z_{\rm spec} = 7.154$ by~\citet{Stark2016}.

\begin{figure*}
\begin{center}

\includegraphics[width = 0.49\textwidth]{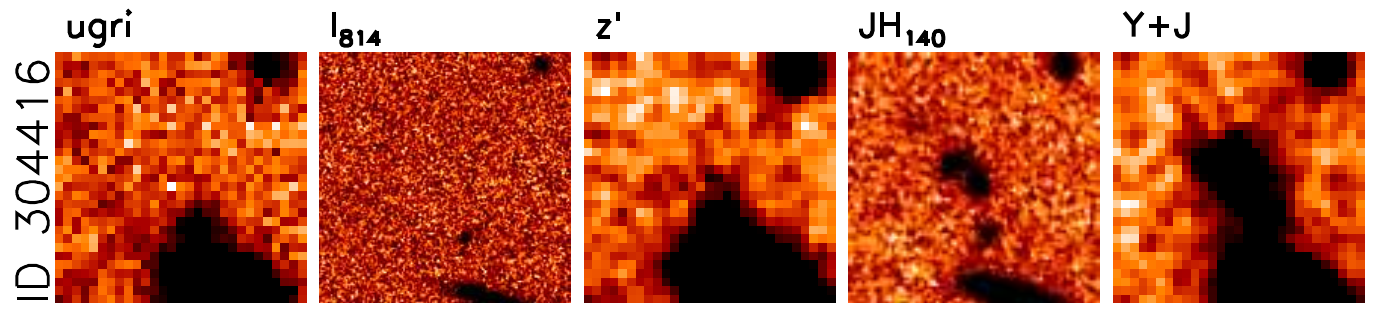}\hspace{0.0cm}
\includegraphics[width = 0.49\textwidth]{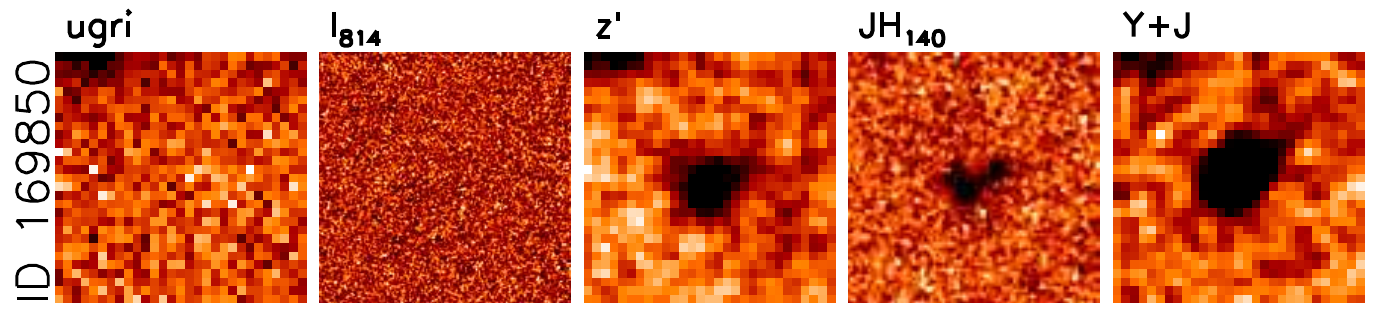}
\includegraphics[width = 0.49\textwidth]{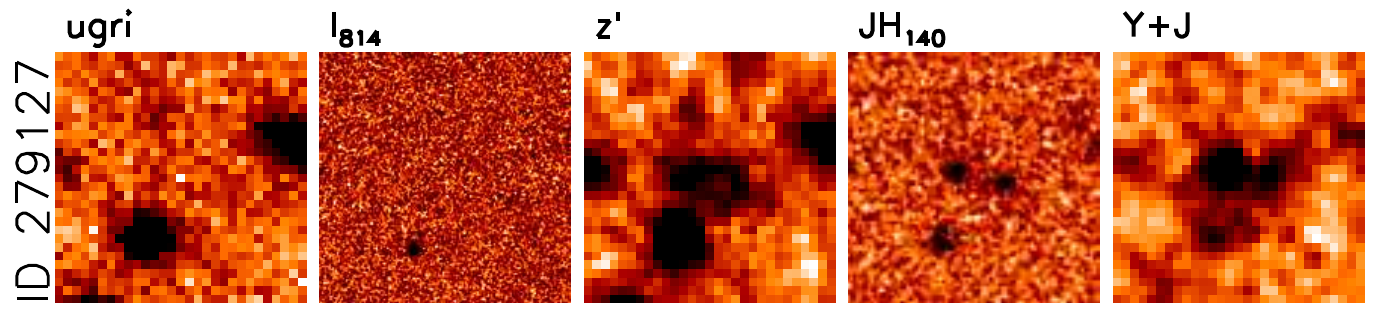}\hspace{0.0cm}
\includegraphics[width = 0.49\textwidth]{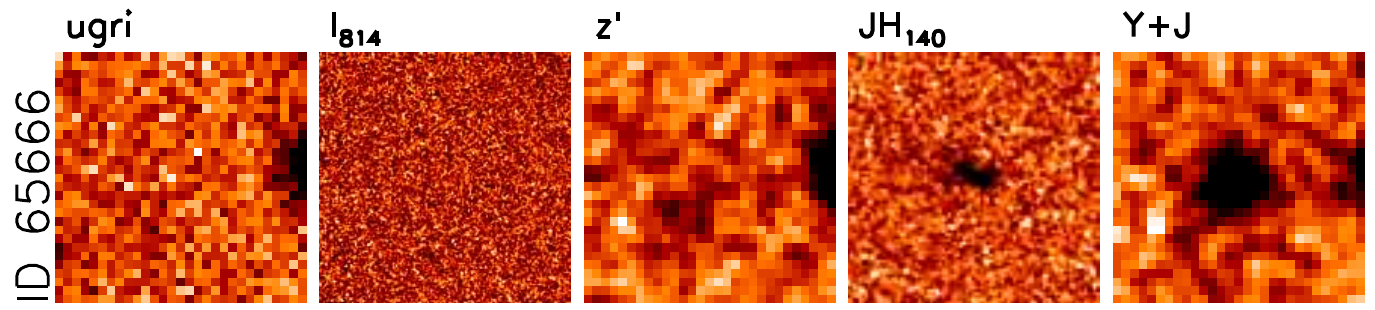}
\includegraphics[width = 0.49\textwidth]{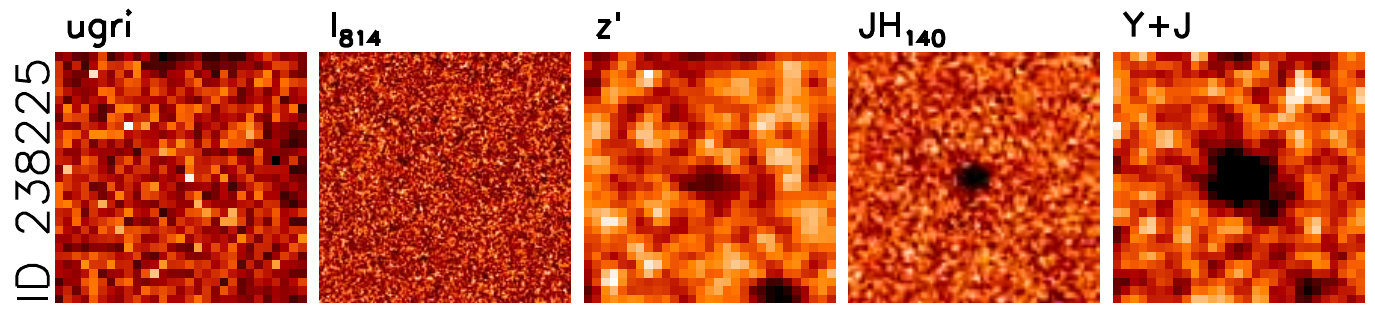}\hspace{0.0cm}
\includegraphics[width = 0.49\textwidth]{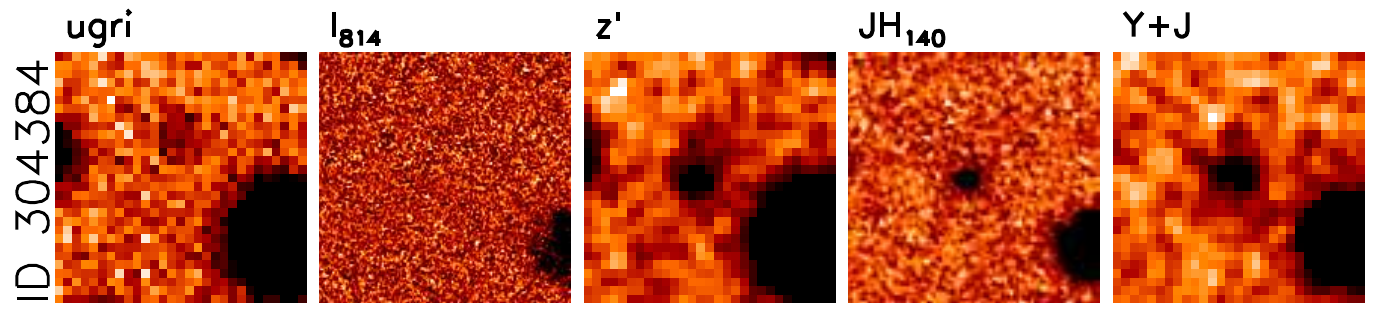}
\includegraphics[width = 0.49\textwidth]{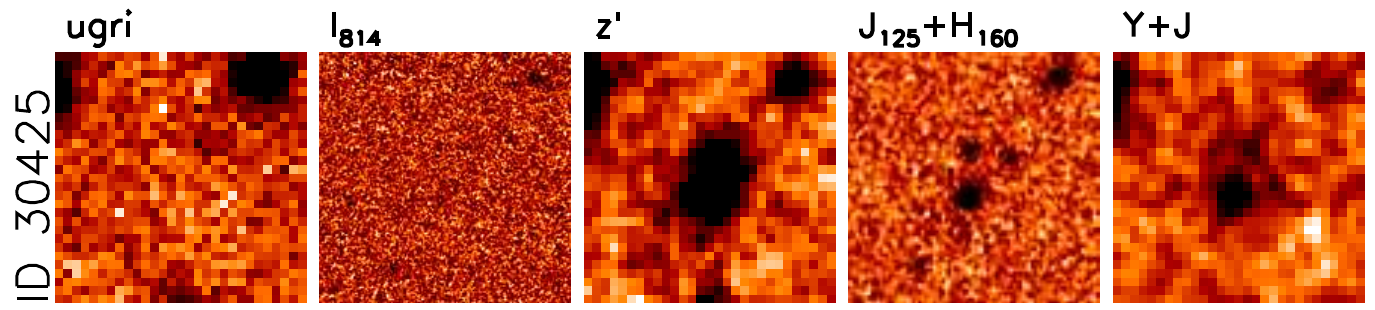}\hspace{0.0cm}
\includegraphics[width = 0.49\textwidth]{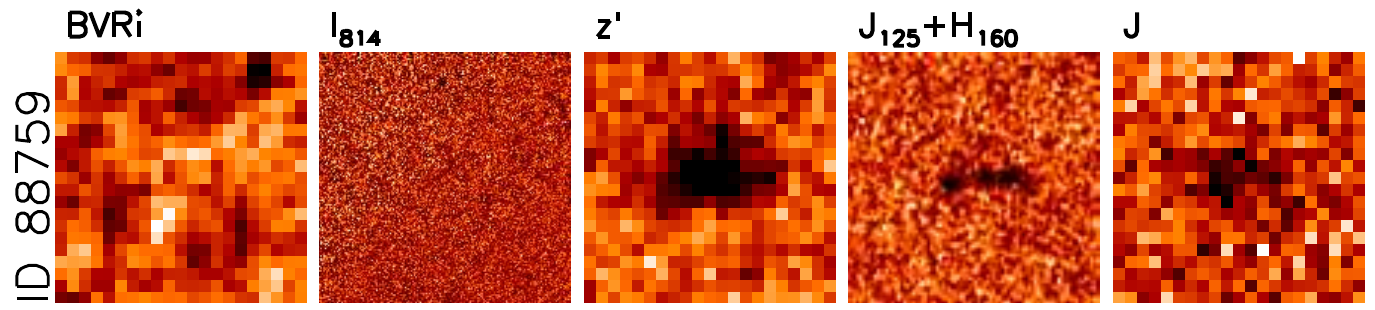}
\includegraphics[width = 0.49\textwidth]{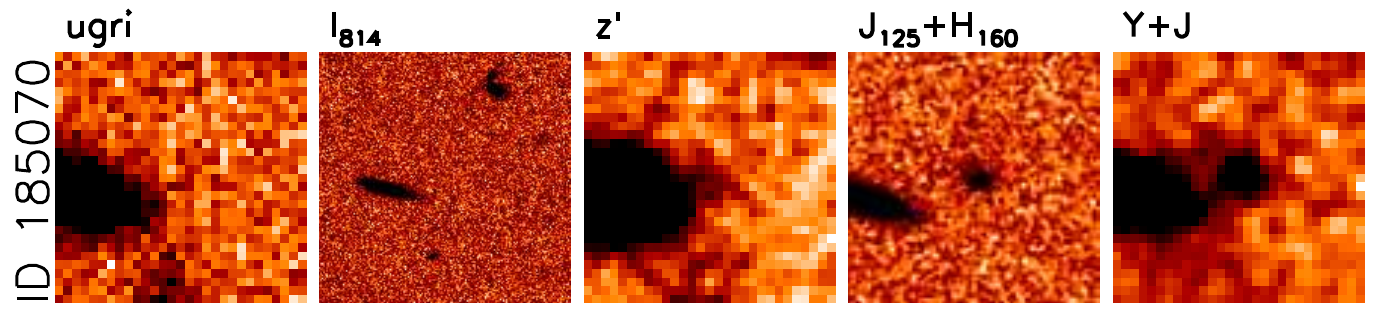}\hspace{0.0cm}
\includegraphics[width = 0.49\textwidth]{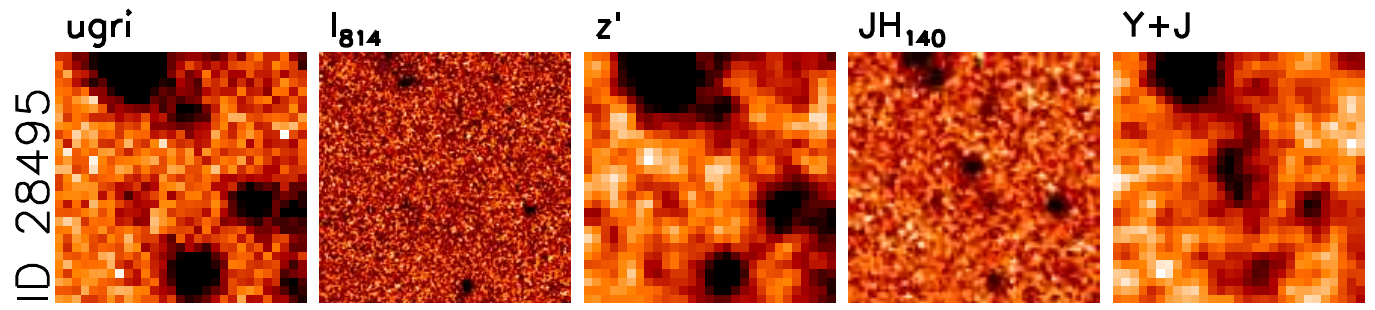}
\includegraphics[width = 0.49\textwidth]{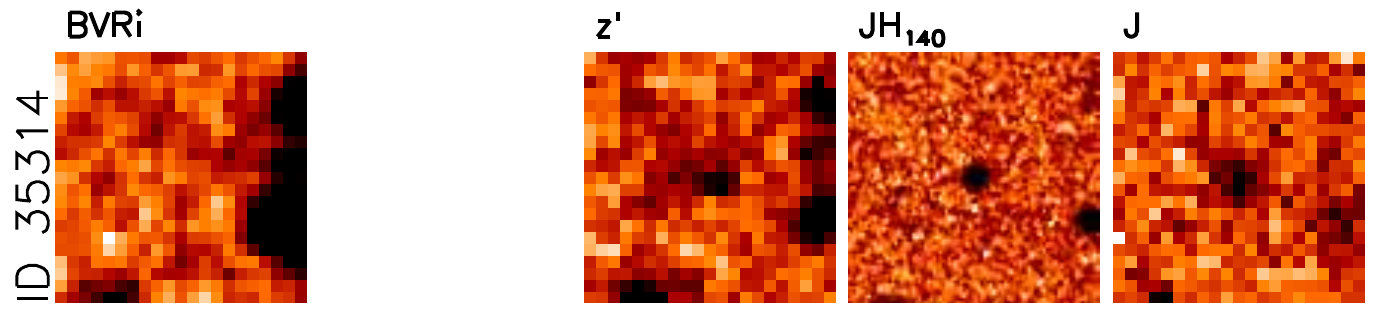}\hspace{0.0cm}
\includegraphics[width = 0.49\textwidth]{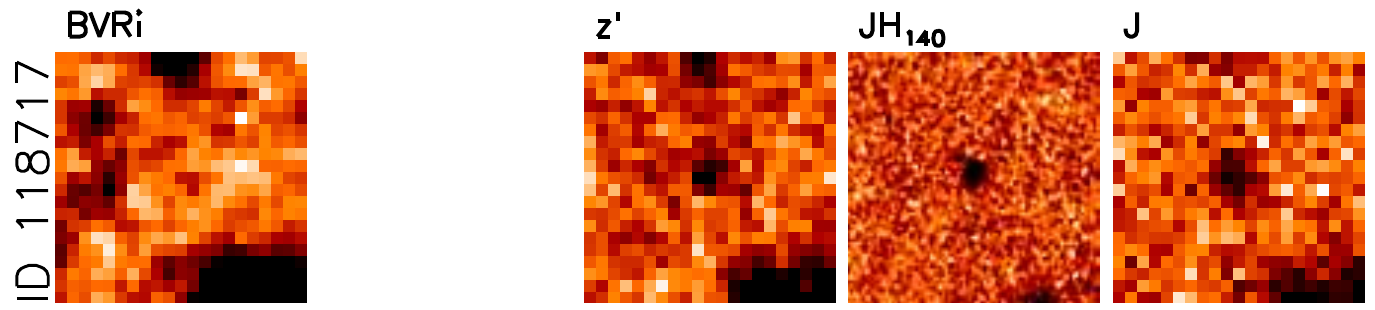}
\includegraphics[width = 0.49\textwidth]{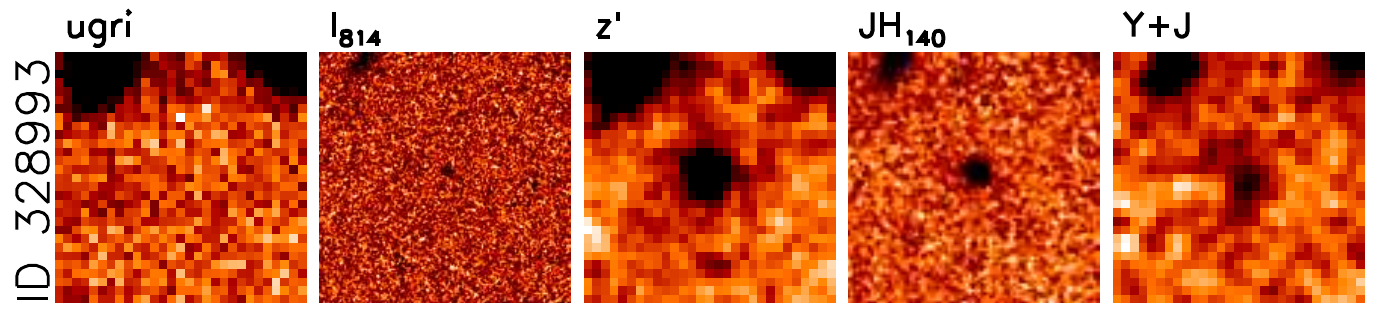}\hspace{0.0cm}
\includegraphics[width = 0.49\textwidth]{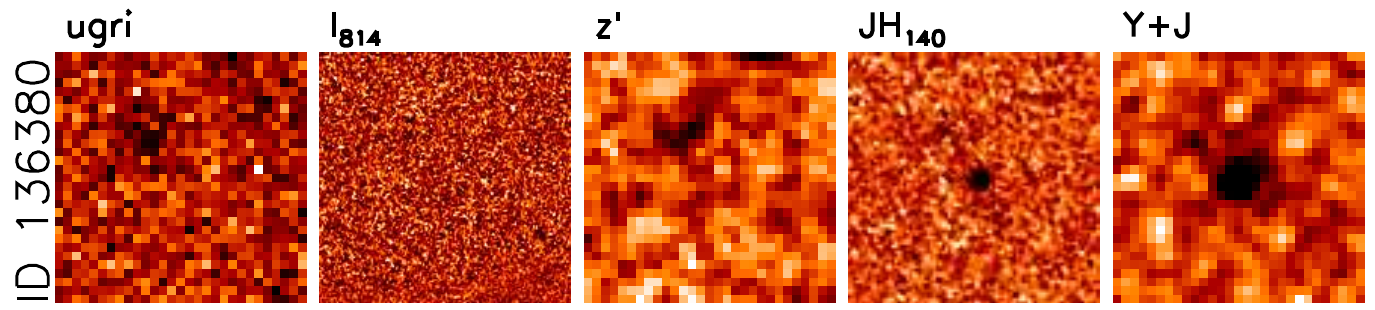}
\includegraphics[width = 0.49\textwidth]{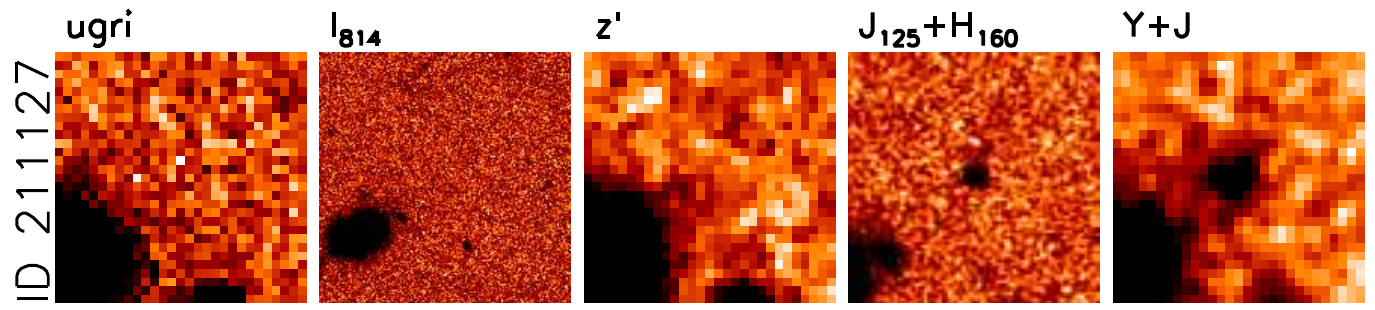}\hspace{0.0cm}
\includegraphics[width = 0.49\textwidth]{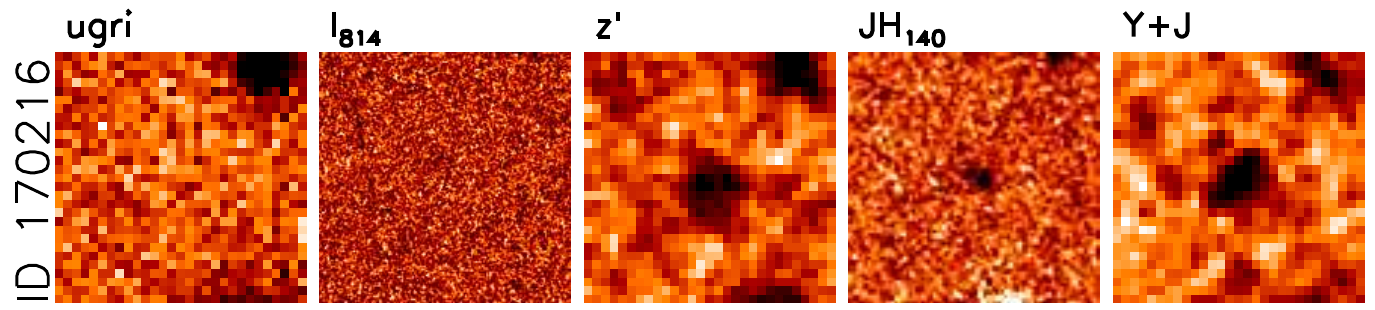}
\includegraphics[width = 0.49\textwidth]{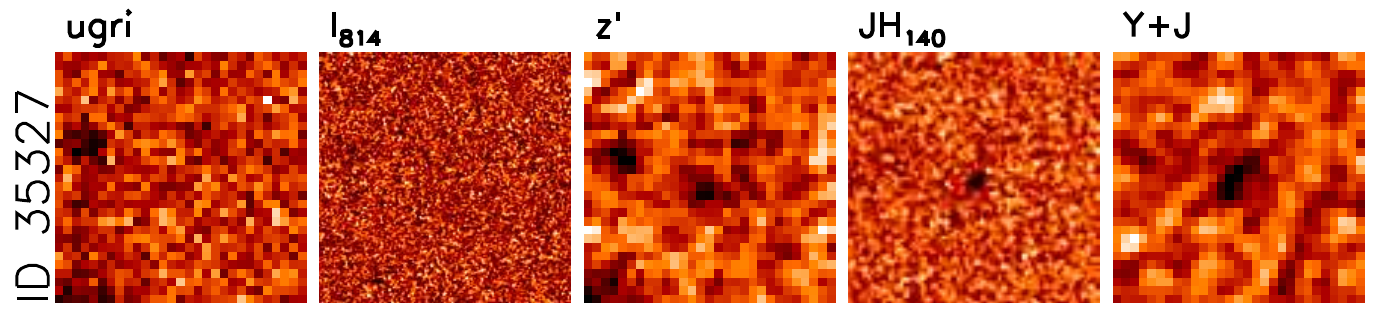}\hspace{0.0cm}
\includegraphics[width = 0.49\textwidth]{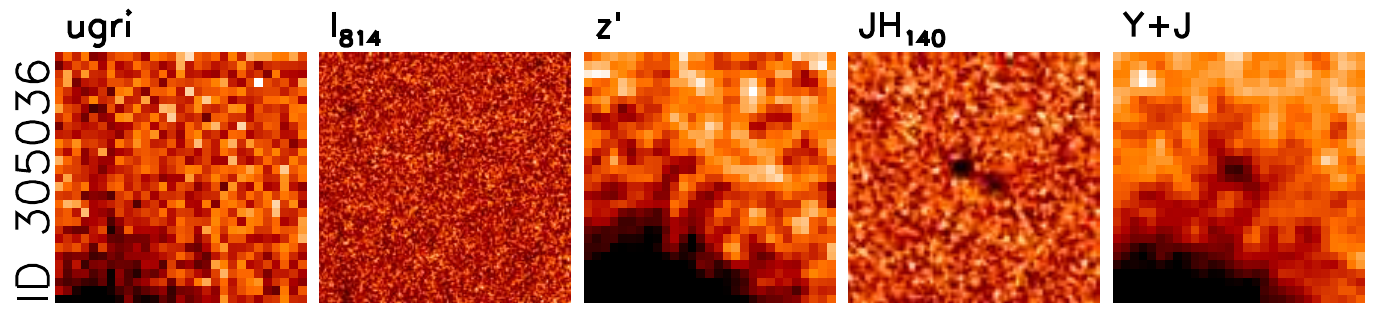}
\includegraphics[width = 0.49\textwidth]{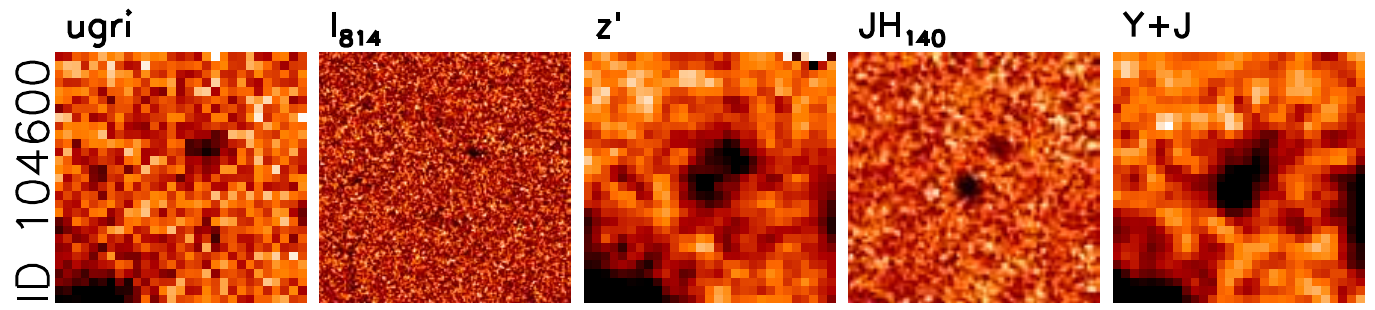}\hspace{0.0cm}
\includegraphics[width = 0.49\textwidth]{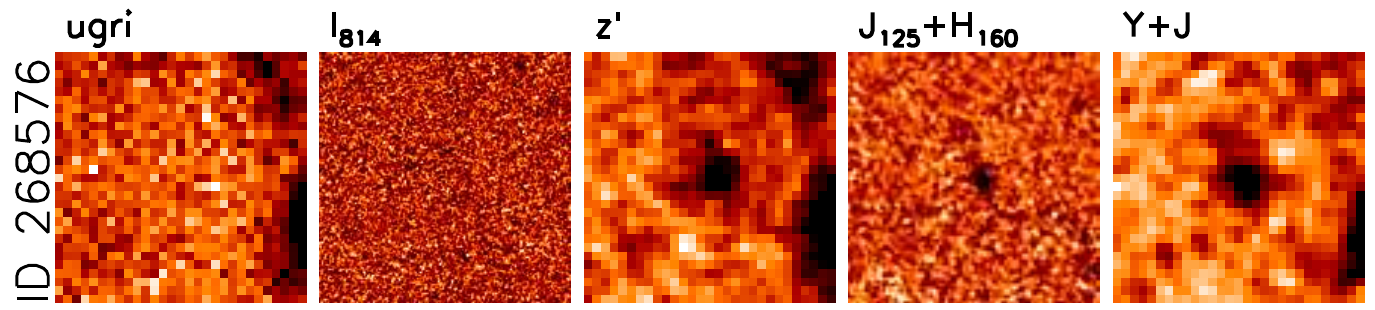}
\includegraphics[width = 0.49\textwidth]{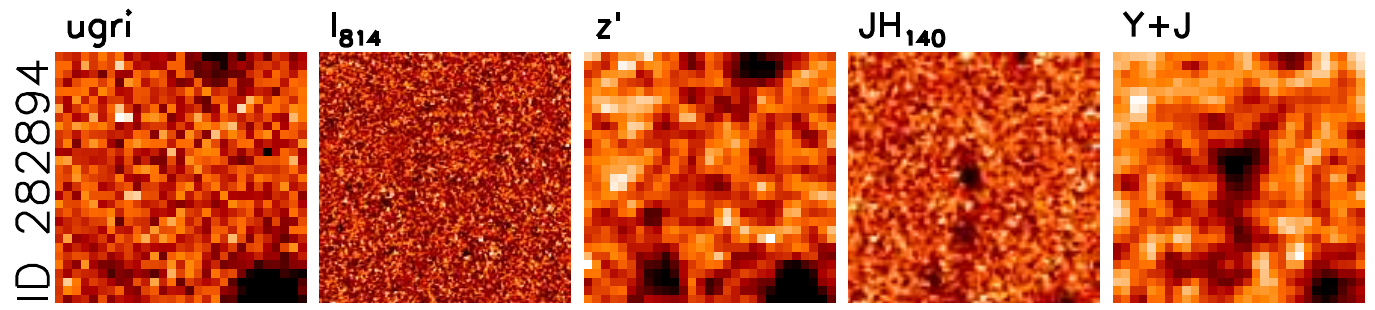}\hspace{0.0cm}
\includegraphics[width = 0.49\textwidth]{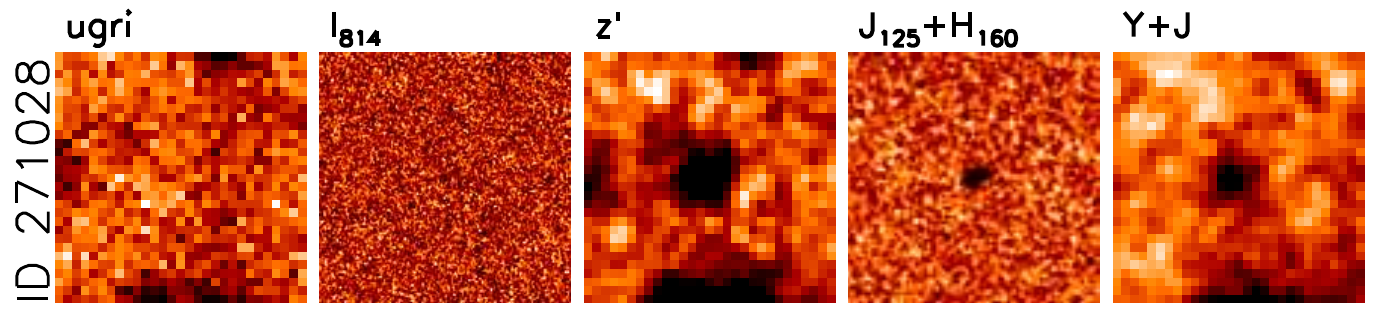}

\caption{Postage-stamp image cut-outs of the 22 detected galaxies in the sample (the final three objects are presented in Fig.~\ref{fig:ct} and discussed in Section~\ref{sect:ct}).
From left to right we show the~\emph{HST}/ACS $I_{814}$-band data (if available), the~\emph{HST}/WFC3 $J_{140}$ or $J_{125} + H_{160}$ data and finally the ground-based $Y+J$ or $J$-band data.
The stamps are ordered by absolute UV magnitude from top left to bottom right, and are $5\times 5$ arcsec with North to the top and East to the left.
The stamps have been scaled linearly with flux, and pixels above the $3\sigma$ level for that stamp have been saturated.
The LAE `Himiko'  is shown with ID$88759$.
An IR `blob' is visible below object ID$170216$, which was not the primary target for that orbit (orbit 12, centred on ID$169850$).
}
\label{fig:groundhst}
\end{center}
\end{figure*}

Postage-stamp cutouts for the sample are shown in Fig.~\ref{fig:groundhst}.
The stamps show the~\emph{HST}/WFC3 data compared to the ground-based near-infrared selection image ($Y+J$ in the UltraVISTA/COSMOS field and the $J$-band in the UDS/SXDS), a stack of the optical data, the Subaru $z'$-band imaging in the field and where available, $I_{814}$ imaging from~\emph{HST}/ACS.
The stamps are  $10 \times 10$ arcsec to include nearby objects and to highlight the improvement in resolution of the~\emph{HST} imaging, however smaller postage-stamps (of 3 arcsec across) scaled by surface brightness are shown in Fig.~\ref{fig:hstsb}.
For three of the orbits obtained in our Cycle 22 program, we found no detection at the expected coordinate of the LBG candidate.
These three objects are shown separately in Fig.~\ref{fig:ct}, and are discussed in Section~\ref{sect:ct}.

\subsection{Components at low-redshift}\label{sect:lowz}

As shown in Fig.~\ref{fig:groundhst} and Fig.~\ref{fig:hstsb}, several of the galaxies selected as a single object in the ground-based data break-up into several discrete components with the improved resolution of~\emph{HST}.
To investigate whether these fainter components are individually consistent with being at $z > 6$, we utilized the available deep ground- and space-based optical imaging to measure the strength of the Lyman-break at each object position.
The majority of the sample is covered by~\emph{HST}/ACS $I_{814}$ imaging from either the COSMOS or CANDELS survey (see Section~\ref{sect:hstdata}), which provides a higher resolution optical image with which to directly compare our~\emph{HST}/WFC3 results.
The $I_{814}$-band data reaches a $2\sigma$ depth of $\sim 28.6$ ($0.4$ arcsec diameter circular aperture), and hence provides stringent constraints on the $I_{814} - JH_{140}$ colour.
We also visually inspected a stack of the ground-based optical data available in each field ($u^*gri$ bands in UltraVISTA/COSMOS, $BVRi$ in UDS/SXDS), to identify any counterparts visible at a $2 \sigma$ limiting depth of $m_{\rm AB} > 28$ (see table 1 of~\citealp{Bowler2014}).
While the~\emph{HST}/ACS data is formally deeper than the ground-based optical stack, the ground-based stack samples a bluer wavelength range (Fig.~\ref{fig:filters}) and the coarser pixel scale and seeing can highlight low surface-brightness features in the data (as is evident comparing this image to the $I_{814}$ imaging).

As shown in Fig.~\ref{fig:groundhst}, for the vast majority of our sample we find no detection in the available optical data.
For a small sub-set however, we find optical detections within $1$-arcsec of the LBG candidate.
We find a compact detection in the $I_{814}$ image at the position of the central galaxy component for one object in our sample, ID328993, with no detection in the ground-based optical stack.
This detection is consistent with the low photometric redshift of this object ($z_{\rm phot} = 6.11^{+0.13}_{-0.14}$) due to the width of the $I_{814}$ filter, as shown in Fig.~\ref{fig:filters}, which extends to $\sim 9000$\AA.
For five of the galaxies in our sample we find weak optical detections offset from the central position, but within $r \simeq 1\,$arcsec of the high-redshift LBG.
In three of these cases (ID136380, ID28495 and ID104600) , the optical detection is coincident with a faint~\emph{HST}/WFC3 source detected in our initial {\sc SExtractor} catalogues, and hence we attribute these sources to low-redshift galaxies close to the line-of-sight of the central high-redshift LBG.
We therefore mask the faint source to the North of ID136380, one of the two fainter components of ID28495 (that to the South-East) and the North-West component of ID104600 in all further analysis.
While the low-redshift objects are extremely faint for ID136380 and ID28495 ($I_{814} \sim 27.6$), in ID104600 the companion object is relatively bright ($I_{814} \sim 27.0$) and has comparable flux to the high-redshift LBG in the deep $z'$-band imaging.
Hence the photometric redshift of ID104600 ($z_{\rm phot} = 6.47 \pm 0.07$) is likely to be biased low by the additional $z$-band flux wrongly attributed to the high-redshift component.
Finally, two of the ground-based selected objects show a detection in the optical stack at a radius of $\simeq 1$ arcsec from the central position (ID304384, ID279127), with no counterpart in the~\emph{HST}/WFC3 data.
For these two galaxies, the detection is of an extremely faint ($m_{\rm AB} > 28$) object that is not coincident with the central WFC3 detection and is not visible at longer wavelengths.

These optical detections show that for $23$ percent of the sample (5/22) we find extremely faint ($m_{\rm AB} \ge 27$) low-redshift galaxies within $\simeq 1\,$arcsec of the central object.
Only for a single object in our sample (ID104600) do we find that the red-optical and potentially the near-infrared photometry has been contaminated by this interloper, hence confirming that our $z \simeq 7 $ LBG selection from ground-based data is providing a clean sample of genuine high-redshift galaxies.
Excluding the identified faint low-redshift companions, we find no significant optical flux for the remainder of the~\emph{HST}/WFC3 detected components of our ground-based sample, which places strong limits on the optical to near-infrared colour.
For an object measured at the $5\sigma$ limit in the near-infrared, the $2\sigma$ limit of the ACS data implies $I_{814} - JH_{140} > 1.7$, therefore excluding a low-redshift contaminant galaxy (e.g.~\citealp{Ouchi2010} impose a $z - y > 1.5$ colour-cut and~\citealp{Bouwens2015} use $z_{850} - Y_{105} \gtrsim 1$ in the selection of $z \simeq 7$ LBGs).
The multiple components identified in our~\emph{HST} data all show a clear $I_{814} - JH_{140}$ colour that is to be expected for $z > 6$ LBGs.
However as we do not have photometry in additional filters from~\emph{HST}, with which to perform a full SED analysis, it is possible that some of the fainter components could be low-redshift interlopers.
If this is the case, the interloper components must be faint relative to the genuine high-redshift candidate, in order that the ground-based photometry be well fit as a $z > 6$ object in our SED analysis.
We consider the probability of chance alignments of red, low-redshift interlopers with our high-redshift sample using deep photometry from the CANDELS survey compiled by~\citet{Galametz2013}.  
From the catalogue we identified all of the objects that had a strong optical to near-infrared break of $I_{814} - JH_{140} > 1.5$, were compact in the WFC3 data and had a best-fitting photometric redshift of $z < 4$ (which is the majority of such objects).
The resulting catalogue showed a surface density of $\simeq 1.1\,{\rm arcmin}^{-2}\,{\rm mag}^{-1}$, with an approximately uniform distribution in the magnitude range $m_{\rm AB} \simeq 23$--$27$.
The photometric-redshift distribution of the interlopers peaks at $z = 1.5$ as expected.
Using this surface density we calculate that the probability of a chance alignment of an interloper galaxy (in the magnitude range $m_{\rm AB} = 25.5$--$27$) to within $1\,{\rm arcsec}$ of one of our sample of high-redshift objects, is $0.0015\,$.
If we consider our full sample of 22 high-redshift LBGs, the probability of a low-redshift interloper appearing at $r \le 1\,{\rm arcsec}$ to one or more $z \simeq 7$ galaxies is $0.03$.
Hence this effect cannot account for the observed incidence of multiple components in our sample.
Finally, this calculation is likely an upper limit, as the red colour of the brighter interloper galaxies would boost the observed $K$ and~\emph{Spitzer}/IRAC fluxes, potentially leading to a poor high-redshift galaxy fit and exclusion of these systems from our initial sample.

\subsection{Cross-talk in the VISTA/VIRCAM data}\label{sect:ct}

\begin{figure*}
\begin{center}
\includegraphics[width = \textwidth]{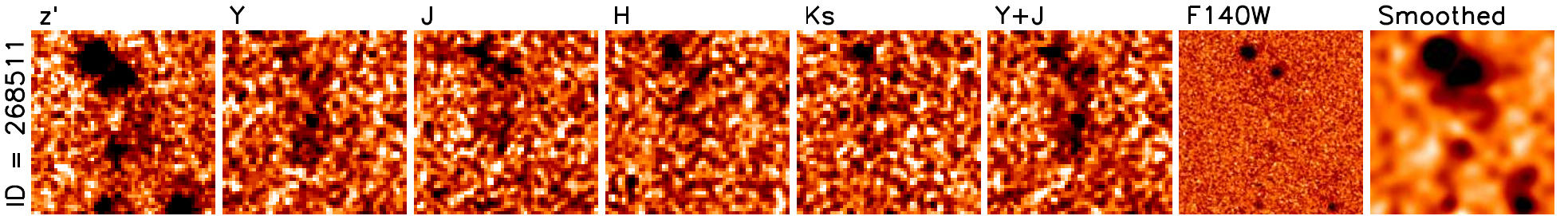}\hspace{1cm}
\includegraphics[width = \textwidth]{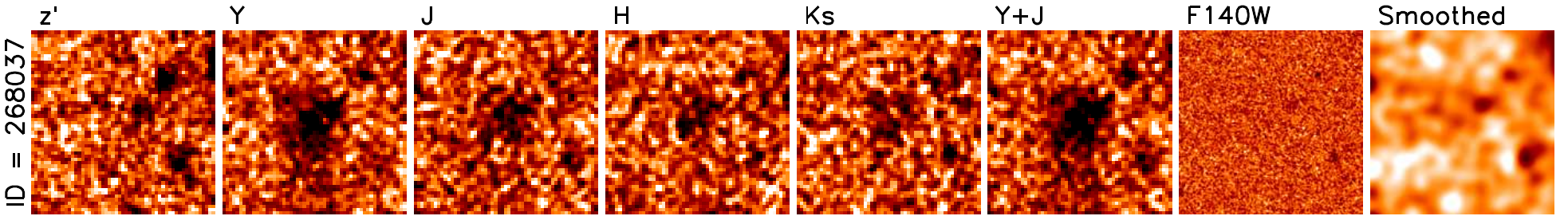}\hspace{1cm}
\includegraphics[width = \textwidth]{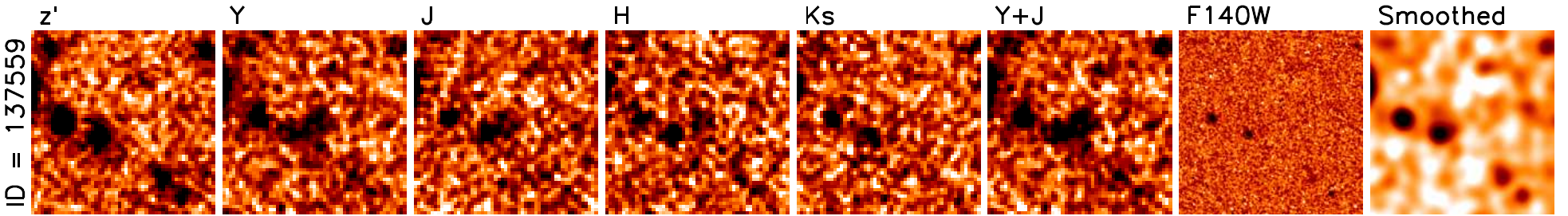}\hspace{1cm}

\caption{Postage-stamp images of the $z \simeq 7$ candidates that appear to be cross-talk artefacts in the VISTA/VIRCAM data in the UltraVISTA/COSMOS field.
On the far left we show the ground-based Subaru $z'$-band image, followed by the VISTA $Y$, $J$, $H$, $K_s$ and $Y+J$ data. 
The original $J_{140}$ data from~\emph{HST}/WFC3 is shown to the right of the ground-based data, and the final stamp shows this data smoothed to ground-based resolution.
In all three cases no object is detected in the original or smoothed $J_{140}$ imaging.
The cross-talk artefact can have an extended appearance as observed in the top two cases, or a more compact morphology as shown for the final object.
In all cases, the cross-talk artefacts were identified at a constant multiple of 128 pixels (on the native $0.34$ arcsec/pixel scale) from a saturated star in the VISTA/VIRCAM data.
}
\label{fig:ct}
\end{center}
\end{figure*}

For three of the LBG candidates imaged as part of our Cycle 22~\emph{HST} program no object was detected at the expected coordinates.
These objects (ID269511, ID268037 and ID137559) are shown in Fig~\ref{fig:ct}, where the ground-based imaging in the $z'$ and near-infrared bands used for selection are compared to the newly available~\emph{HST}/WFC3 $JH_{140}$ data.
With the aim of detecting any diffuse emission that could have been resolved-out by the fine resolution and pixel size of~\emph{HST}, we smoothed the $JH_{140}$ data to ground-based resolution with a convolution kernel obtained using {\sc GALFIT}~\citep{Peng2010}.
These objects are exclusively detected in the near-infrared imaging from VISTA, with no counterpart in the optical or~\emph{Spitzer} data.
Furthermore, ID258511 and ID268037 have an unusual extended appearance in the $YJHK_{s}$ bands in the new DR3 UltraVISTA data.
The undetected LBG candidates are at the faint-end of the~\citet{Bowler2014} sample, however a weak detection in the smoothed $JH_{140}$ data was expected.
Further investigations into the position of these objects and other near-infrared only detected sources selected from our available catalogues in the field, showed that they were preferentially located in vertical lines across the $Y+J$ mosaic.
The vertical alignment is in coincidence with bright saturated stars, with a separation of a multiple of $\simeq 43.5$~arcsec.
The position and regular spacing of the artefact suggest that it is similar to the cross-talk artefact found in imaging from the Wide Field Camera on UKIRT~\citep{Dye2006, Warren2007}.
The separations of the VISTA/VIRCAM artefact is $128$ pixels in the native pixels scale ($0.34\,{\rm arcsec}/{\rm pixel}$) or $43.52\,$arcsec, extending both north and south of saturated stars on the CCD.
Cross-talk occurs during the read-out of each detector, which has dimensions of $2048 \times 2048$ pixels~\citep{Sutherland2015}.
Each detector has sixteen parallel readout channels, of width $128$ pixels, exactly as we find for the artefact, which we conclude arrises from an inter-channel electronic cross-talk introduced in the read-out process.

The cross-talk in the VISTA/VIRCAM instrument is significantly fainter than that in the UKIRT imaging, with a magnitude difference of $>12$ mag compared to the stellar magnitude, or $\simeq 0.001$ percent of the flux of the origin star.
Our investigations show that the detectible artefact is produced by the brightest stars in the image, which have total magnitudes of $m_{\rm AB} \simeq 12$--$14\,$mag.
We created a mask using a catalogue of bright stars from the full $1.5\,{\rm deg}^2$ of UltraVISTA data, and masking out regions at the appropriate separations.
At each expected cross-talk position, a region of dimensions $4\, \times 7\, {\rm arcsec}$ was masked to account for the uncertainties in the centroid of the saturated stars.
The existence of the cross-talk artefact is strongly confirmed by the application of the mask, as it excludes only $1$ per cent of the total image area, but over half of the near-infrared only sample of objects used to investigate the cross-talk locations were flagged as a potential artefact.
Applying such a mask to the~\citet{Bowler2014} sample results in the removal of all three~\emph{HST}/WFC3 undetected objects described above, which we exclude from the revised sample discussed in the remainder of this work.
Implications for the luminosity function at $z \simeq 7$ are discussed in Section~\ref{sect:lf}, however we note here that the galaxies identified as artefacts are at the very faint end of the sample and were exclusively detected in the VISTA near-infrared imaging.
As the remainder of the~\citet{Bowler2014} sample are significantly brighter, they are also detected in the Subaru $z'$-band data and/or the~\emph{Spitzer}/IRAC data in addition to the near-infrared imaging.

The application of the cross-talk mask as described above also highlighted another object (ID279127) in the~\citet{Bowler2014} sample as a potential cross-talk artefact.
Galaxy ID279127 is detected in the~\emph{HST}/WFC3 imaging, the Subaru $z'$-band data and in the~\emph{Spitzer}/IRAC data however, indicating that the cross-talk in this case is coincident with a genuine galaxy.
Comparing the UltraVISTA photometry to the~\emph{HST}/WFC3 results we find that the cross-talk artefact makes a negligible difference to the photometry for this object, and hence the rest-frame UV properties are unaffected.

\section{The Final $z \sim 7$ LBG Sample}\label{sect:sample}

After removing the three objects identified as a likely cross-talk artefact in the VISTA/VIRCAM imaging we now present a detailed analysis of our final sample of 22 bright ($M_{\rm UV} \lesssim -21$) LBGs at $z \simeq 7$ shown in Fig.~\ref{fig:hstsb}.

\subsection{Properties of the final sample}

In Table~\ref{table:sample} we present the photometric redshifts and derived galaxy properties for the sample, calculated using the third data-release of UltraVISTA (deeper $YJHK_{s}$ data) and deeper VIDEO $Y$-band data in the UDS/SXDS.
The photometric redshift methodology was identical to that in~\citet{Bowler2014} and is described in Section~\ref{sect:initialsample}.
The new photometric redshifts agree within the errors with those determined in~\citet{Bowler2014}, confirming that the sample is robustly at $z > 6$.
We do not fit the observed photometry with galaxy SEDs that include Lyman-$\alpha$ emission in this work, as the derived equivalent width ($EW_{0}$) is highly degenerate with the photometric redshift in broad-band filters.
The inclusion of strong Lyman-$\alpha$ emission in the SEDs can result in higher redshifts by up to $\delta z \sim 0.4$ (see table 3 of~\citealp{Bowler2014}) when compared to results from fitting with line-free SEDs.
To provide a complete analysis of the available~\emph{HST}/WFC3 imaging for our sample of bright $z \sim 7$ LBGs, we include four objects in our further analysis that have a photometric redshift without Lyman-$\alpha$ emission at $z < 6.5$.
These include the spectroscopically confirmed LAEs `Himiko' and `CR7' and two additional galaxies with $z_{\rm phot} \simeq 6.6$ with Lyman-$\alpha$ included ($z_{\rm phot} \simeq 6.1$ with line-free model fitting).
We find that our results for the size-luminosity relation are unchanged if these four objects are removed.

Using the new near-infrared data we are able to provide improved constraints on the rest-frame UV slopes of the sample.
We calculate the slope ($\beta_{\rm UV}$) by fitting a power law ($F_{\lambda} \propto \lambda^{\beta}$) to the $YJHK_{s}$ photometry.
As the $Y$-band can be contaminated by the Lyman break or potential Lyman-$\alpha$ emission at $z > 6.8$, we also provide the $\beta_{\rm UV}$ values calculated with only the $JHK_s$ bands for comparison.
The resulting $\beta_{\rm UV}$ measurements are shown in Table~\ref{table:sample}.
While there is still significant scatter for the faintest objects in the sample, the brightest galaxies appear consistently blue, showing $\beta_{UV} \sim -2.0$ as found in samples of fainter $ z \simeq 7 $ LBGs~\citep{Dunlop2013, Bouwens2014beta}.
This is in contrast to a similar analysis at $z \simeq 6$ which found a slightly redder rest-frame UV slope in a sample of bright LBGs ($\beta_{\rm UV} = -1.8 \pm 0.1$;~\citealp{Bowler2015}; see also~\citealp{Jiang2013a}).
A redder slope of $\beta \simeq -1.6$ would also be predicted by extrapolating the colour-magnitude relation measured at $z \simeq 7$~\citep{Bouwens2014beta} and by similarly bright galaxies at $z \sim5$~\citep{Rogers2014}.
For two bright LBGs with $M_{\rm UV} \simeq -22.5$, ID238225 and ID304384 however, we do find a slightly redder slope of $ \beta \simeq -1.7$ as expected.
Interestingly, these two objects appear as single components in the~\emph{HST}/WFC3 data, and are therefore unusual at the bright end of our sample.
While a larger sample is clearly needed, this could suggest that the bright, single components galaxies in our sample have a more evolved stellar population compared to the bluer clumpy/merger-like systems that are undergoing a particularly violent burst of star-formation.

\begin{table*}
\caption{The basic properties of our final sample of 22 LBGs at $z \simeq 7$.
From left to right we show the ID number (identical to that in~\citealp{Bowler2014}) and then the best-fitting photometric redshift, followed by the total magnitude measured from the $JH_{140}$ data ($J_{125} + H_{160}$ imaging for objects in CANDELS or $YJ_{110}$ for `CR7').
For `CR7' (ID30425) and `Himiko' (ID88759) we show the spectroscopic redshifts from~\citet{Sobral2015} and~\citet{Ouchi2009} respectively.
The total magnitude was measured in a $2$ arcsec diameter aperture for the majority of galaxies, and a $3$ arcsec diameter aperture for ID304416, ID169850, ID279127 and Himiko.
We then present the absolute UV magnitude (at 1500\AA), the ${\rm SFR}_{\rm UV}$ calculated using the~\citet{Madau1998} prescription and the rest-frame UV slope $\beta_{UV}$ determined using filters $YJHK$ and $JHK$ in columns 6 and 7.
Finally we show the~\emph{Spitzer}/IRAC magnitudes from our deconfusion analysis, and the PSF corrected half-light radius in ${\rm kpc}$ derived from our curve-of-growth analysis.
Where we have classified the galaxy as a multi-component system it is highlighted with a dagger.
}
\begin{tabular}{l c c c r c c c c c l}
\hline
ID & $z$ & $m_{\rm AB}$ & $M_{\rm UV} $ & ${\rm SFR}_{\rm UV} $ & $\beta_{YJHK}$ & $\beta_{JHK}$ & $[3.6] - [4.5]$ & $[3.6\mu m]$ &  $[4.5\mu m]$ & $r_{1/2}$ \\
& & $/{\rm mag}$  & $/{\rm mag}$  & $/{M \rm}_{\sun}/{\rm yr}$ & & & $/{\rm mag}$  & $/{\rm mag}$  & $/{\rm mag}$ &  $/{\rm kpc}$\\
\hline

UVISTA-304416 & $  6.85_{-  0.09}^{+  0.10}$ & $ 23.85_{ -0.09}^{+  0.10}$ & $-23.16_{ -0.09}^{+  0.10}$ & $      55$& $  -1.9_{-    0.2}^{+   0.2}$& $  -2.1_{-    0.2}^{+   0.3}$& $\phantom{-}      0.28_{     -0.33}^{+      0.37}$& $     24.41_{     -0.22}^{+      0.28}$&$     24.14_{     -0.09}^{+      0.10}$&$2.2^{+0.1}_{-0.1}$\textdagger\\
UVISTA-169850 & $  6.64_{-  0.08}^{+  0.07}$ & $ 24.03_{ -0.10}^{+  0.12}$ & $-22.92_{ -0.10}^{+  0.12}$ & $      44$& $  -2.0_{-    0.1}^{+   0.2}$& $  -2.1_{-    0.2}^{+   0.3}$& $     -0.67_{     -0.37}^{+      0.37}$& $     24.03_{     -0.18}^{+      0.22}$&$     24.70_{     -0.16}^{+      0.18}$&$2.6^{+0.2}_{-0.2}$\textdagger\\
UVISTA-279127 & $  6.53_{-  0.09}^{+  0.06}$ & $ 24.35_{ -0.15}^{+  0.18}$ & $-22.62_{ -0.15}^{+  0.18}$ & $      33$& $  -2.3_{-    0.3}^{+   0.2}$& $  -2.9_{-    0.4}^{+   0.4}$& $     -0.32_{     -0.39}^{+      0.42}$& $     24.24_{     -0.22}^{+      0.28}$&$     24.57_{     -0.14}^{+      0.17}$&$3.0^{+0.4}_{-0.3}$\textdagger\\
UVISTA-65666 & $  7.04_{-  0.15}^{+  0.16}$ & $ 24.48_{ -0.10}^{+  0.11}$ & $-22.43_{ -0.10}^{+  0.11}$ & $      28$& $  -2.0_{-    0.3}^{+   0.2}$& $  -2.3_{-   0.4}^{+   0.4}$& $\phantom{-}      0.77_{     -0.64}^{+      1.08}$& $     25.27_{     -0.19}^{+      0.24}$&$     24.50_{     -0.13}^{+      0.15}$&$1.6^{+0.1}_{-0.1}$\textdagger\\
UVISTA-238225 & $  6.94_{-  0.18}^{+  0.13}$ & $ 24.59_{ -0.12}^{+  0.12}$ & $-22.41_{ -0.12}^{+  0.12}$ & $      27$& $  -1.7_{-    0.3}^{+   0.2}$& $  -1.7_{-    0.4}^{+   0.4}$& $     -1.31_{     -0.69}^{+      0.61}$& $     24.19_{     -0.22}^{+      0.28}$&$     25.50_{     -0.33}^{+      0.47}$&$0.8^{+0.2}_{-0.2}$\\
UVISTA-304384 & $  6.56_{-  0.13}^{+  0.15}$ & $ 24.50_{ -0.10}^{+  0.11}$ & $-22.40_{ -0.10}^{+  0.11}$ & $      27$& $  -1.7_{-    0.4}^{+   0.3}$& $  -2.2_{-    0.6}^{+   0.4}$& $     -0.30_{     -0.41}^{+      0.45}$& $     24.32_{     -0.23}^{+      0.30}$&$     24.62_{     -0.15}^{+      0.18}$&$1.3^{+0.2}_{-0.2}$\\
UVISTA-30425 & $  6.604$ & $ 24.65_{ -0.08}^{+  0.07}$ & $-22.22_{ -0.08}^{+  0.07}$ & $      23$& $  -2.5_{-    0.3}^{+   0.4}$& $  -3.1_{-
    0.6}^{+   0.6}$& $     -1.08_{     -0.22}^{+      0.22}$& $     23.49_{     -0.09}^{+      0.10}$&$     24.57_{     -0.12}^{+      0.13}$&$1.8^{+0.1}_{-0.1}$\textdagger\\
UDS-88759 & $  6.595$ & $ 24.64_{ -0.15}^{+  0.16}$ & $-22.17_{ -0.15}^{+  0.16}$ & $      22$& $  -1.7_{-    0.3}^{+   0.3}$& $  -1.9_{- 0.3}^{+   0.4}$& $     -0.72_{     -0.36}^{+      0.31}$& $     23.92_{     -0.10}^{+      0.11}$&$     24.65_{     -0.20}^{+      0.25}$&$3.2^{+0.4}_{-0.3}$\textdagger\\
UVISTA-185070 & $  6.78_{-  0.19}^{+  0.12}$ & $ 24.91_{ -0.11}^{+  0.12}$ & $-22.03_{ -0.11}^{+  0.12}$ & $      19$& $  -1.6_{-    0.3}^{+   0.2}$& $  -1.7_{-    0.4}^{+   0.3}$& $     -1.36_{     -0.29}^{+      0.28}$& $     23.92_{     -0.10}^{+      0.11}$&$     25.27_{     -0.16}^{+      0.19}$&$1.3^{+0.2}_{-0.2}$\\
UVISTA-28495 & $  7.12_{-  0.12}^{+  0.14}$ & $ 25.01_{ -0.21}^{+  0.28}$ & $-22.02_{ -0.21}^{+  0.28}$ & $      19$& $  -1.9_{-    0.2}^{+   0.3}$& $  -2.1_{-   0.5}^{+   0.4}$& $\phantom{-}      0.57_{     -0.35}^{+      0.43}$& $     24.79_{     -0.27}^{+      0.35}$&$     24.22_{     -0.08}^{+      0.08}$&$1.4^{+0.4}_{-0.4}$\textdagger\\
UDS-35314 & $  6.70_{-  0.10}^{+  0.12}$ & $ 25.07_{ -0.14}^{+  0.17}$ & $-22.01_{ -0.14}^{+  0.17}$ & $      19$& $  -2.5_{-    0.4}^{+   0.3}$& $  -2.6_{-0.5}^{+   0.4}$& $     -0.69_{     -1.16}^{+      0.81}$& $     24.60_{     -0.25}^{+      0.32}$&$     25.29_{     -0.23}^{+      0.30}$&$0.7^{+0.2}_{-0.2}$\\
UDS-118717 & $  6.51_{-  0.13}^{+  0.06}$ & $ 24.87_{ -0.13}^{+  0.13}$ & $-21.91_{ -0.13}^{+  0.13}$ & $      17$& $  -1.7_{-    0.3}^{+   0.4}$& $  -1.7_{-  0.4}^{+   0.4}$& $     -0.08_{     -0.45}^{+      0.45}$& $     24.26_{     -0.19}^{+      0.23}$&$     24.34_{     -0.21}^{+      0.27}$&$0.8^{+0.2}_{-0.2}$\\
UVISTA-328993 & $  6.11_{-  0.14}^{+  0.13}$ & $ 24.70_{ -0.12}^{+  0.14}$ & $-21.90_{ -0.12}^{+  0.14}$ & $      17$& $  -1.8_{-    0.5}^{+   0.5}$& $  -2.7_{-    0.9}^{+   0.8}$& $     -0.10_{     -0.29}^{+      0.30}$& $     24.14_{     -0.17}^{+      0.19}$&$     24.25_{     -0.11}^{+      0.12}$&$1.2^{+0.2}_{-0.2}$\\
UVISTA-136380 & $  7.09_{-  0.12}^{+  0.10}$ & $ 25.34_{ -0.19}^{+  0.23}$ & $-21.77_{ -0.19}^{+  0.23}$ & $      15$& $  -2.5_{-    0.4}^{+   0.3}$& $  -2.8_{-    0.6}^{+   0.5}$& $\phantom{-}      1.04_{     -0.66}^{+      0.83}$& $     26.16_{     -0.40}^{+      0.63}$&$     25.12_{     -0.21}^{+      0.26}$&$0.7^{+0.3}_{-0.3}$\\
UVISTA-211127 & $  7.02_{-  0.08}^{+  0.08}$ & $ 25.46_{ -0.18}^{+  0.21}$ & $-21.55_{ -0.18}^{+  0.21}$ & $      12$& $  -2.5_{-    0.3}^{+   0.3}$& $  -2.4_{-    0.4}^{+   0.5}$& $\phantom{-}      1.00_{     -0.41}^{+      0.57}$& $     25.11_{     -0.34}^{+      0.50}$&$     24.10_{     -0.07}^{+      0.07}$&$0.5^{+0.2}_{-0.3}$\\
UVISTA-170216 & $  6.52_{-  0.17}^{+  0.14}$ & $ 25.41_{ -0.20}^{+  0.24}$ & $-21.48_{ -0.20}^{+  0.24}$ & $      11$& $  -2.0_{-    0.5}^{+   0.4}$& $  -2.0_{-    0.6}^{+   0.7}$& $     -0.66_{     -0.69}^{+      0.73}$& $     24.71_{     -0.32}^{+      0.46}$&$     25.37_{     -0.27}^{+      0.37}$&$1.1^{+0.3}_{-0.3}$\textdagger\\
UVISTA-35327 & $  6.71_{-  0.21}^{+  0.20}$ & $ 25.50_{ -0.34}^{+  0.50}$ & $-21.47_{ -0.34}^{+  0.50}$ & $      11$& $  -2.2_{-    0.7}^{+   0.7}$& $  -1.8_{-    1.0}^{+   1.0}$& --& --&--&$0.8^{+0.8}_{-0.8}$\\
UVISTA-305036 & $  6.91_{-  0.30}^{+  0.16}$ & $ 25.52_{ -0.24}^{+  0.30}$ & $-21.42_{ -0.24}^{+  0.30}$ & $      11$& $  -1.4_{-    0.3}^{+   0.2}$& $  -1.4_{-    0.4}^{+   0.3}$& $\phantom{-}      0.52_{     -0.43}^{+      0.55}$& $     24.82_{     -0.32}^{+      0.45}$&$     24.30_{     -0.10}^{+      0.11}$&$1.6^{+0.3}_{-0.3}$\textdagger\\
UVISTA-104600 & $  6.47_{-  0.07}^{+  0.07}$ & $ 25.41_{ -0.31}^{+  0.45}$ & $-21.35_{ -0.31}^{+  0.45}$ & $      10$& $  -1.8_{-    0.3}^{+   0.2}$& $  -2.0_{-    0.3}^{+   0.3}$& $     -0.00_{     -0.21}^{+      0.22}$& $     23.98_{     -0.14}^{+      0.16}$&$     23.98_{     -0.06}^{+      0.07}$&$0.2^{+0.4}_{-0.2}$\\
UVISTA-268576 & $  6.59_{-  0.23}^{+  0.12}$ & $ 25.69_{ -0.22}^{+  0.26}$ & $-21.29_{ -0.22}^{+  0.26}$ & $       9$& $  -2.1_{-    0.4}^{+   0.4}$& $  -1.6_{-    0.7}^{+   0.6}$& $<     -1.43$& $     24.66_{     -0.22}^{+      0.28}$&$>     26.09$&$0.8^{+0.2}_{-0.2}$\\
UVISTA-282894 & $  7.02_{-  0.13}^{+  0.13}$ & $ 25.73_{ -0.28}^{+  0.38}$ & $-21.12_{ -0.28}^{+  0.38}$ & $       8$& $  -1.9_{-    0.3}^{+   0.4}$& $  -1.5_{-    0.5}^{+   0.6}$& $\phantom{-}      0.13_{     -0.59}^{+      0.81}$& $     25.40_{     -0.40}^{+      0.64}$&$     25.27_{     -0.17}^{+      0.20}$&$0.7^{+0.5}_{-0.4}$\\
UVISTA-271028 & $  6.11_{-  0.15}^{+  0.17}$ & $ 25.87_{ -0.25}^{+  0.32}$ & $-20.67_{ -0.25}^{+  0.32}$ & $       5$& $  -1.8_{-    0.3}^{+   0.3}$& $  -1.9_{-    0.5}^{+   0.4}$& $>     -0.03$& $>     25.35$&$     25.38_{     -0.24}^{+      0.31}$&$0.6^{+0.3}_{-0.6}$\\

\hline
\end{tabular}\label{table:sample}
\end{table*}

\subsection{Nebular emission}\label{sect:nebular}

\begin{figure}
\includegraphics[width = 0.49\textwidth]{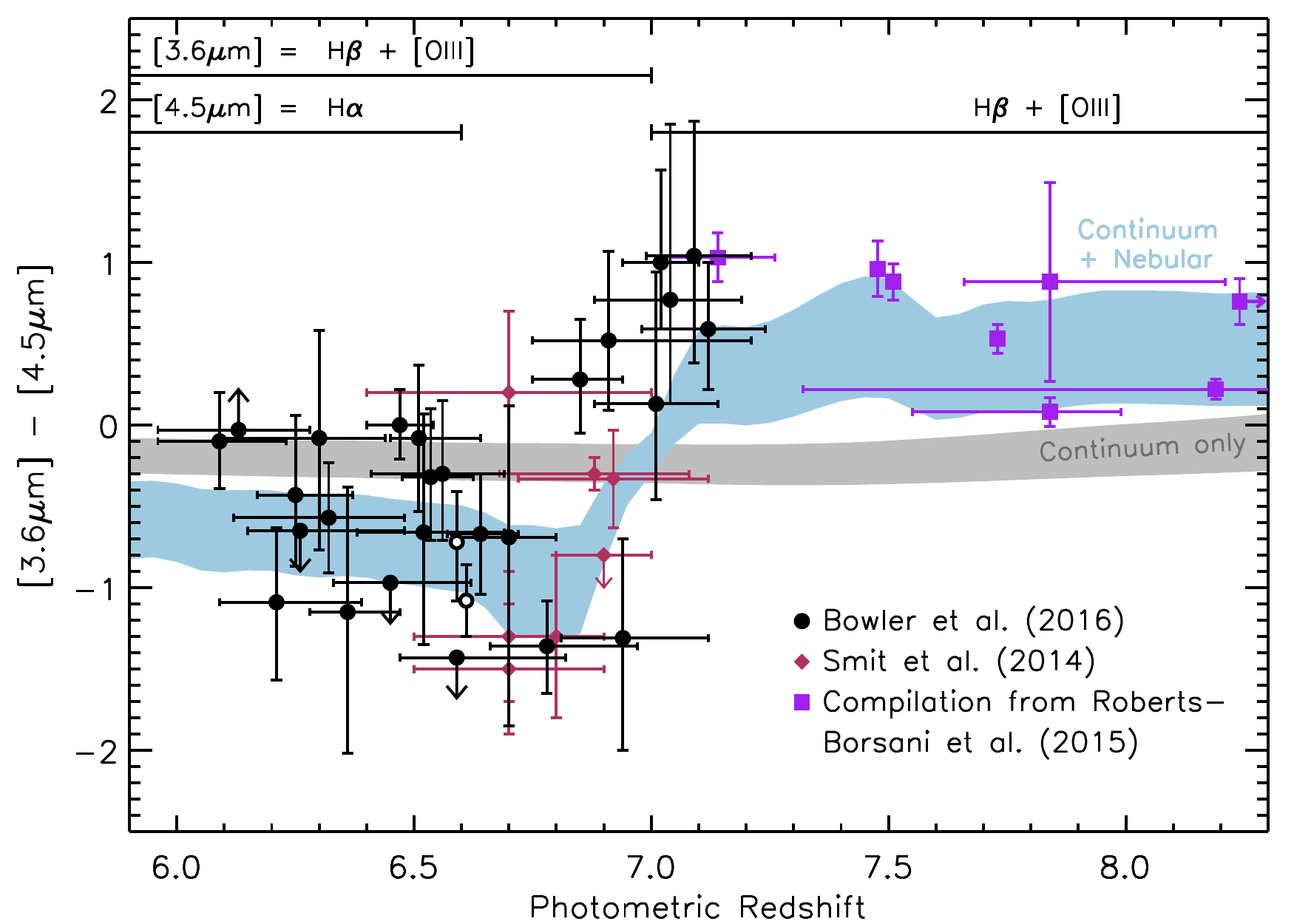}
\caption{The IRAC \chone -- \chtwo~colour against photometric redshift for the full~\citet{Bowler2014} sample of bright $z \simeq 7$ LBGs (black circles).
The shaded regions show the predicted colour for a range of~\citet{Bruzual2003} SED models with (blue) and without (grey) the inclusion of rest-frame optical nebular emission lines with a combined H$\beta$ + [{\sc OIII}] equivalent width in the range $637 < EW_{0} < 1582\,$\AA.
The nebular emission lines present in each~\emph{Spitzer} filter as a function of redshift is illustrated in the upper region of the plot.
The open circles show the colours of the LAEs `Himiko' and `CR7', which are spectroscopically confirmed to be at $z = 6.6$.
We also show the colours of galaxies presented by~\citet{Smit2014} (red diamonds) and in a compilation by~\citet{RobertsBorsani2015} (purple squares).
To focus on the range of interest for this work, we plot the $z = 8.68$ object presented in~\citet{RobertsBorsani2015} and~\citet{Zitrin2015} at $z = 8.25$ with a rightward arrow.
Limit arrows are shown when the LBG was undetected in either the \chone~or \chtwo~bands at $< 2\sigma$ significance.
}\label{fig:nebular}
\end{figure}

In the last few years the availability of deep~\emph{Spitzer}/IRAC photometry over the multi-wavelength survey fields imaged by~\emph{HST} has allowed the first investigation into the rest-frame optical emission in LBGs at $z > 5$ (e.g. in the GOODS fields;~\citealp{Stark2013}).
In addition to constraining the galaxy mass, the rest-frame optical wavelength range observed in the \chone~and \chtwo~bands also includes several strong nebular emission lines such as H$\alpha$, H$\beta$ and [{\sc OIII}] $\lambda \lambda 4959, 5007$.
With increasing redshift, these emission lines pass through the~\emph{Spitzer}/IRAC bands producing characteristic offsets in the \IRACcolour colour as compared to a continuum only model (see Fig.~\ref{fig:nebular}).
The presence of strong rest-frame optical nebular emission lines in the SEDs of $3 < z < 7$ LBGs has now been demonstrated by several studies~\citep{Stark2013, deBarros2014, Smit2014, Smit2016, MarmolQueralto2015}.
These previous results however, have focused on relatively faint LBGs selected from within the CANDELS fields or cluster-lensing fields imaged by~\emph{HST}.

Using the deep~\emph{Spitzer}/SPLASH data over the UltraVISTA/COSMOS and UDS/SXDS fields, we can investigate the prevalence of nebular emission in our sample of bright $ z\simeq 7$ objects.
In~\citet{Bowler2014} we performed a similar analysis using the shallower SPLASH data available at that time.
Here we exploit the final SPLASH imaging using the deconfusion analysis described in~\citet{McLeod2015} and discussed in Section~\ref{sect:spitzer}.
The majority of the galaxies in our sample are strongly detected in the~\emph{Spitzer}/IRAC data, without the need for stacking or strong gravitational lensing (e.g.~\citealp{Smit2014}).
Furthermore, from the~\emph{Spitzer} photometry presented in Table~\ref{table:sample} it is evident that the IRAC colour is either significantly redder or bluer than that predicted by a continuum only model (\IRACcolour $\simeq 0.0$).
In Fig.~\ref{fig:nebular} we show the \IRACcolour colour as a function of photometric redshift for our sample.
The observed IRAC colours are fully consistent with the contamination of these photometric bands by strong rest-frame optical nebular emission lines, which are redshifted through these filters from $ z \simeq 6.5$ to $ z \simeq 7.5$, resulting in a change in sign of the \IRACcolour colour.
We also show $z \simeq 7$ results from~\citet{Smit2014}, derived from a sample of lensed galaxies (with intrinsic $M_{\rm UV} \sim -20.5$) and from higher redshift objects presented in a compilation by~\citet{RobertsBorsani2015} which includes the the objects at $z = 7.51$~\citep{Finkelstein2013}, $z = 7.73$~\citep{Oesch2015} and $z = 8.68$~\citep{Zitrin2015}.
The~\citet{RobertsBorsani2015} galaxies were selected based on their red IRAC colour, whereas the~\citet{Finkelstein2013} and~\citet{Oesch2015} objects were initially selected based on their rest-frame UV emission.
We also show the expected redshift-dependent \IRACcolour colour predicted by~\citet{Bruzual2003} models with and without rest-frame optical emission lines from H$\alpha$, H$\beta$ and [{\sc OIII}].
The range in rest-frame H$\beta + $\oiii equivalent widths shown ($637$\AA $< EW_{0} < 1582\,$\AA), was chosen to match the upper and lower limits found by~\citet{Smit2014}.
While there is still scatter in the \IRACcolour colours derived for our sample as shown in Fig.~\ref{fig:nebular}, we find a clear change in the sign of the colour with photometric redshift.
Below $z \simeq 6.8$ we find exclusively blue \IRACcolour $\lesssim 0$ values, corresponding to either no rest-frame optical emission lines or a stronger contribution of H$\beta$ + \oiii~in the \chone~band compared to H$\alpha$ in the \chtwo~band.
The \IRACcolour colour agrees well with previous results from~\citet{Smit2014} and other studies at $z > 7$ compiled by~\citet{RobertsBorsani2015}.
Around the sharp colour change at $z =6.8$--$ 7.1$ we find evidence for even stronger rest-frame emission lines than our most extreme model shown in Fig.~\ref{fig:nebular}, implying $EW_{0}$(H$\beta + $\oiii)$> 1600\,$\AA.
The extreme LAEs `Himiko' and `CR7' show similar colours to our sample, however as illustrated by the smaller error bars on their photometry, they are particularly bright in the \chone~band, as you would expect for the increased nebular emission from these galaxies.
These results show that strong ($EW_{0} $(H$\beta + $\oiii)$> 600\,$\AA) rest-frame optical emission lines appear to be common in bright $ z \sim 7$ LBGs as well as in fainter galaxies at this epoch~\citep{Smit2014, deBarros2014, Stark2013}.
Furthermore, the surprisingly high success rate at detecting Lyman-$\alpha$ emission in  $z > 7.5$ LBGs that show red IRAC colours~\citep{RobertsBorsani2015, Zitrin2015, Oesch2015, Finkelstein2015} suggests that the galaxies in the sample presented here could also show detectible Lyman-$\alpha$ emission.

We note that the photometric redshifts for our sample were calculated without any nebular emission lines in the spectrum, using the optical and near-infrared photometry only.
With the addition of a strong Lyman-$\alpha$ emission line in the galaxy SED, the spectroscopic redshift could be higher (typically $\Delta z \simeq 0.1$, to a maximum of $\Delta z \simeq 0.4$;~\citealp{Bowler2014}) than the line-free photometric redshift, due to the additional flux present in the broad-band photometry.
Given the presence of two extreme LAEs over the fields (Himiko and CR7) detected within the narrow-band NB921 from Subaru, which selects Lyman-$\alpha$ emitters in the redshift range $z = 6.6 \pm 0.05$, it is feasible that similar LAEs exist in the field at $z > 6.6$.
In Fig.~\ref{fig:nebular} we find that several objects (including our brightest object ID304416) with \IRACcolour $> 0.0$ at $z_{\rm phot} \sim 6.9$ are offset from the predicted redshift-colour region (shown as the shaded blue curve), suggesting that the photometric redshift could be underestimated due to contamination by strong Lyman-$\alpha$ emission.

\begin{figure*}
\begin{center}

\includegraphics[width = 0.15\textwidth, trim = 0.6cm 0.2cm 0.6cm 0.4cm ]{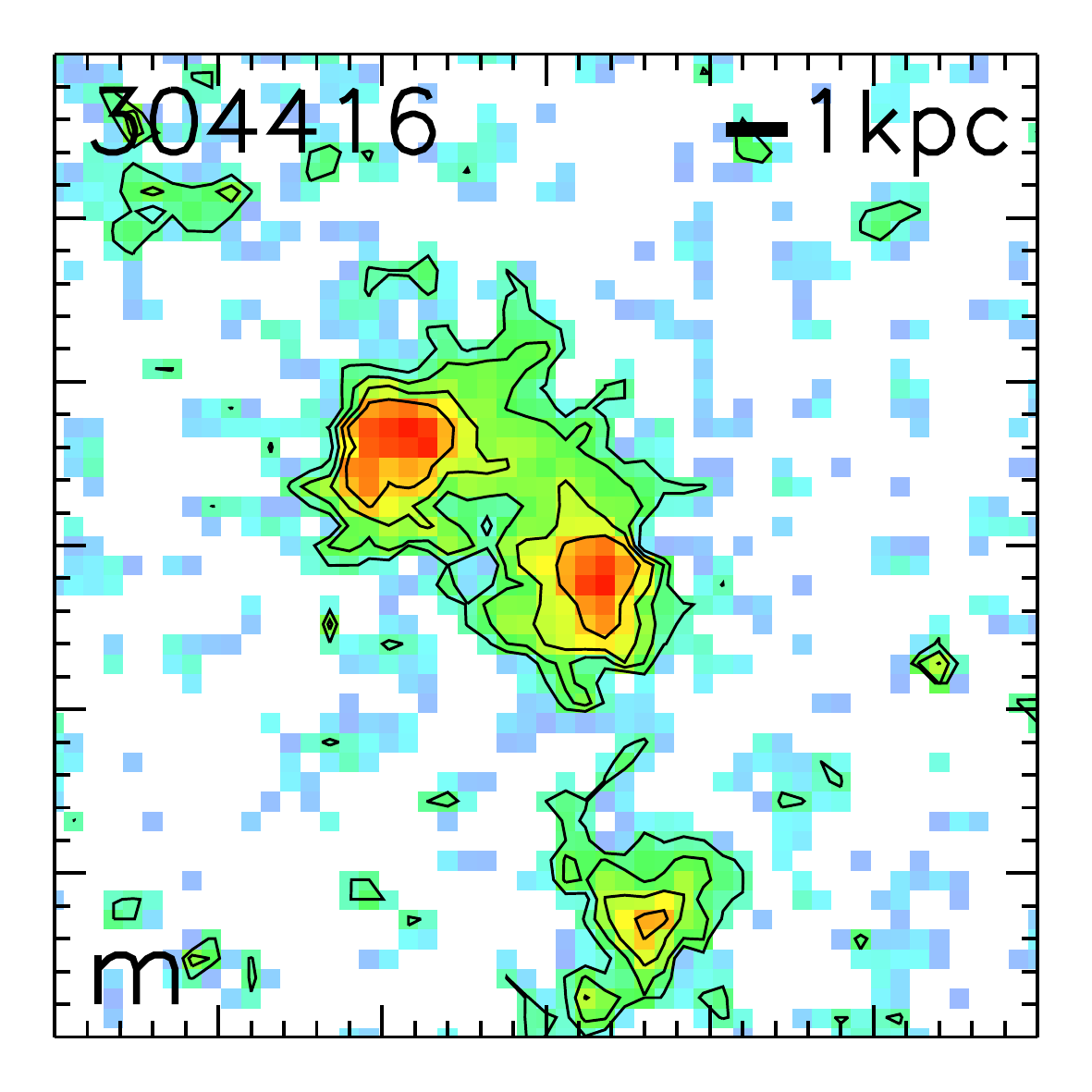}\hspace{0.05cm}
\includegraphics[width = 0.15\textwidth, trim = 0.6cm 0.2cm 0.6cm 0.4cm ]{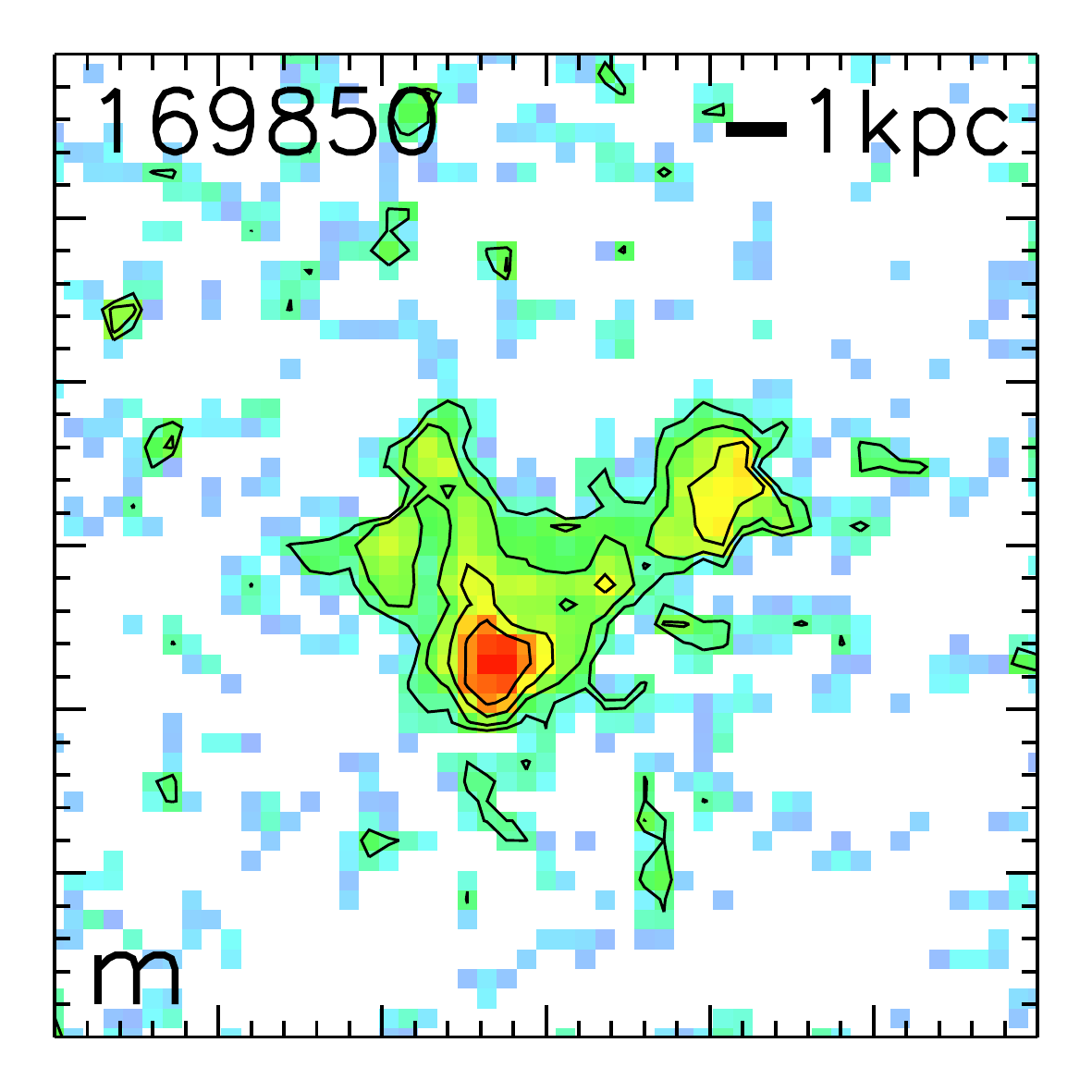}\hspace{0.05cm}
\includegraphics[width = 0.15\textwidth, trim = 0.6cm 0.2cm 0.6cm 0.4cm ]{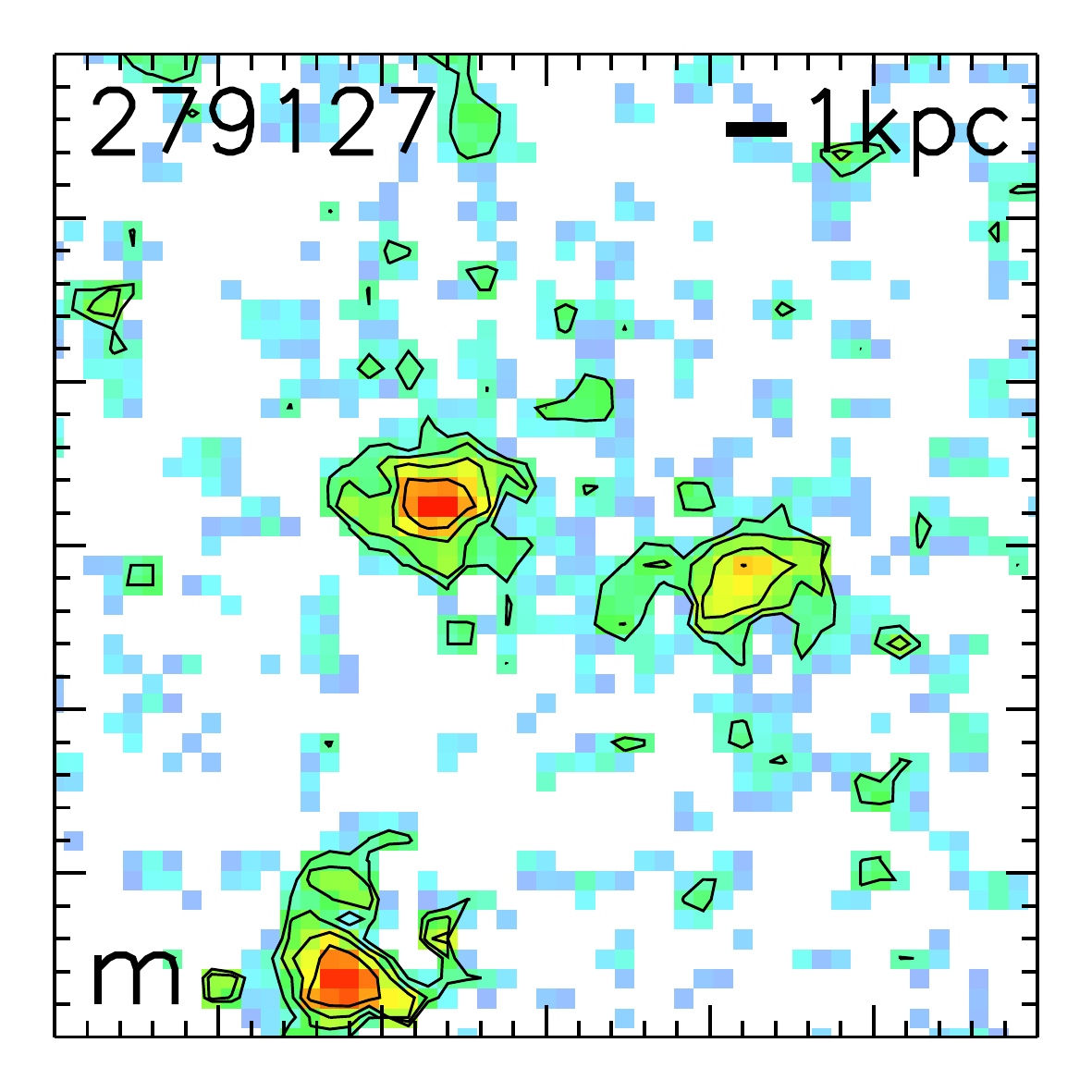}\hspace{0.05cm}
\includegraphics[width = 0.15\textwidth, trim = 0.6cm 0.2cm 0.6cm 0.4cm ]{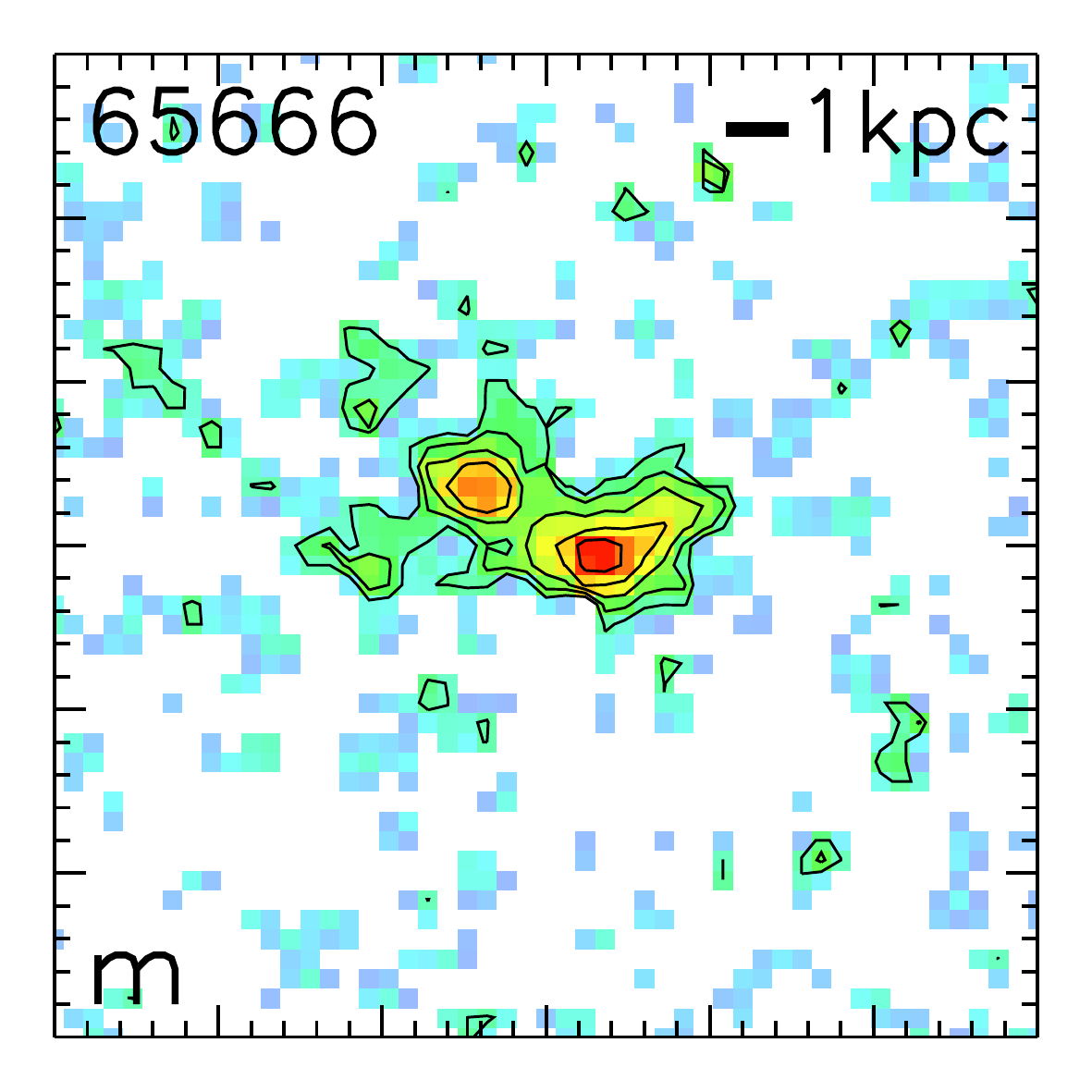}\hspace{0.05cm}
\includegraphics[width = 0.15\textwidth, trim = 0.6cm 0.2cm 0.6cm 0.4cm ]{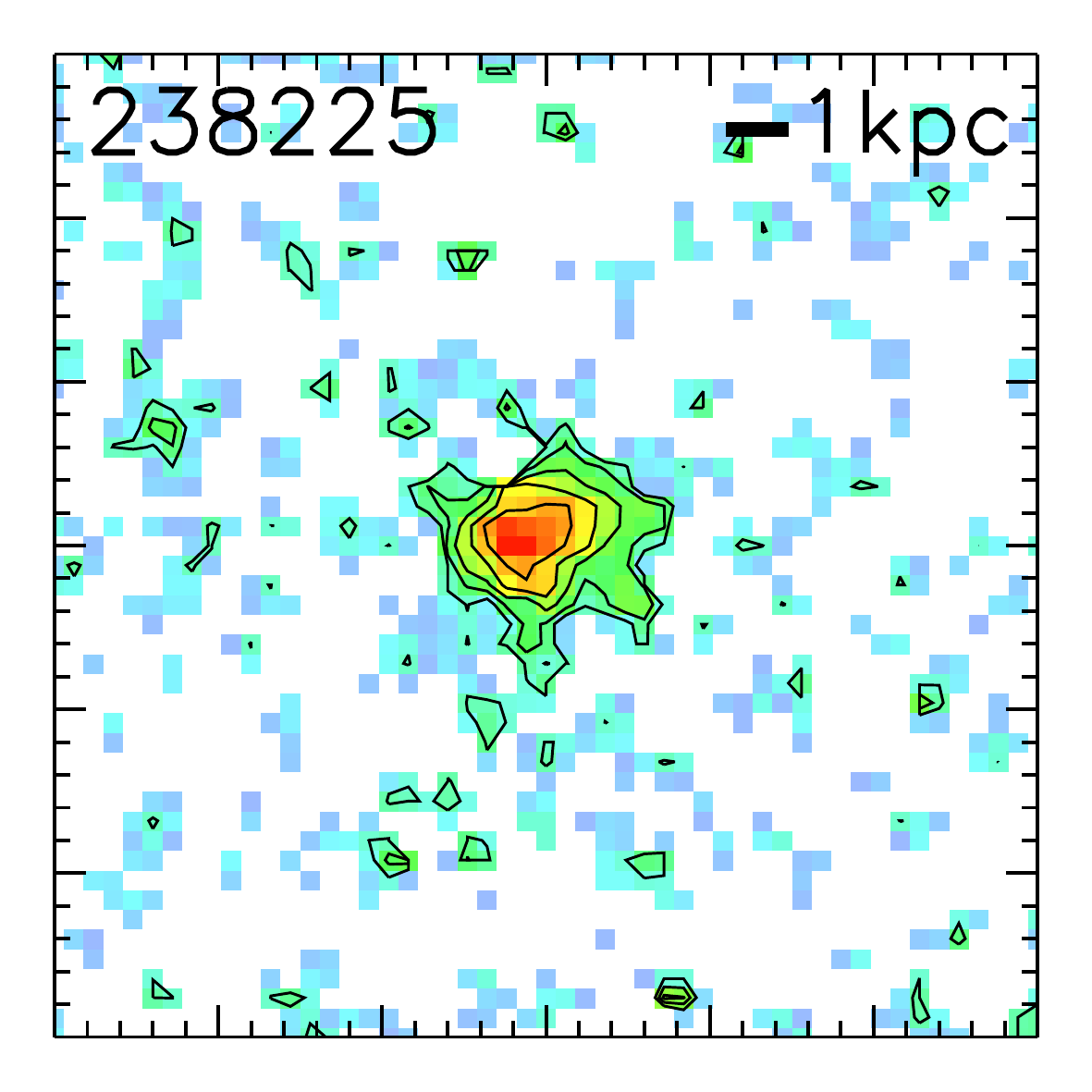}\hspace{0.05cm}
\includegraphics[width = 0.15\textwidth, trim = 0.6cm 0.2cm 0.6cm 0.4cm ]{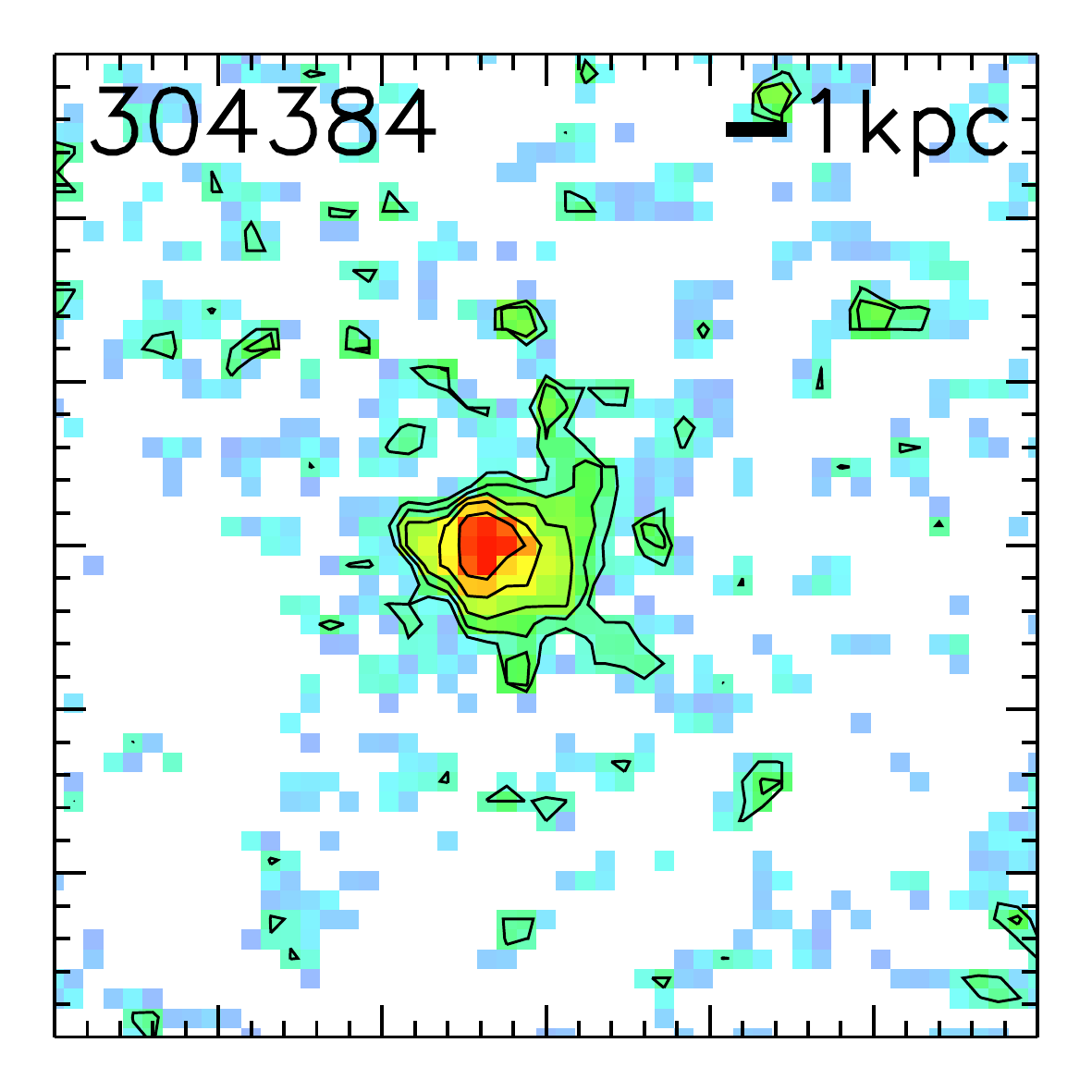}\\
\includegraphics[width = 0.15\textwidth, trim = 0.6cm 0.2cm 0.6cm 0.4cm ]{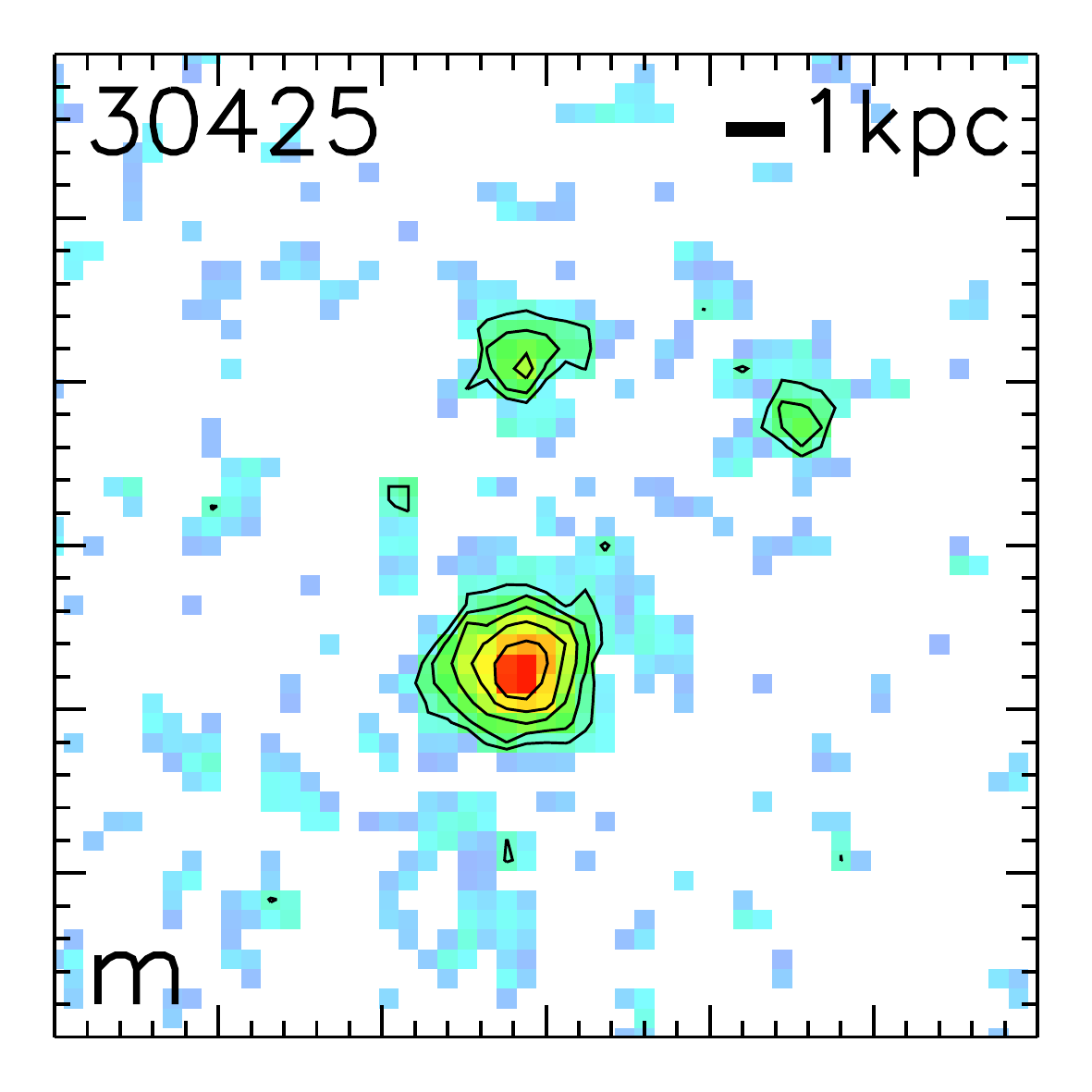}\hspace{0.05cm}
\includegraphics[width = 0.15\textwidth, trim = 0.6cm 0.2cm 0.6cm 0.4cm ]{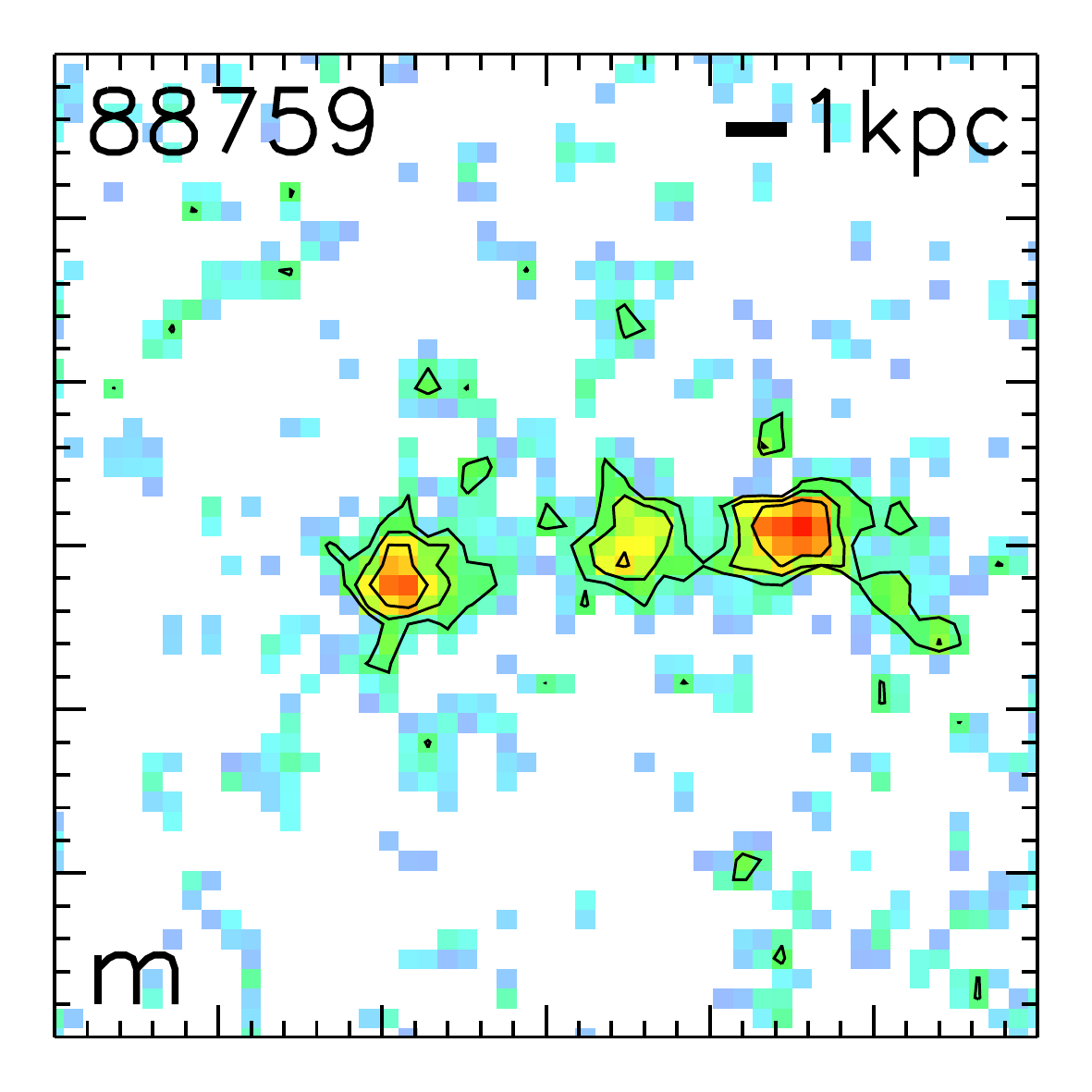}\hspace{0.05cm}
\includegraphics[width = 0.15\textwidth, trim = 0.6cm 0.2cm 0.6cm 0.4cm ]{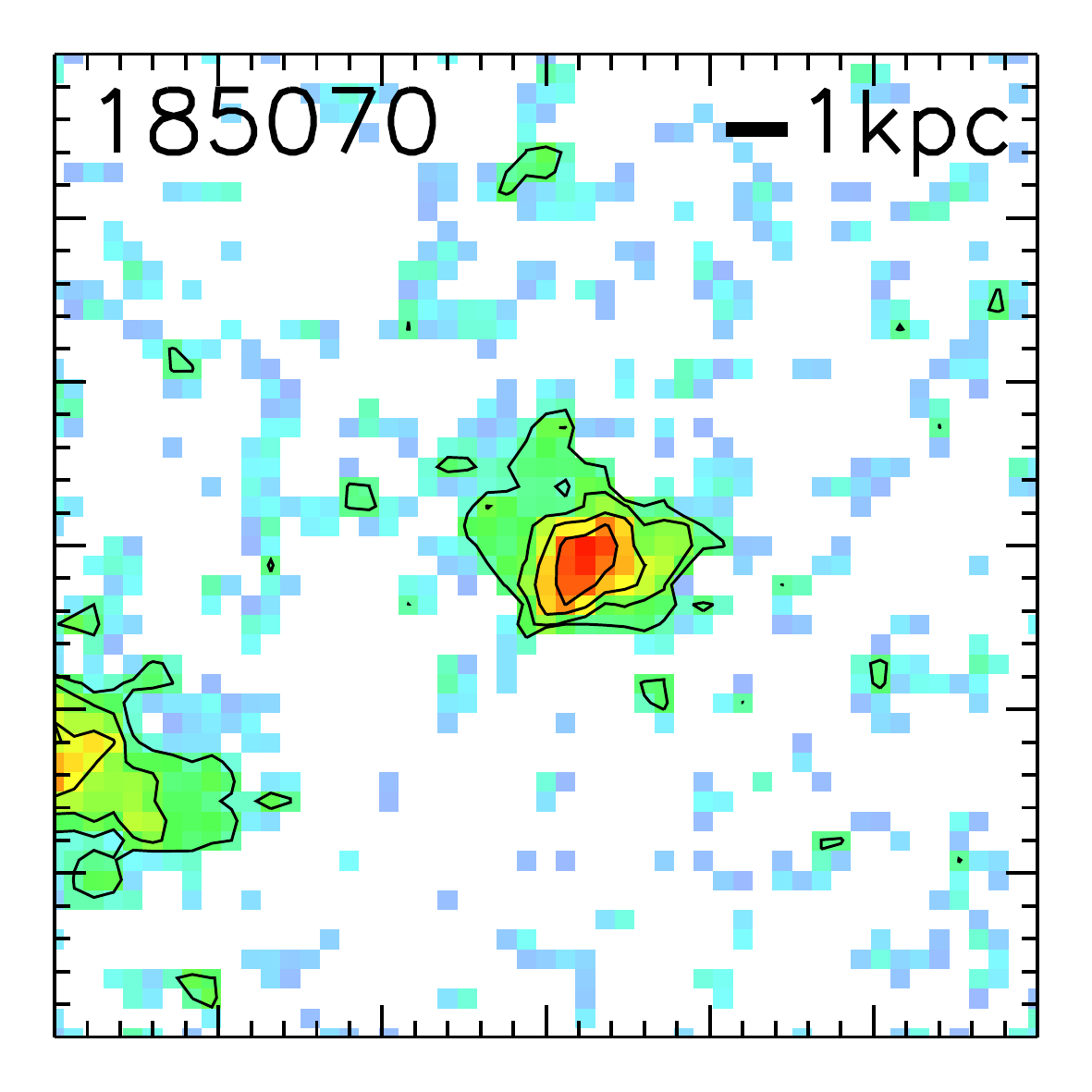}\hspace{0.05cm}
\includegraphics[width = 0.15\textwidth, trim = 0.6cm 0.2cm 0.6cm 0.4cm ]{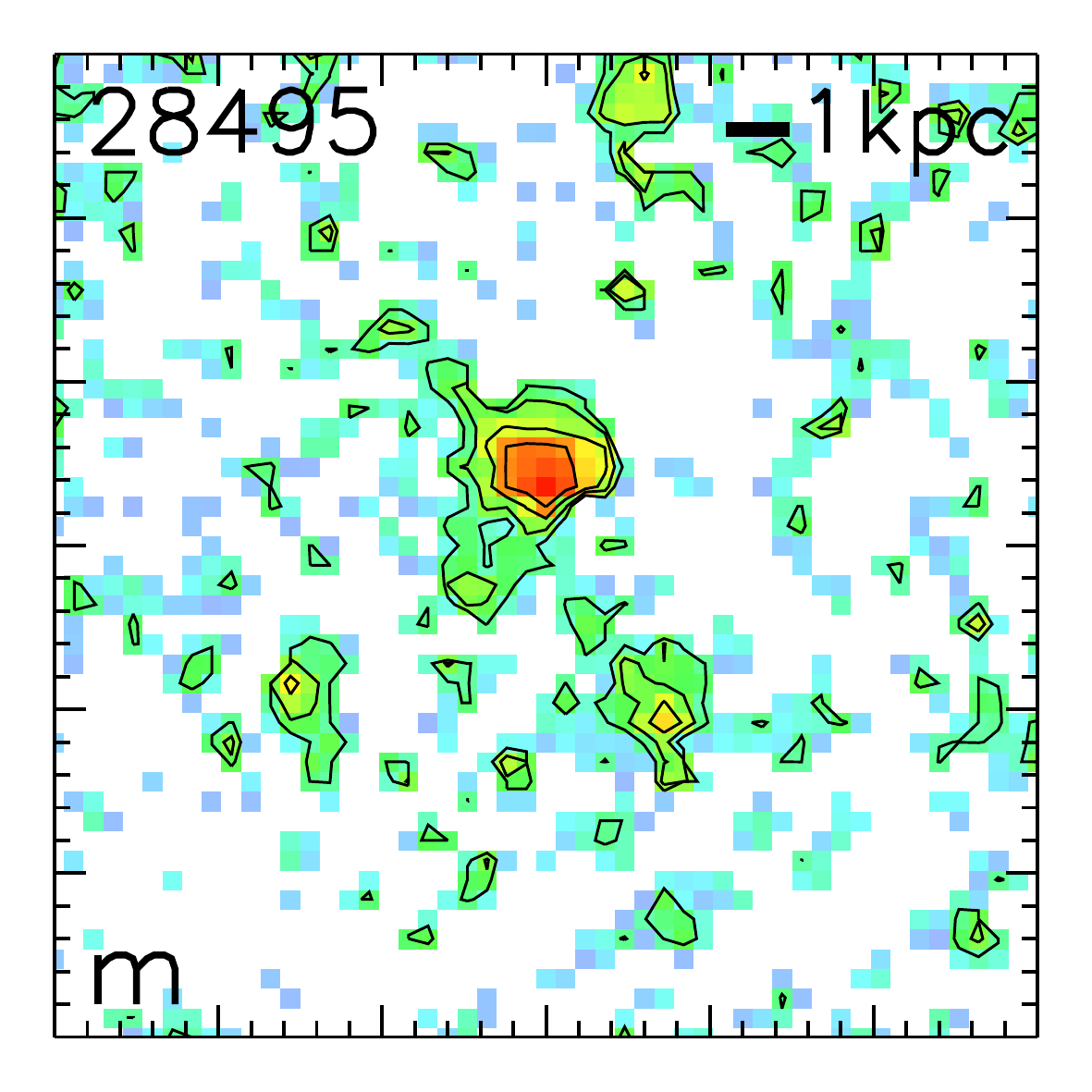}\hspace{0.05cm}
\includegraphics[width = 0.15\textwidth, trim = 0.6cm 0.2cm 0.6cm 0.4cm ]{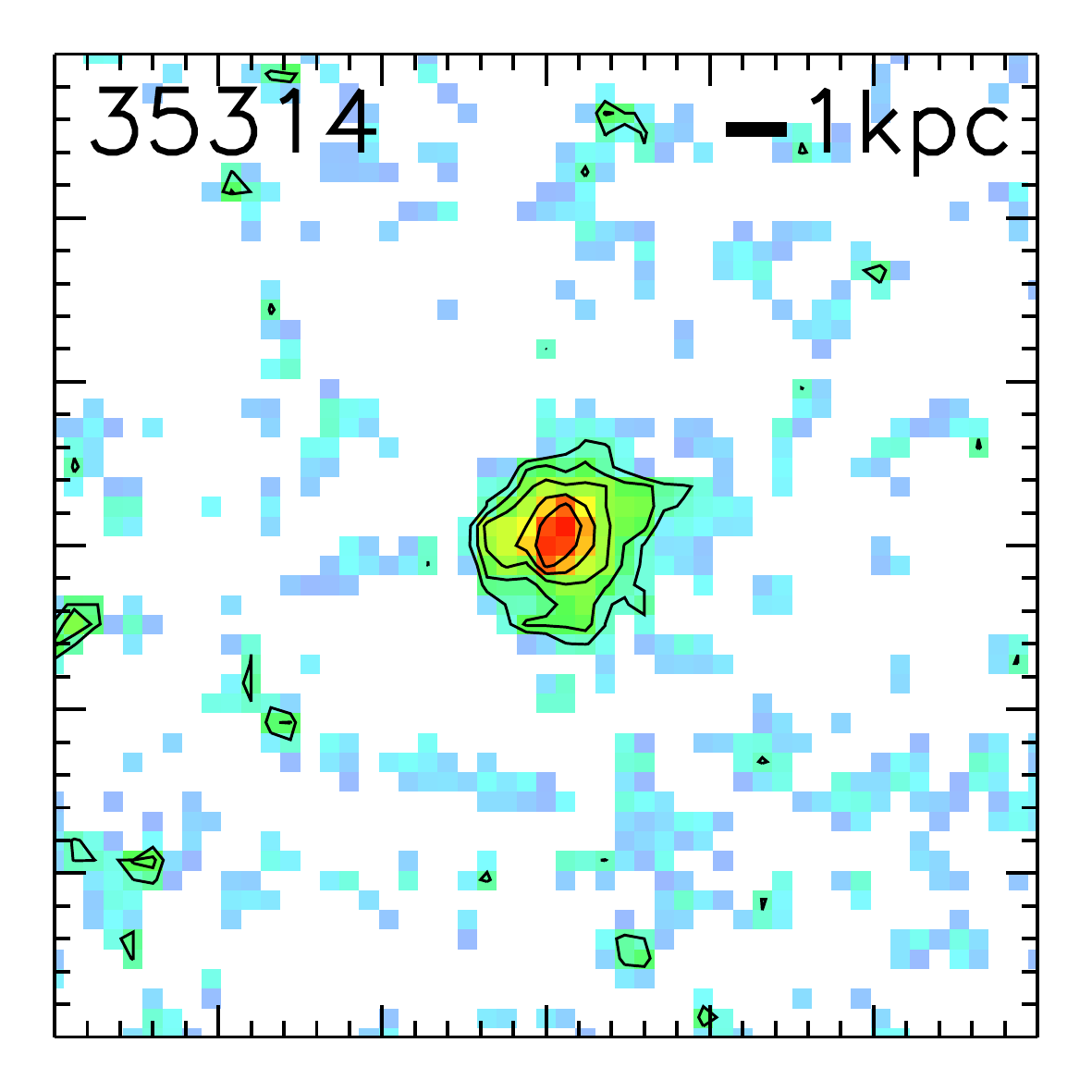}\hspace{0.05cm}
\includegraphics[width = 0.15\textwidth, trim = 0.6cm 0.2cm 0.6cm 0.4cm ]{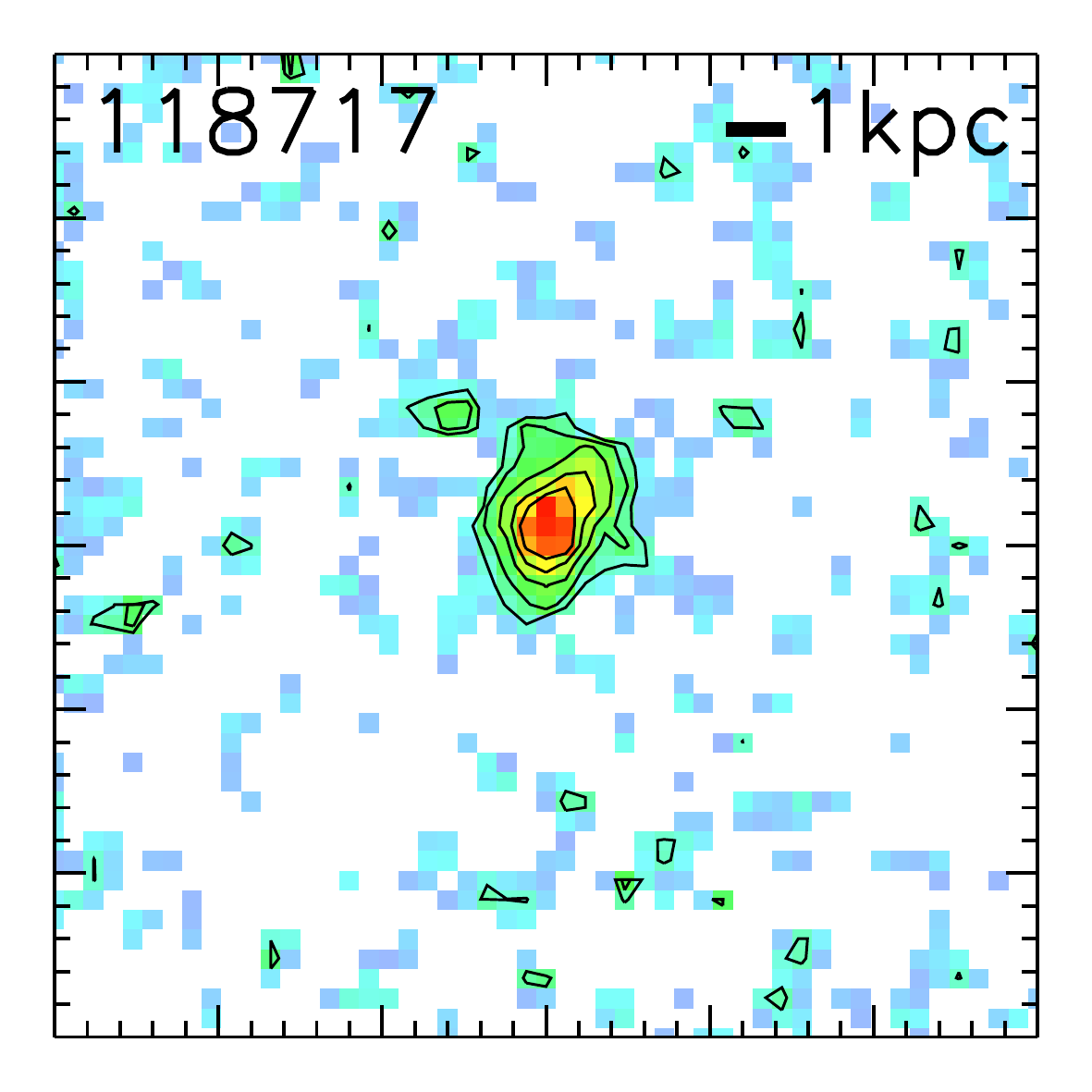}\\
\includegraphics[width = 0.15\textwidth, trim = 0.6cm 0.2cm 0.6cm 0.4cm ]{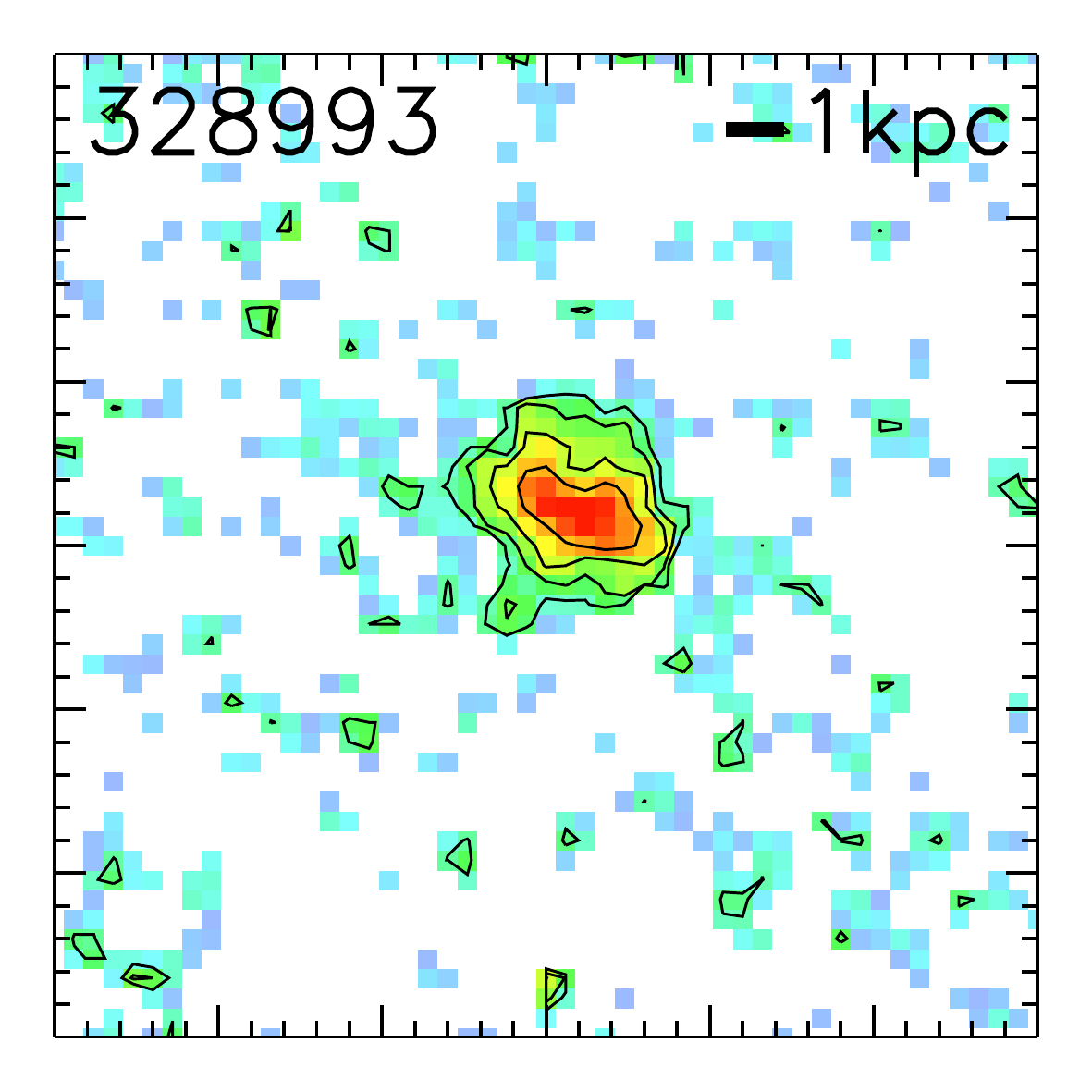}\hspace{0.05cm}
\includegraphics[width = 0.15\textwidth, trim = 0.6cm 0.2cm 0.6cm 0.4cm]{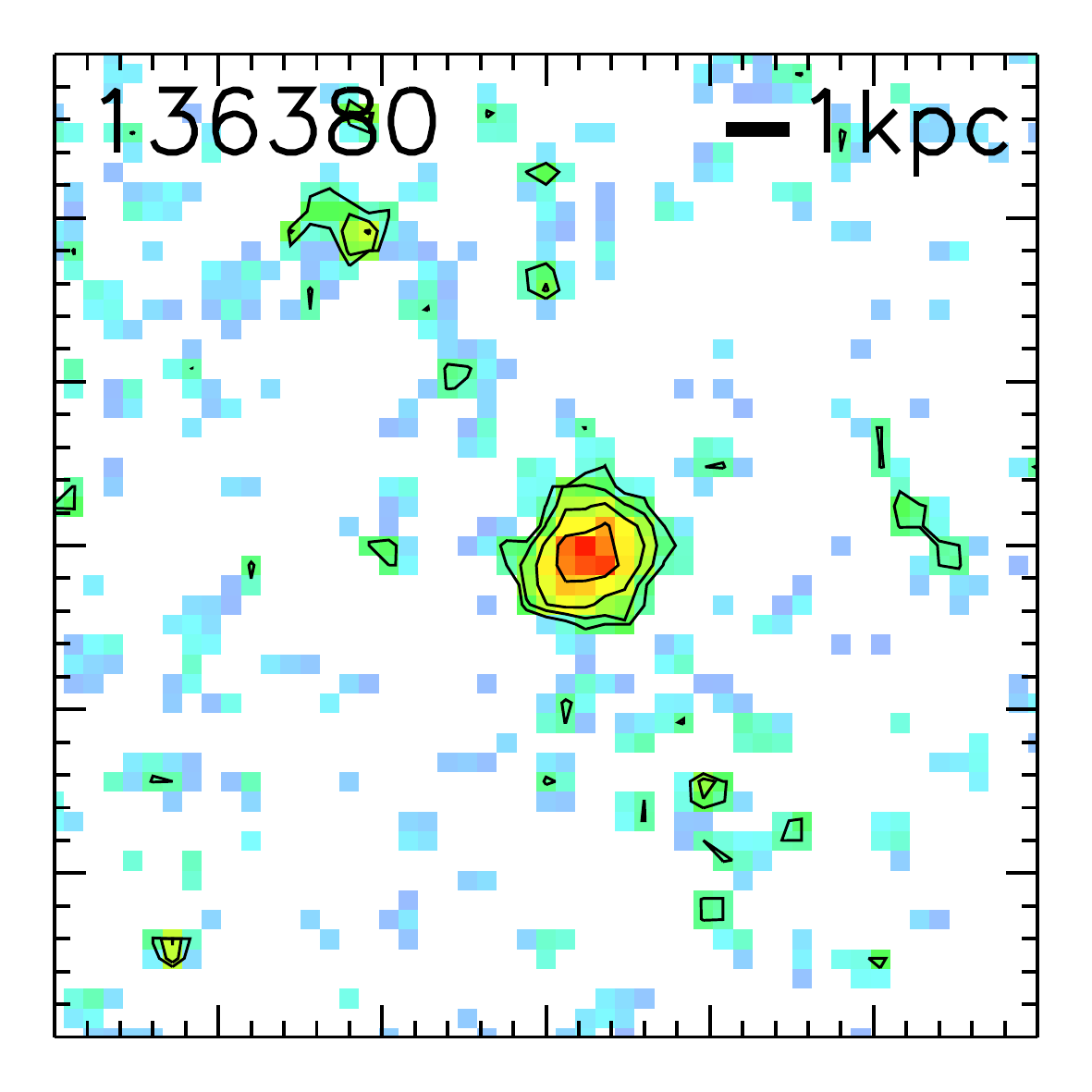}\hspace{0.05cm}
\includegraphics[width = 0.15\textwidth, trim = 0.6cm 0.2cm 0.6cm 0.4cm ]{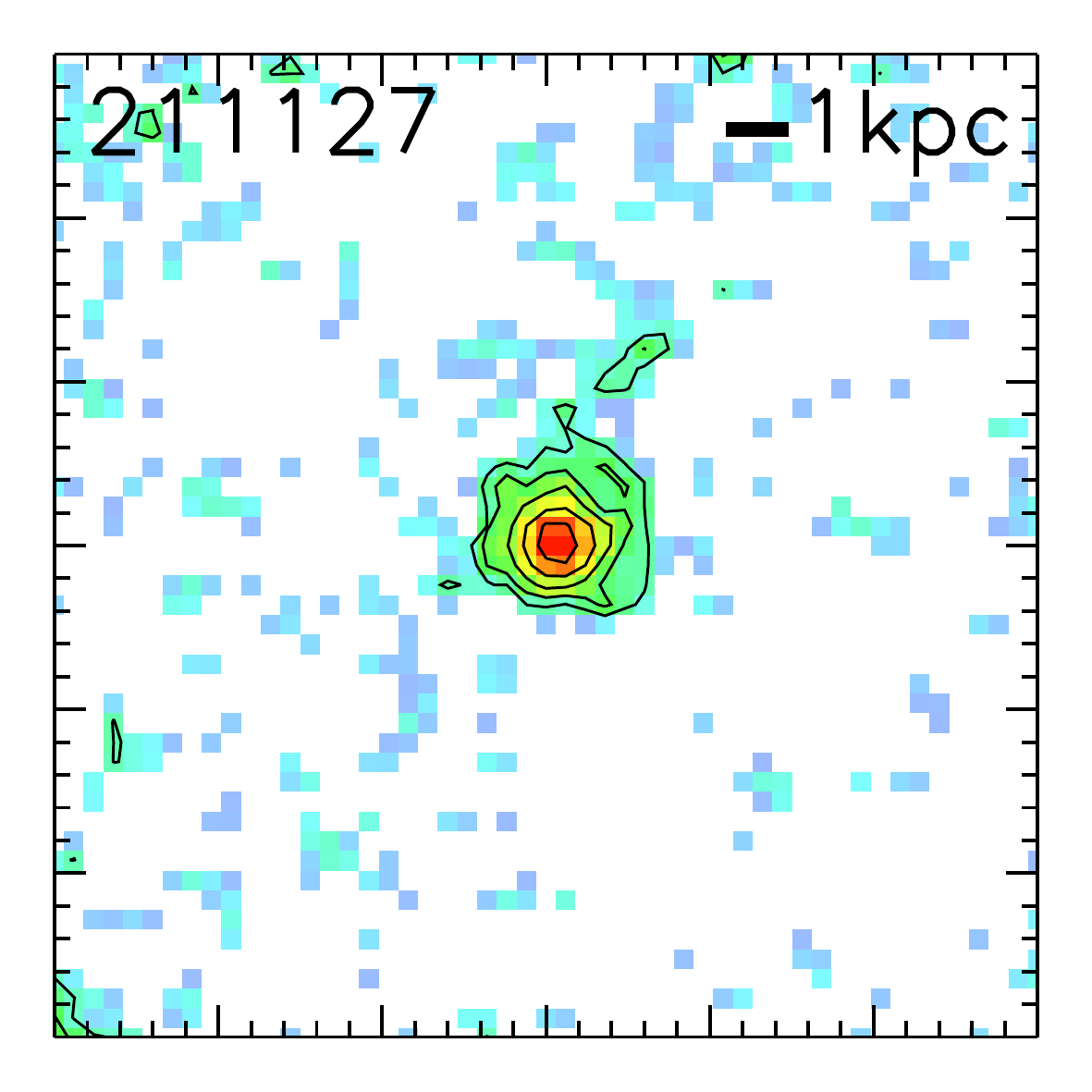}\hspace{0.05cm}
\includegraphics[width = 0.15\textwidth, trim = 0.6cm 0.2cm 0.6cm 0.4cm ]{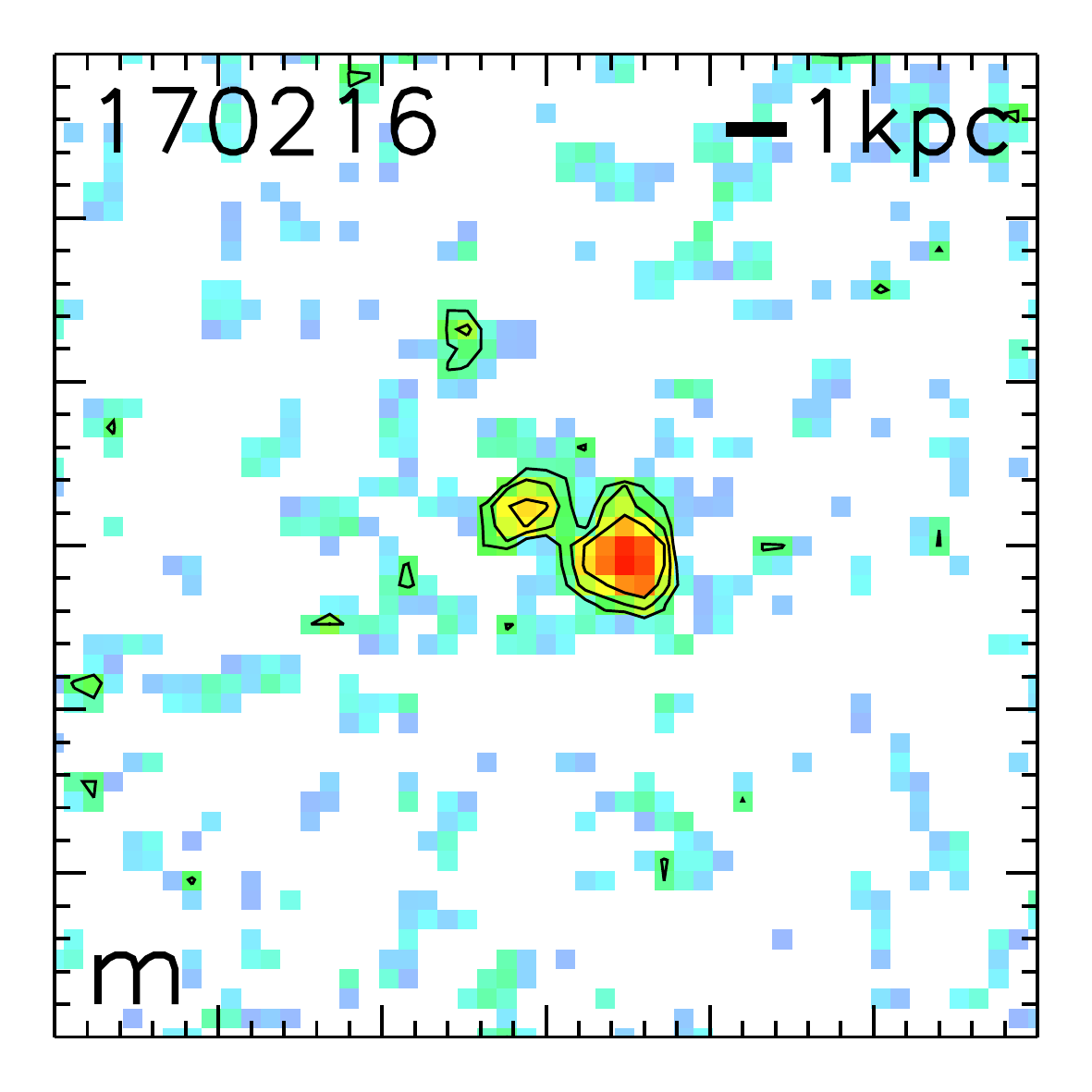}\hspace{0.05cm}
\includegraphics[width = 0.15\textwidth, trim = 0.6cm 0.2cm 0.6cm 0.4cm ]{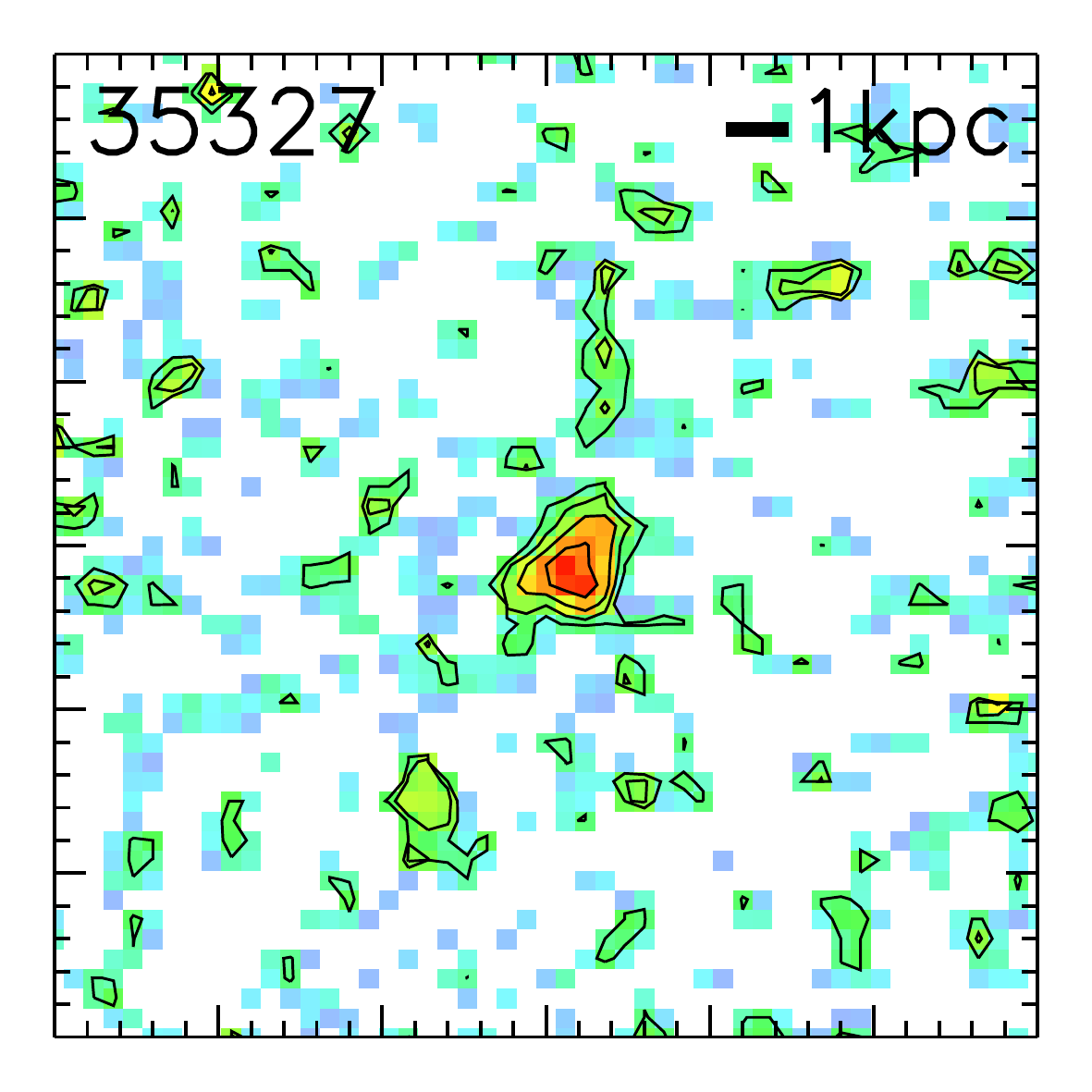}\hspace{0.05cm}
\includegraphics[width = 0.15\textwidth, trim = 0.6cm 0.2cm 0.6cm 0.4cm ]{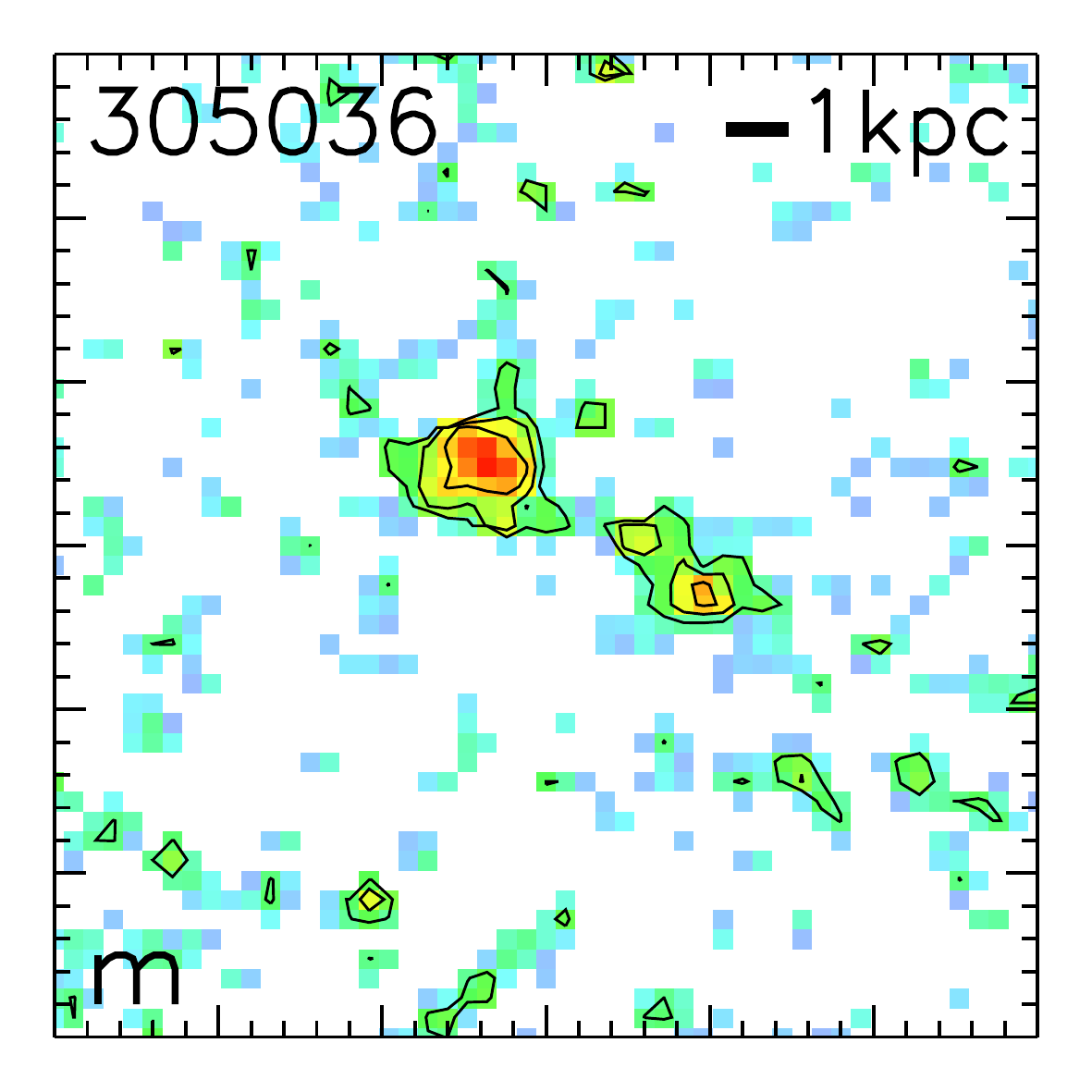}\\
\includegraphics[width = 0.15\textwidth, trim = 0.6cm 0.2cm 0.6cm 0.4cm ]{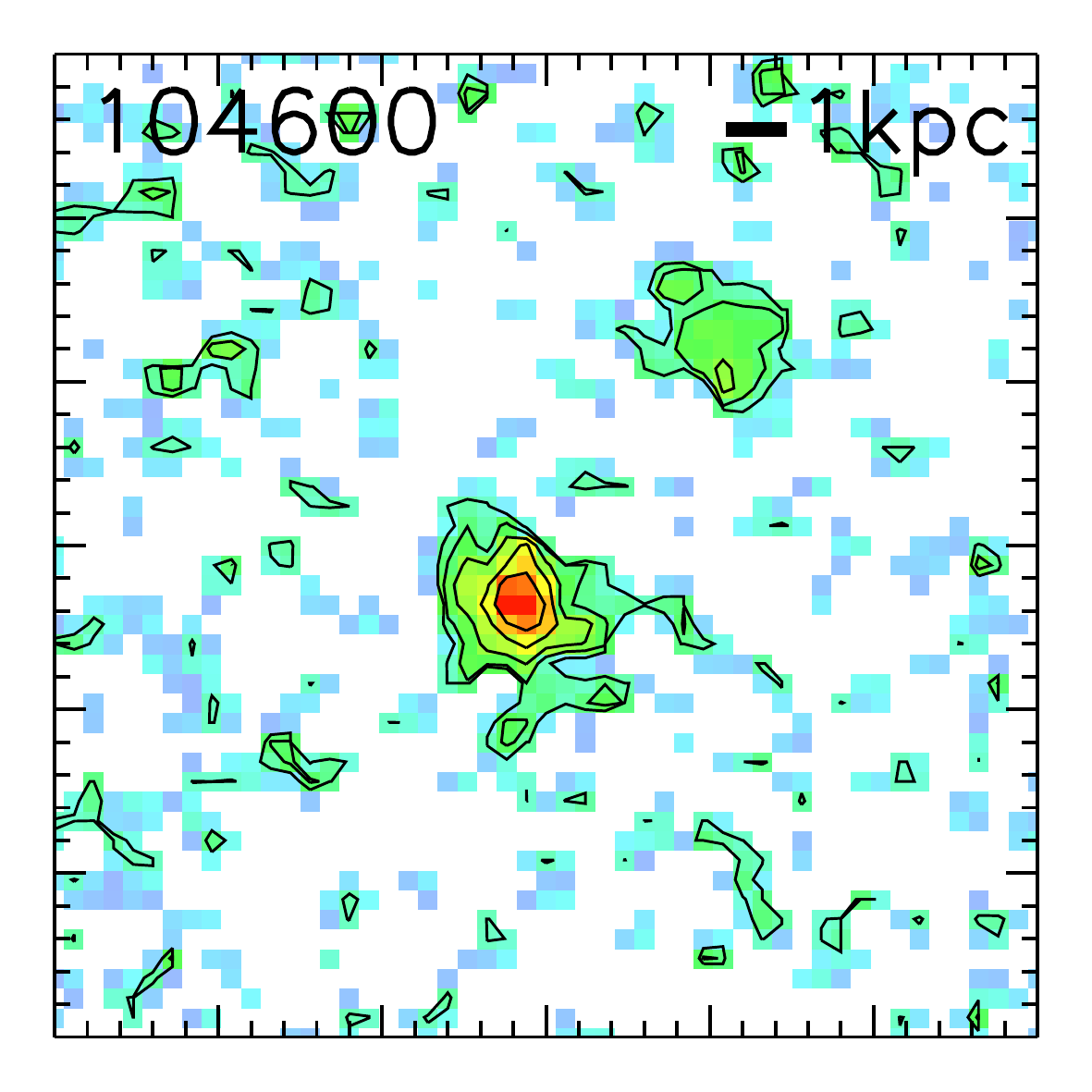}\hspace{0.05cm}
\includegraphics[width = 0.15\textwidth, trim = 0.6cm 0.2cm 0.6cm 0.4cm ]{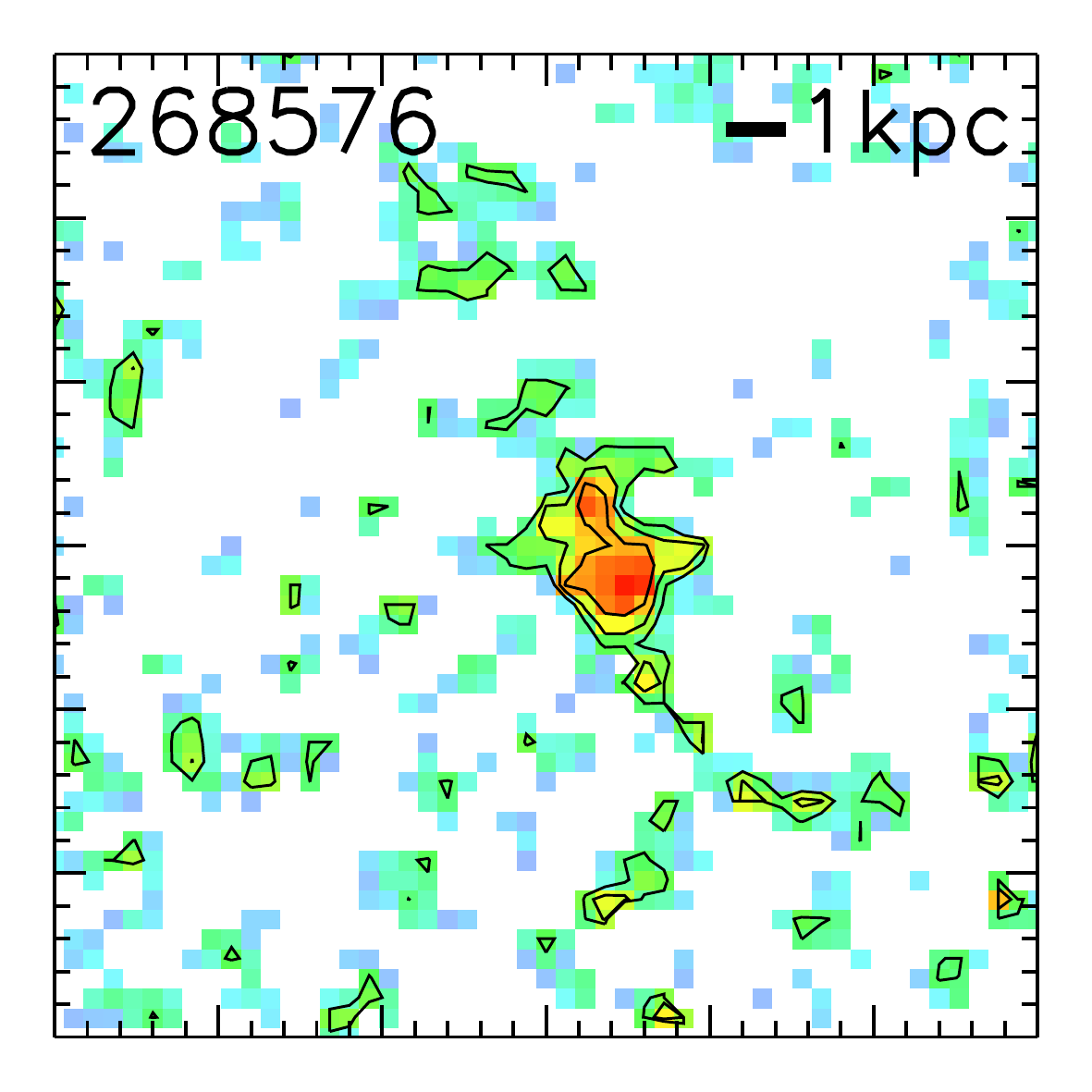}\hspace{0.05cm}
\includegraphics[width = 0.15\textwidth, trim = 0.6cm 0.2cm 0.6cm 0.4cm ]{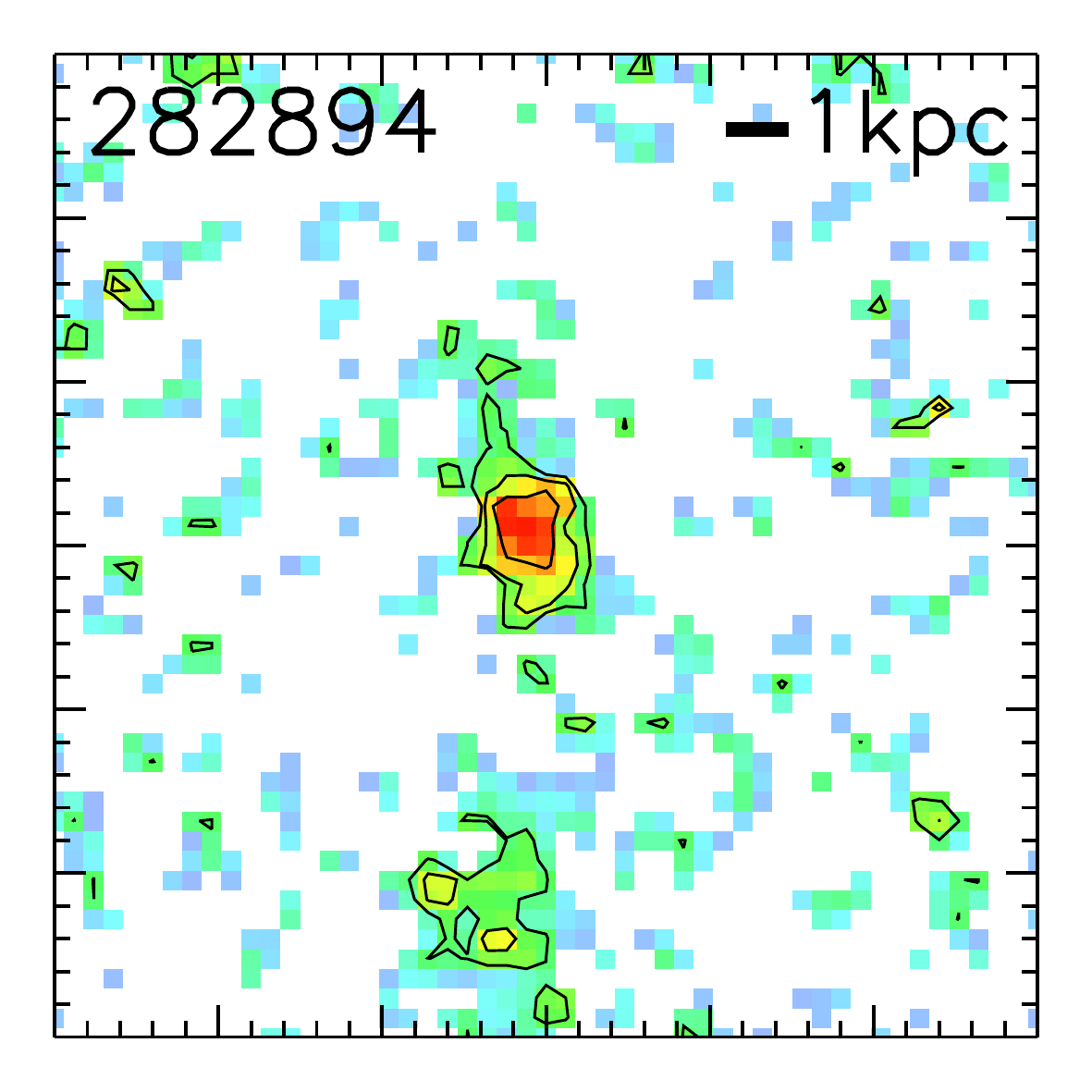}\hspace{0.05cm}
\includegraphics[width = 0.15\textwidth, trim = 0.6cm 0.2cm 0.6cm 0.4cm ]{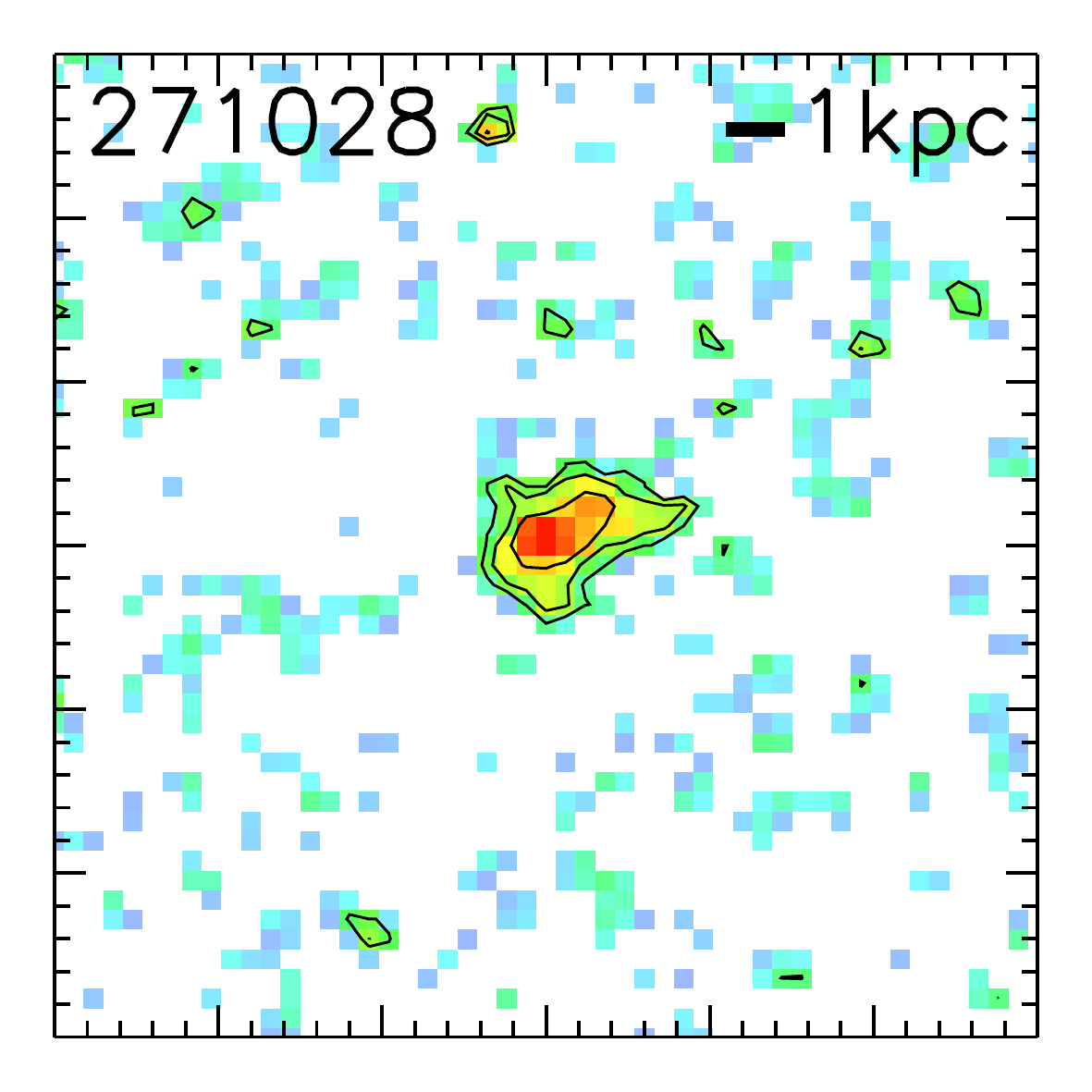}
\caption{Postage-stamp~\emph{HST}/WFC3 images of the final sample of 22 LBGs at $ z\simeq 7$ (initially selected in~\citealt{Bowler2014}).
The galaxies are ordered by absolute UV magnitude, with the brightest object in the upper left ($M_{\rm UV} = -23.1$) and the faintest in the lower right ($M_{\rm UV} = -20.7$).
The colour scale for each stamp has been scaled between a minimum surface brightness of $26\,{\rm mag}/{\rm arcsec}^2$ and the peak surface brightness (typically $22\,{\rm mag}/{\rm arcsec}^2$) to highlight any extended emission.
Contours are shown in $0.5\,{\rm mag}$ intervals bright-ward of $25\,{\rm mag}/{\rm arcsec}^2$.
Each stamp is $3\,$arcsec on the side, with North to the top and East to the left.
The images are centred on the coordinates determined from the ground-based selection, where each stamp displays a single ground-based object.
The physical distance in ${\rm kpc}$ has been calculated according to the redshift of each galaxy and is displayed as a scale-bar in the upper right of each stamp.
Galaxies that show multiple distinct components at $z > 6$ are highlighted with the letter `m' in the lower left-hand corner of the stamp.
}
\label{fig:hstsb}
\end{center}
\end{figure*}

\subsection{Visual morphologies}

The final sample of 22 LBGs is shown in Fig.~\ref{fig:hstsb}, ordered by absolute UV magnitude.
The stamps are $3\,$arcsec across, or approximately $16\,{\rm kpc}$ at the median redshift of our sample.
The four brightest galaxies in the sample all have an elongated and clumpy appearance, with the brightest clumps in ID304416 and ID169850 separated by $\simeq 5\,{\rm kpc}$.
The extent of these very bright LBGs is similar to that of `Himiko'~\citep{Ouchi2009, Ouchi2013}, however the brightest galaxies here are up to $\Delta {m}_{\rm AB} = 1.0$ mag brighter in the rest-frame UV.
As well as the extended clumpy objects, we also find galaxies with apparently distinct components (at least to the limiting surface brightness of our data), for example ID279127, ID30425/CR7 and ID28495.
These objects were all selected as a single galaxy with the resolution of the ground-based data used in~\citet{Bowler2014}, and the close separation ($< 10\,{\rm kpc}$) suggests that they are associated with the same galaxy (e.g. clumps in an extended structure) or in the process of merging.
The remainder of the sample appear as isolated single-component objects, with several appearing elongated (e.g. ID268576, ID271028) or showing extended emission (e.g. ID185070, ID238225).

The diversity of morphologies observed in our sample is similar to that found in lower redshift star-forming galaxies.
For example~\citet{Law2012} found a range of structures in at $z \simeq 2$--$3$ LBGs including single nucleated sources ($\sim 40$ percent of the sample), groups of distinct components ($\sim 20$ percent) and highly irregular extended objects ($\sim 40$ percent).
The results of our analysis show that at $z \simeq 7$ the brightest galaxies tend to be clumpy/merger-like systems, with the four brightest objects all showing two or more distinct components.
Fainter galaxies are instead generally single components, although with some signs of an irregular morphology~\citep{Oesch2010,Jiang2013b}.
Our results are similar to those found by~\citet{Kawamata2015} in a study of LBGs found within the HFF program, who found that LBGs at $z \sim 8$ with multiple cores tended to be at the bright-end of their sample.
The depth of our WFC3 data is sufficient to detect if the galaxies in our sample with $24.5 < m_{\rm AB} < 25.0$ consisted of multiple components, however the most clumpy LBGs are clearly preferentially found at the bright-end of our sample.
At the faint end of our sample however, our observations are only sensitive to multiple-component systems without extended/low surface brightness emission (see the Appendix), and hence deeper data are required to robustly determine the morphologies of $M_{\rm UV} \simeq -21$ LBGs at $z \simeq 7$.
We note here that the LBGs in this sample were selected independently of the surface brightness profile due to the dominant effect of the relatively large ground-based PSF, and hence we are not strongly biased to compact systems.

\subsection{Interaction/merger fraction}

Visibly disturbed, elongated or clumpy galaxies at high-redshift have often been interpreted as merging systems~\citep{Conselice2014}.
The merger fraction at intermediate redshift has been calculated by using non-parametric morphology measurements such as the Concentration-Asymmetry-Clumpiness (CAS;~\citealp{Conselice2003}) or the Gini/M20 parameters~\citep{Lotz2004}, which when combined with an estimate of the timescale for merger activity to be visible ($200$--$800$Myrs;~\citealp{Lotz2010, Lotz2010a}), can be converted into a merger rate.
At very high redshifts however, the small size of the galaxies with respect to the PSF results in biases in these parameters which depend sensitively on the depth of the imaging~\citep{Jiang2013b, Curtis-Lake2016}, and hence visual inspection is often employed to estimate the merger/interaction fraction.
In our sample of bright $z \simeq 7$ LBGs, we find 9 galaxies that show distinct components that could be attributed to a merging system (7 excluding the known LAEs `Himiko' and `CR7').
These 9 objects are highlighted in Table~\ref{table:sample}.
In addition, several of the single component objects show extended emission or an elongation, and hence we place a lower limit on the merger fraction of $> 40$ percent from visual inspection.
For fainter LBGs at $z \simeq 7$ however, the disturbed fraction is significantly lower ($< 10$ percent) despite the data being sufficiently deep to detect additional components (e.g.~\citealt{Oesch2010}; see also~\citealp{Conselice2009}).
It is clear in the postage-stamp images shown in Fig.~\ref{fig:hstsb} that the very brightest galaxies in our sample all show a clumpy and extended morphology.
Determining the merger fraction for these extremely bright LBGs therefore results in a merger or interaction fraction of $100$ percent at $M_{\rm UV} \lesssim -22.5$.
For these LBGs, the individual components are as bright, or brighter, than `normal' $z \simeq 7$ LBGs ($M_{\rm UV} = -22.3$ and $-22.4$ for ID304416, and $M_{\rm UV} = -22.6$ and $-21.5$ for ID169850, $M_{\rm UV} = -21.8$ and $-21.5$ for ID65666 and $M_{\rm UV} = -21.7$ and $-21.6$ for ID279127).
Hence, if the multi-component galaxies are the result of a merger of two or more typical LBGs, then the SFR has been significantly boosted as a result of the interaction (the component luminosities are in the range $L \simeq2.5$--$7\,L^*$).

The derived merger fraction from our data is comparable to that found at $z \simeq 2$--$3$, where a merger fraction of around $40$--$50$ has been derived for massive galaxies ($M_{\star} > 10^{10}\, {\rm M}_{\sun}$;~\citealp{Mortlock2013}).
~\citet{Law2012} found that $70$ percent of a sample of $z = 1.5$--$3.2$ LBGs showed either multiple components ($20$ percent of the sample) or a highly disturbed or irregular morphology.
At $z \simeq 6$, both~\citet{Jiang2013b} and~\citet{Willott2013} found a similarly high fraction of disturbed systems in $M_{\rm UV} < -20.5$ LBGs, with the brightest galaxies ($M_{\rm UV} < -21$) in the~\citet{Jiang2013b} sample showing a merger fraction of $60$ percent.
We find therefore that the visually derived merger fraction of bright $z \simeq 7$ LBGs is similarly high compared to that found at lower redshifts.
This is in agreement with the results of~\citet{Curtis-Lake2016}, who finds no evidence for an evolution in the fraction of fainter galaxies ($-22 \lesssim M_{\rm UV} < -20$) showing irregular features from $z = 4$--$6$.

\section{Galaxy Luminosity Function}\label{sect:lf}

The new~\emph{HST}/WFC3 imaging of the~\citet{Bowler2014} sample presented in this work provides both a measurement of the galaxy size and an independent measurement of the total galaxy flux.
The total galaxy luminosity is crucial when determining the bright end of the rest-frame UV luminosity function.
Despite the ground-based selected objects appearing to fragment into discrete components under~\emph{HST} resolution, we argue based on the following reasons that the clumps are physically associated and should therefore be treated as a single galaxy when calculating the LF.
Firstly, the small separations of the different components ($< 10\,{\rm kpc}$) are dramatically lower than the separations between similarly bright galaxies in the field (typically $> 10\,{\rm arcmin}$), showing that the observed objects are very unlikely to be simply chance alignments of fainter galaxies.
Furthermore, the deep optical imaging can strongly rule out that the brightest components are low-redshift interlopers (Section~\ref{sect:lowz}).
Secondly, the individual components of the clumpy LBGs in the sample are brighter than typical LBGs at $z \simeq 7$ despite having similar sizes.
This suggests that there is a physical association between the star-forming components (whether via a merging process or a vigorously star-forming clumpy system) that has increased the SFR of the system as compared to the field.
Finally, despite the galaxy sample including extremely bright $z \simeq 7 $ LBGs, the surface brightness limit of the data is insufficient to detect diffuse emission such as that observed in $z \simeq 2$ LBGs (e.g.~\citealp{Law2012}).
Hence clumpy star-forming galaxies as observed at lower redshifts would naturally appear as discrete bright clumps when observed in the rest-frame UV at higher redshift.

\subsection{Galaxy total magnitudes and $M_{\rm UV}$}\label{sect:totalmag}

To explore the extent of the rest-frame UV emission, and to extract the total magnitude for each galaxy in our sample, we measured the curve-of-growth using circular apertures. 
Due to the clumpy nature of the LBGs in this work, we calculated our own weighted galaxy centroid from the data following the procedure employed by {\sc SExtractor}, which uses the barycentre or first order moment of the profile defined as:
\begin{equation}\label{equation}
x = \sum_i I_{i}\, x_i / \sum_i I_{i}, \: {\rm and }\: \: \: y = \sum_i I_{i}\, y_i / \sum_i I_{i}
\end{equation}
\noindent
where the sum in over all pixels assigned to an object by {\sc SExtractor}.
In the case of single component LBGs, we find excellent agreement between our centroid and that calculated by {\sc SExtractor}.
For galaxies with an irregular, clumpy light distribution, the barycentre matches the ground-based centroid (which is effectively a weighted centroid of the full system) to within the astrometric error of $\lesssim 0.1$ arcsec, as expected from the relatively coarse pixel size and larger FWHM of the ground-based imaging.
Inspecting the curve-of-growth for each object we found that a $2$ arcsec diameter circular aperture was required to measure the total magnitudes of the majority of the galaxies in our sample, as supported by our stacked results shown in Section~\ref{sect:stack}.
As motived by studying the COG, we used larger 3 arcsec diameter apertures for galaxies ID304416, ID169850, ID279127 and ID88759/Himiko, due to the extended and clumpy nature of these objects.
The total and absolute magnitudes of the galaxies in our sample are shown in Table~\ref{table:sample}.
We note that the brightest galaxies in our sample have ${m}_{\rm AB} \lesssim 24$ and are therefore exceptionally bright for $z \simeq 7$ galaxies without the boost of strong gravitational lensing.
The absolute UV magnitude ($M_{\rm UV}$) was calculated from the best-fitting galaxy SED at a rest-frame wavelength of $1500$\AA~(using a top-hat filter of width $100\,$\AA), after the SED was scaled to match the total magnitude measured from the~\emph{HST}/WFC3 data in the appropriate~\emph{HST} filter.

The new photometry can be compared to our previous measurements from the ground-based data (using 1.8 arcsec diameter circular apertures, corrected to total assuming a point source;~\citealp{Bowler2014}).
The comparison shows that the ground-based measurement underestimated the total magnitude by $\Delta m = 0.1$ mag, rising to $\sim 0.2$ mag for the two brightest galaxies in the sample.
As is visible in the stack of the ground-based data shown in Fig.~\ref{fig:groundstack}, the brighter galaxies in our sample are resolved in the VISTA and UKIRT data, and hence it is unsurprising that we underestimated the total magnitudes assuming a point-source correction.
For four of the sample (ID305036, ID104600, ID211127, ID271028), we find that the ground-based magnitude is significantly brighter than the~\emph{HST} measurement, by $\Delta {m}_{\rm AB} \simeq 0.5$--$1.0\,$ mag.
In the case of object ID104600, the flux is boosted due to contamination from a nearby low-redshift object (to the NW, separated by $1\,$arcsec), which we have masked when performing the~\emph{HST}/WFC3 measurement.
The other three objects all have low-redshift galaxies (ID277727, ID271028) or stars (ID305036) close to the line-of-sight, which has resulted in contamination of the ground-based photometry (see Figure~\ref{fig:groundhst}) and an overestimated near-infrared magnitude.

As noted in~\citet{Bowler2014} and~\citet{Bouwens2015} the use of small apertures (e.g. $< 0.6$ arcsec diameter) can underestimate the total magnitudes of high-redshift galaxies.
For example,~\citet{Bouwens2015} has suggested that~\citet{McLure2013}, and other works that use similarly small apertures such as~\citet{Lorenzoni2012}, could underestimate the total magnitudes of the brightest galaxies in their samples ($M_{\rm UV} \sim -21$) by $\simeq 0.2$ mag.
For the brightest galaxies, which are clearly extended in the~\emph{HST}/WFC3 imaging in this work (and in other works e.g.~\citealp{Oesch2010, Willott2013, Jiang2013b}), we expect this problem to be exacerbated.
We compare our total magnitudes to those obtained by {\sc SExtractor} using small (0.6 arcsec diameter) circular apertures, corrected to total assuming a point source, and MAG\_AUTO which uses scalable elliptical Kron apertures.
While using a small circular aperture unsurprisingly underestimates the total magnitude of the brightest LBGs (by $\Delta {\rm m}_{\rm AB} > 0.5$ mag), we find that the measurements made using MAG\_AUTO are in good agreement with our COG measurements, suggesting that analyses that use Kron-type elliptical apertures (e.g.~\citealp{Bouwens2015, Finkelstein2015}) are less prone to bias in the total magnitudes.

\subsection{The luminosity function}

Using the independent measurement of galaxy flux from~\emph{HST} we have found that the measurement procedure used in~\citet{Bowler2014} can underestimate the total magnitude of the galaxies in this sample by around $0.1$--$0.2$ mag.
Furthermore, the new data has resulted in the discovery of an artefact in the VISTA/VIRCAM data and also the identification of contaminated flux measurements for several galaxies.
These effects must be taken into account in the determination of the rest-frame UV luminosity function from the UltraVISTA/COSMOS and UDS/SXDS datasets at $z \simeq 7$.
Here we present an updated LF derived from the~\citet{Bowler2014} sample using the new~\emph{HST}/WFC3 data.

We calculated the rest-frame UV LF following the $1/V_{\rm max}$ methodology~\citep{Schmidt1968} presented in~\citet{Bowler2014, Bowler2015}, where we refer the reader for a more comprehensive description.
In brief, we use the best-fitting SED model for each galaxy to calculate the maximum redshift at which it would be retained in the sample, and derive from this the $V_{\rm max}$.
The incompleteness of our sample due to the finite depth and photometric redshift uncertainty was calculated at each redshift and absolute magnitude using injection and recovery simulations.
In the final LF calculation, we included the UltraVISTA/COSMOS DR2 strips (area $0.62\,{\rm deg}^2$) and the full UDS/SXDS field (area $0.74\,{\rm deg}^2$).
The near-infrared data in the UltraVISTA/COSMOS DR1 gaps region (area $0.29\,{\rm deg}^2$) was excluded from the analysis, as it was too shallow to significantly affect the derived number densities.
We corrected for the moderate gravitational lensing of our LBGs by low-redshift galaxies close to the line-of-sight using the approach presented in~\citet{Bowler2014, Bowler2015}.
The de-lensing method uses the photometric redshift and $i$-band luminosity of nearby galaxies to approximate their gravitational lensing potential via the Faber-Jackson relation.
The magnification was typically $0.1$ mag ($\mu \sim 1.1$), rising to a maximum of $0.3$ mag ($\mu \sim 1.3$).

\begin{figure}

\includegraphics[width = 0.49\textwidth]{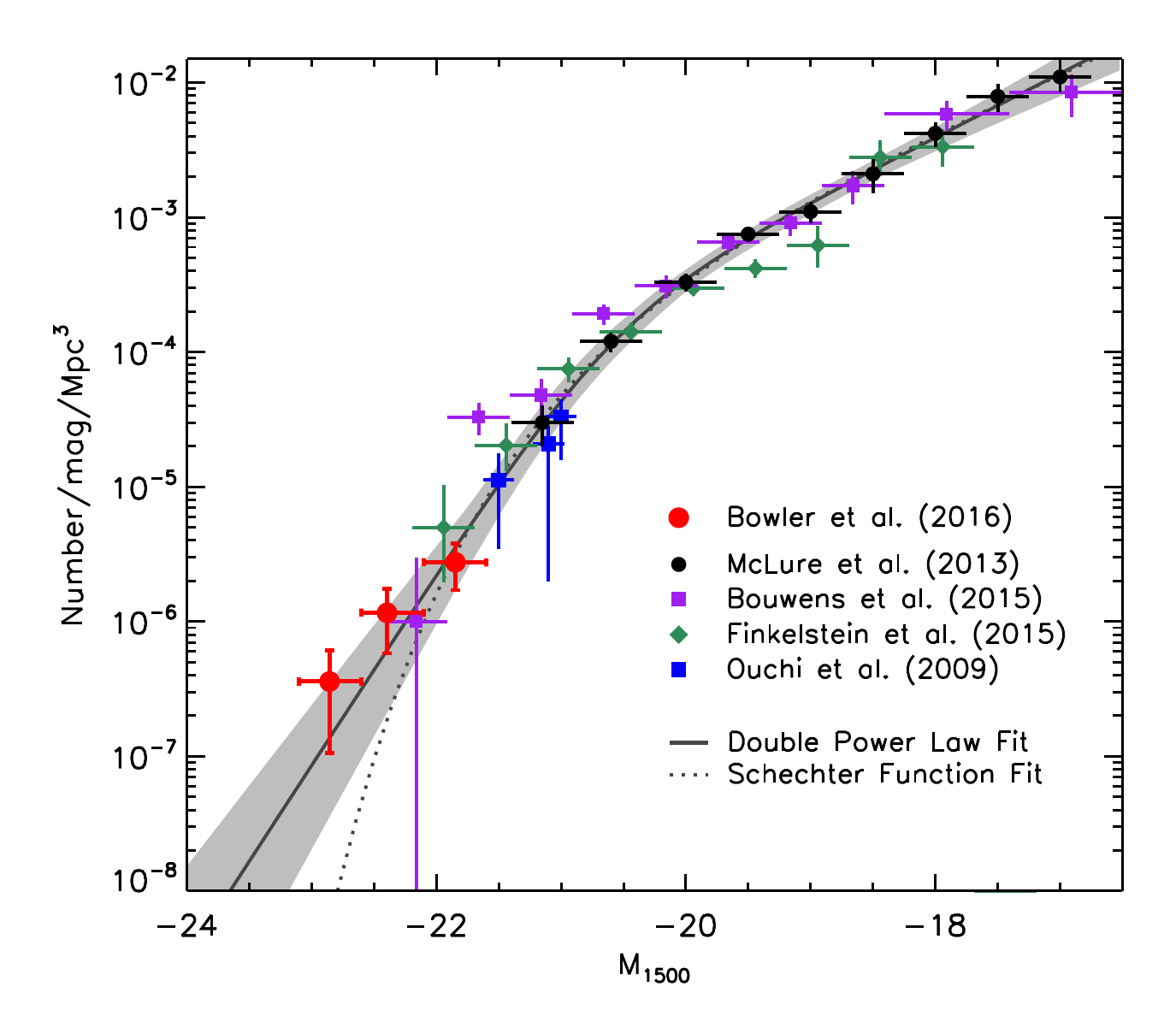}
\includegraphics[width = 0.49\textwidth]{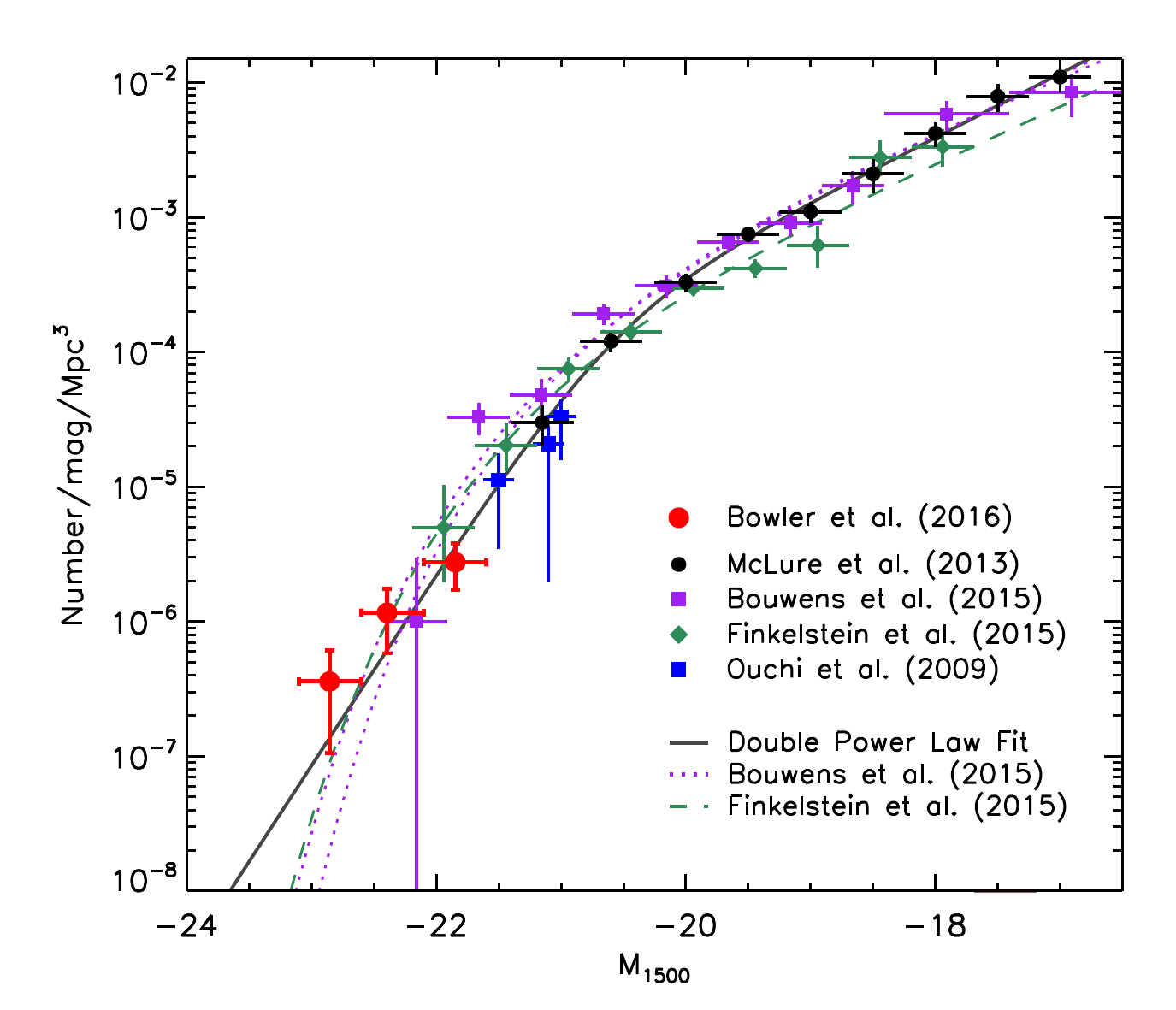}

\caption{The rest-frame UV LF at $z \simeq 7$ from the UltraVISTA/COSMOS DR2 and UDS/SXDS datasets as derived in~\citet{Bowler2014}, updated using the results of this work from new~\emph{HST}/WFC3 imaging (red circles).
We show the results of previous works using ground-based data from~\citet{Ouchi2009} (blue squares), and from a combination of~\emph{HST} surveys such as CANDELS from~\citet{Bouwens2015} (purple squares),~\citet{Finkelstein2015} (green diamonds) and~\citet{McLure2013} (black circles).
The best-fitting DPL and Schechter functions to our results and those of~\citet{McLure2013} are shown in the upper plot as solid and dotted lines respectively.
The one-sigma confidence limit on the best-fitting DPL is shown as the grey shaded region. 
In the lower plot we also show the best-fitting Schechter function derived by~\citet{Bouwens2015} and~\citet{Finkelstein2015} as the purple dotted and green dashed lines respectively.
}\label{fig:lf}
\end{figure}

The brightest galaxies in our sample were the most robust to magnitude errors and the number densities derived for the two brightest bins are effectively unchanged from the~\citet{Bowler2014} analysis.
The absolute magnitudes of the objects were typically underestimated from the ground-based data by $0.1$ mag and $0.2$ mag in the central and brightest bin respectively.
In the faint bin presented in~\citet{Bowler2014} we now find that two of the nine galaxies were cross-talk artefacts and two of the galaxies had overestimated fluxes due to contamination from nearby galaxies or stars.
We therefore remove the cross-talk artefacts, and update the absolute magnitudes of the galaxies in our sample using the new~\emph{HST}/WFC3 photometry.
The resulting binned LF points are centred on slightly brighter magnitudes than presented in~\citet{Bowler2014}.
These points are calculated from a sample that includes the spectroscopically confirmed $z = 6.6$ galaxies `Himiko' and `CR7', to provide number counts for all known $z \simeq 7$ rest-frame UV bright galaxies in the two fields.
These objects are relatively faint in our sample, occupying the faintest bin in our LF analysis.
As a result, if they are excluded from our analysis we find no significant changes to our LF results, with the DPL becoming a slightly better fit to the data (over a Schechter function), due to the small drop in the number density in our faintest bin ($\simeq 20$ percent).
The updated LF is shown in Fig.~\ref{fig:lf} where we compare to previous results at $z \simeq 7$ derived from the Subaru Deep Field~\citep{Ouchi2009} and from a compilation of~\emph{HST} surveys including the UDF and CANDELS datasets~\citep{Bouwens2015, Finkelstein2015, McLure2013}.
The results presented in~\citet{Finkelstein2015} have been corrected to account for the different cosmology, and the two brightest points from~\citet{McLure2013} are plotted $0.15$ and $0.1$ mag brighter to account for the extended size relative to the assumed point source correction.
The binned points are presented in Table~\ref{table:points}.
We fit both a double-power law and a Schechter function to the updated LF points and those determined by~\citet{McLure2013}, who used a similar methodology to this work.
The resulting functional fits are shown in the upper panel of Fig.~\ref{fig:lf} and presented in Table~\ref{table:fits}.
The updated best-fitting parameters agree well with the previous results presented in~\citet{Bowler2014, Bowler2015}, and consistently show that the DPL is formally the best-fitting function with a reduced $\chi^2 = 0.6$ (compared to $0.9$ for the Schechter function).

\subsubsection{Comparison to previous studies}

We find a good agreement between our updated $z \simeq 7$ LF points with the results of~\citet{Finkelstein2015} in the region of overlap, and with the brightest point from~\citet{Bouwens2015}.
The DPL fit is also in good agreement with the data points from~\citet{Ouchi2009} around the knee of the function, and with previous results at fainter magnitudes where we reproduce the steep faint-end slope of $\alpha \simeq -2$ found by previous studies (e.g.~\citealp{McLure2013}).
At $M_{\rm UV} \simeq -21.75$ however, we find a tension between our determination of the LF and the results of~\citet{Bouwens2015}.
The~\citet{Bouwens2015} point at $M_{\rm UV} = -21.66$ is significantly in excess of our best-fitting function at this magnitude and the ground-based results from~\citet{Ouchi2009}, who found $< 1/3$ the number density of galaxies at a similar luminosity in an analysis of the Subaru Deep Field.
As shown in~\citet{Bowler2015} at $z \simeq 6$, cosmic variance is significant at the bright-end of the LF at $z > 5$, and hence part of the discrepancy could be a result of the strong field-to-field variation between the $200\,{\rm arcmin}^2$ CANDELS fields.
For LBG candidates at the bright-end of the~\citet{Bouwens2015} $z \simeq 7$ sample however, there is an additional uncertainty that must be considered due to the lack of deep $Y$-band imaging over the full survey area used in their analysis.
For three of the five relatively wide-area CANDELS fields studied by~\citet{Bouwens2015} there is no $Y$-band imaging available from~\emph{HST}/WFC3.
Imaging in the $Y$-band is essential to constrain the position and strength of the Lyman-break from $z \simeq 7$--$8$, and to exclude low-redshift galaxy and cool galactic brown dwarf contaminants which can show identical optical-to-near-infrared colours in the absence of the $Y$-band data~\citep{Bowler2012,Finkelstein2015}.
While there exists some relatively shallow $Y$-band imaging in the CANDELS COSMOS and UDS fields from ground-based surveys, there is no space- or ground-based $Y$-band imaging available in the Extended Groth Strip (EGS) making this field in particular vulnerable to contamination.
The number of bright $z \simeq 7 $ LBGs found by~\citet{Bouwens2015} in the EGS is more than double the average number found in the other fields, and hence we suggest that the origin of the high number density derived around $M_{\rm UV} \sim -21.7$ could be due to contamination by low-redshift galaxies or brown dwarfs in the CANDELS `wide' fields utilized (also see the discussion in~\citealp{Finkelstein2015}).

The effect of the uncertain number counts at the bright-end of the LF determined from the CANDELS data can be seen in the best-fitting Schechter function parameters.
The recent determinations of the rest-frame UV LF at $z = 7 $ by~\citet{Bouwens2015} and~\citet{Finkelstein2015} both found a brighter characteristic magnitude of the best-fitting Schechter function than previous studies~\citep{McLure2013, Schenker2013, Bouwens2011a}, finding an approximately constant value of $M_{\rm UV} \simeq - 21$ from $z \simeq 5$ to $z \simeq 7$.
However, if the CANDELS `wide' fields that have limited or no $Y$-band data are excluded,~\citet{Bouwens2015} finds $M^* = -20.61 \pm 0.31$, which is in better agreement with our results and previous studies.
Using the wider area ground-based data, we find a characteristic magnitude of $M^* = -20.49 \pm 0.17$ (assuming a Schechter function fit), which is in good agreement with our previous results~\citep{Bowler2015} and supports a brightening of the characteristic magnitude by $\Delta M^* \simeq 0.5 \,$mag from $z = 7$ and $z = 5$.
While the LF points derived by~\citet{Finkelstein2015} appear to match our data at the bright end, this is at the expense of a satisfactory fit to the majority of the other data points at $M_{\rm UV} > -20$.
Here the degeneracy between the faint-end slope and the characteristic magnitude results in a similarly bright $M^* = -21.03^{+0.37}_{-0.50}$ to that found by~\citet{Bouwens2015}.
This comparison illustrates the importance of wide area, ground-based data in constraining the bright end of the LF at high redshift.

In conclusion, we find that a double-power law remains the best-fitting functional form to the data at $z \simeq 7$.
Furthermore, our results favour a smooth evolution in the characteristic magnitude from $M_{\rm UV} \simeq - 21$ at $z = 5$~\citep{vanderBurg2010, Bouwens2015} to $M_{\rm UV} \simeq - 20$ at $z = 8$~\citep{Oesch2012b, McLure2013, Schenker2013,Schmidt2014}.
While we find evidence for strong evolution in the number densities of galaxies around the knee of the rest-frame UV LF from $z = 5$ to $z = 7$, we find little evolution for the brightest $M_{\rm UV} < -22$ LBGs, which have an approximately constant number density of $\phi \simeq 1 \times 10^{-6} \,{\rm mag}^{-1}\,{\rm Mpc}^{-3}$ over this redshift range~\citep{Bowler2015}. 

\begin{table}
\caption{The best fitting DPL and Schechter function parameters obtained by fitting our new LF determination and the~\citet{McLure2013} results.
The upper part of the table shows the results when ground-based objects are considered as one galaxy in the analysis.
In contrast, the lower part shows the results when multiple component objects are instead split into several components that are treated as distinct galaxies.
The DPL is formally the best-fitting function in both cases.
The columns show the characteristic absolute magnitude ($M^*$) followed by the corresponding characteristic number density ($\phi^*$) and faint-end slope ($\alpha$.
For the DPL parameterisation we also show the best-fitting bright-end slope, $\beta$.
}
\begin{tabular}{l c c c c}
\hline

& $M^*$ & $\phi^*$ & $\alpha$ & $\beta$ \\
& $/{\rm mag}$ & $/{\rm mag}/{\rm Mpc}^3$ & & \\
\hline
DPL & $-20.60_{ -0.27}^{+  0.33}$ & $2.3_{ -0.9}^{+  1.8}  \times 10^{-4}$ &  $ -2.19_{ -0.10}^{+  0.12}$ & $  -4.6_{  -0.5}^{+   0.4}$\\
Sch. & $-20.49_{ -0.17}^{+  0.17}$ & $4.2_{ -1.3}^{+  1.7}  \times 10^{-4}$ & $ -2.07_{ -0.09}^{+  0.10}$ &  -- \\
\hline

DPL &  $-20.71_{ -0.22}^{+  0.25}$ & $1.8_{ -0.7}^{+  1.1}  \times 10^{-4}$ & $ -2.22_{ -0.09}^{+  0.10}$ & $  -5.0_{  -0.5}^{+   0.4}$\\
Sch. & $-20.44_{ -0.16}^{+  0.16}$ & $4.5_{ -1.3}^{+  1.8}  \times 10^{-4}$ & $ -2.05_{ -0.09}^{+  0.10}$ &  -- \\

\hline

\end{tabular}\label{table:fits}
\end{table}

\begin{table}
\caption{The binned rest-frame UV LF points for our sample of bright $z \simeq 7$ LBGs initially selected in~\citet{Bowler2014}.
The LF analysis has been updated from that presented in~\citet{Bowler2014} using the results of our~\emph{HST}/WFC3 follow-up.
As in Table~\ref{table:fits}, the upper part shows the results when ground-based selected objects are considered as one galaxy (as shown in Fig.~\ref{fig:lf}).
The lower part shows the results when multiple component objects are instead split into separate `galaxies' (Section~\ref{lf:indi}).
The first and second column show the absolute magnitude range and weighted centroid for each bin, followed by the derived number density.
The absolute magnitudes were calculated at a rest-frame wavelength of $1500$\AA.
}

\begin{tabular}{c c c}

\hline
$M_{UV}$ range & $M_{UV}$ & $\phi$ \\
$/{\rm mag}$ & $/{\rm mag}$ & $/{\rm mag}/{\rm Mpc}^3$  \\
\hline
$ -21.6 < M < -22.1 $ & $    -21.85$ & $2.75 \pm 1.04 \times 10^{-6}$ \\
$ -22.1 < M < -22.6 $ & $    -22.40$ & $1.16 \pm 0.58 \times 10^{-6}$ \\
$ -22.6 < M < -23.1 $ & $    -22.86$ & $3.59 \pm 2.54 \times 10^{-7}$\\
\hline

$ - 21.55 < M < -21.8 $ & $    -21.66$ & $4.25 \pm 1.90 \times 10^{-6}$ \\
$ - 21.8 < M < -22.3 $ & $    -22.03$ & $1.72 \pm 0.70 \times 10^{-6}$ \\
$ - 22.3 < M < -22.8 $ & $    -22.56$ & $1.87 \pm 1.87 \times 10^{-7}$ \\

\hline

\end{tabular}\label{table:points}
\end{table}

\subsection{Individual components}\label{lf:indi}

\begin{figure}

\includegraphics[width = 0.49\textwidth]{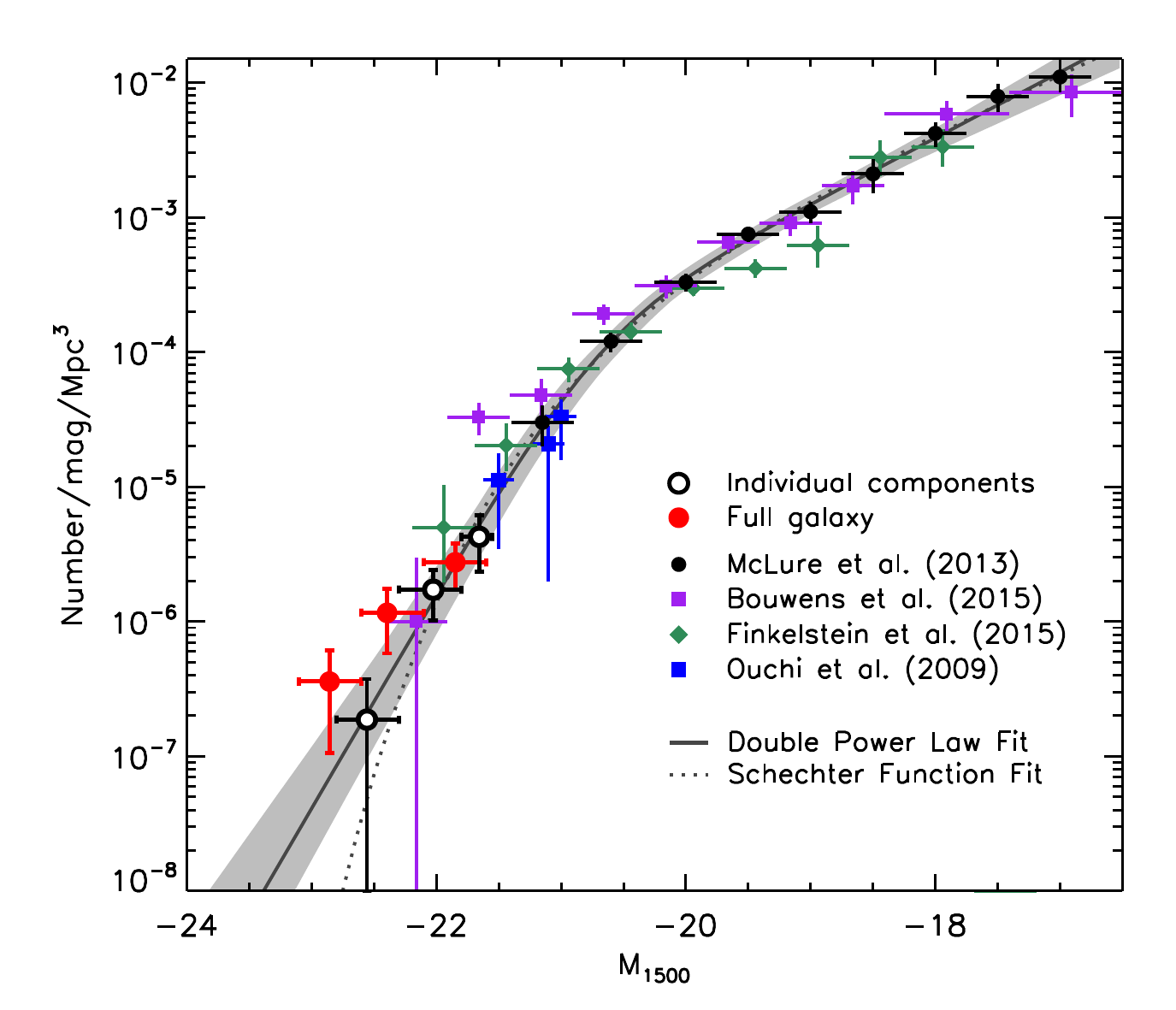}

\caption{The rest-frame UV LF at $z \simeq 7$ from this work calculated from the individual components of each ground-based selected galaxy (open black circles).
The red circles show the results when multi-component galaxies are instead plotted as single objects (as shown in Fig.~\ref{fig:lf}).
Previous LF results are described in the caption to Figure.~\ref{fig:lf}, and the solid black line and dotted line show respectively the best-fitting DPL and Schechter functions when fitted to the LF points derived from the individual component analysis.
}\label{fig:lfindi}
\end{figure}

In the LF analysis presented in Fig.~\ref{fig:lf}, we have classified each ground-based galaxy as a single object due to the close separation of the clumps ($< 10\,{\rm kpc}$) and their consistent photometry (Section~\ref{sect:lowz}).
If we interpret the identifiable clumps as merging galaxies however, then the derived luminosity function points would change.
Using the {\sc SExtractor} de-blended objects and assuming the {\sc MAG\_AUTO} as the total magnitude for each individual clump, the resulting LF points instead follow a steeper decline as shown in Fig.~\ref{fig:lfindi}.
In this case, the brightest ground-based galaxies in our sample, that appear as multiple-component objects at~\emph{HST}/WFC3 resolution, are split into several fainter `galaxies' and hence the number densities at bright magnitudes are reduced.
While the brightest galaxy in our sample (ID304416) splits into two objects of comparable luminosity, the majority of our brighter objects have a dominant component and a fainter part (with $M_{\rm UV} \lesssim -21.5$; e.g. ID169850 and ID65666).
As a consequence the faint-end of our LF determination is in good agreement with the previous results calculated from the full ground-based object.
The results of functional fitting to the individual component results show that a DPL remains the formally best fitting function.
We find that the best-fitting DPL and Schechter function parameters (shown in Table~\ref{table:fits}) are consistent with the previous fits presented in Fig.~\ref{fig:lf}, however for the individual component analysis we find a steeper best-fitting bright-end slope for the DPL fit.
Although the LF determinations agree within the errors, the number counts of the very brightest LBGs ($M_{\rm UV} \simeq -23$) differ significantly between the two approaches, with essentially no `galaxies' brighter than $M_{\rm UV} \simeq -22.6$ when the individual components are considered separately.
When comparing LFs derived from ground and space-based searches at the bright-end therefore, caution must be taken to treat multiple component galaxies in a consistent way.
We note that in the~\citet{Bouwens2015} sample, individually selected objects are grouped into a single galaxy if the centroids are within $0.5\,$arcsec.
This condition would be insufficient for several galaxies in our sample where the components are separated by $\simeq 1\,$arcsec.
An inspection of the~\citet{Bouwens2015} catalogue at $z = 7$ however, shows that there are no galaxy pairs with separations in the range $\simeq 0.5$--$1.5\,$arcsec, indicating that such systems are rare amongst the fainter galaxies found within~\emph{HST} data.
Similarly to obtain meaningful comparisons to galaxy evolution simulations (e.g. via a comparison of the LF; see~\citealp{Bowler2015}) it is clearly important to ensure that merging or clumpy galaxies are selected and characterised using the same methodology as the observations.

\section{Galaxy sizes}\label{sect:sizes}

The~\emph{HST}/WFC3 imaging of $z \simeq 7$ LBGs presented in this paper provides the first high-resolution imaging of a sample of $z > 6.5$ star-forming galaxies in the magnitude range $-23 \lesssim M_{\rm UV} < -21$.
The imaging allows the first robust exploration of the galaxy sizes in a previously unexplored magnitude regime at $z \simeq 7$.
As illustrated in the postage-stamp images in Fig.~\ref{fig:hstsb} however, the sample shows a range of morphologies, including multiple-component, clumpy galaxies.
The measurement of sizes is therefore complicated in comparison to samples of fainter, generally compact and single component galaxies explored previously (e.g~\citealp{Oesch2010}).
In this section we discuss different methods for measuring galaxy sizes for our sample using {\sc SExtractor}, {\sc GALFIT} and a curve-of-growth analysis.

\subsection{Size measurements}

We first explored the standard technique of using {\sc SExtractor} to measure the half-light radius of high-redshift LBGs (e.g.~\citealp{Jiang2013b, Grazian2012, Oesch2010}).
{\sc SExtractor} calculates the half-light radius ($r_{1/2}$) in circular apertures, with reference to the total flux calculated with Kron-type elliptical apertures ({\sc FLUX\_AUTO}).
As discussed by several authors~\citep{Curtis-Lake2016, Huang2013, Grazian2012}, {\sc SExtractor} tends to underestimate the half-light radius at large input galaxy sizes, due to low-surface brightness emission being unaccounted for in the calculation of the total magnitude.
An additional concern for our sample in particular, is the de-blending of clumpy objects into multiple galaxies performed by {\sc SExtractor}.
We require a measurement of the size of the full clumpy galaxy or merging system and hence the fiducial {\sc SExtractor} sizes cannot be  simply utilized in our analysis. 
The sizes of the de-blended individual components of the galaxies by {\sc SExtractor} were retained however, and compared to other methods described below.

We also explored using {\sc GALFIT}, which provides an alternative, parametric, method to determine galaxy sizes by fitting simple 2D galaxy profiles to the data.
{\sc GALFIT} derived sizes are also commonly used at high redshift (e.g.~\citealp{Shibuya2015size, Huang2013, Ono2012, Law2012}).
Typically, single S{\'e}rsic profiles are assumed, where the intensity is parameterised as $I(R) \propto e^{(-k(R/R_{1/2})^{1/n})}$ with a particular S{\'e}rsic index $n$ ($n = 1$ gives an exponential disk profile, and $n = 4$ gives the de Vaucoleurs profile generally found for elliptical galaxies).
As for {\sc SExtractor}, a disadvantage of using {\sc GALFIT} to determine the sizes of galaxies in our sample is that it provides half-light radii for the individual components, rather than the full galaxy.
Furthermore, we do not have sufficient signal-to-noise to constrain the S{\'e}rsic index for each component and hence a S{\'e}rsic index must be assumed as is standard practice at high-redshift (typically $n = 1.0$, e.g.~\citealp{Ono2012} or $n = 1.5$, e.g.~\citealp{Oesch2010, Shibuya2015size}).
Inspecting the residuals from single S{\'e}rsic profile fitting with {\sc GALFIT} with a fixed $n = 1.5$, we find that this profile does not provide a good fit to the full sample due to the extended clumpy emission (see also~\citealp{Ribeiro2016}).
Instead we seek a non-parametric method to measure the individual galaxy sizes.

We therefore choose to estimate the sizes for our sample using a non-parametric half-light radius measure, obtained from the curve-of-growth for each galaxy.
We measure the flux in progressively larger circular apertures centred on the barycentre of the system (Equation~\ref{equation}), simply identifying the radius of an aperture that contains half of the total flux, where the total flux was measured in apertures of diameter $2$ or $3$ arcsec as motivated in Section~\ref{sect:totalmag}.
The COG method is very similar to that employed by {\sc SExtractor}, however it allows full flexibility to mask components identified to be at low-redshift, to centre the apertures appropriately and to determine the total flux of the galaxy in larger apertures.
Low-redshift galaxies close to the line-of-sight to the central LBGs were masked using the {\sc SExtractor} segmentation map, where individual {\sc SExtractor} objects were associated with the ground-based $z \simeq 7$ galaxy within a radius of $1$ arcsec.
Extended flux beyond the {\sc SExtractor} detection threshold in the outskirts of low-redshift galaxies in the stamp was masked by growing the segmentation map by 2 pixels.
We estimated and subtracted the median sky background determined from within an annulus of diameter 3 to 5 arcsec in the masked stamp.
The uncertainty in the COG $r_{1/2}$ was calculated by perturbing the total magnitude according to the magnitude error which, because of the large radii used to contain the total flux, is the dominant source of uncertainty in the measurement.
Comparing our non-parametric $r_{1/2}$ from {\sc SExtractor} and the COG method we find good agreement for individual components and find no evidence for a bias.
Finally, the non-parametric measurement of the half-light radius from the COG and {\sc SExtractor} must be corrected for the effect of the PSF.
To allow a direct comparison with previous results~\citep{Oesch2010, Jiang2013b}, we correct for the PSF by subtracting the $r_{1/2}$ obtained from the PSF in quadrature.
The validity of this approximation is discussed further in Section~\ref{sect:stack}.
A high signal-to-noise point-spread function was created by stacking unsaturated stars in our data.
The $J_{125} + H_{160}$ stack and the $JH_{140}$ imaging have comparable PSFs, with half-light radii of $0.13\,$ arcsec and $0.14\,$ arcsec measured by {\sc SExtractor} and the COG analysis respectively.

\subsection{Individual galaxy sizes}

\begin{figure}
\includegraphics[width = 0.49\textwidth]{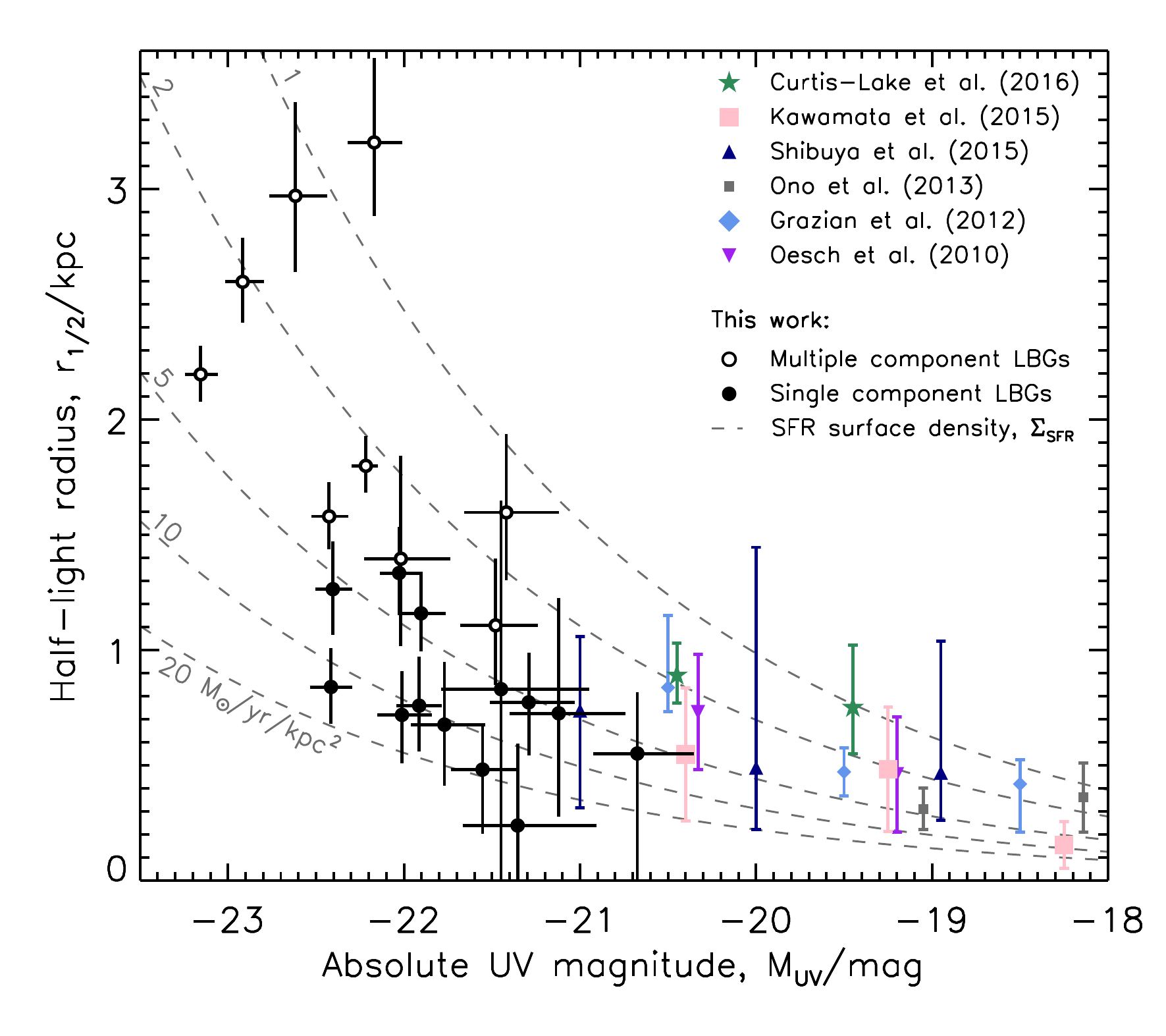}
\caption{The half-light radii of our galaxy sample, plotted against the absolute UV magnitude.
The sizes were measured using a COG analysis and corrected for the affect of the PSF in quadrature.
Galaxies with multiple components suggestive of a clumpy or merger-like system are highlighted as open circles.
At $M_{\rm UV} > -21$ we show the previous results from~\citet{Oesch2010} (purple triangles),~\citet{Ono2012} (grey squares),~\citet{Grazian2012} (blue diamonds),~\citet{Shibuya2015size} (navy triangles),~\citet{Kawamata2015} (salmon squares) and~\citet{Curtis-Lake2016} (green stars).
Lines of constant SFR surface density are shown as the dashed lines, ranging from $\Sigma_{\rm SFR} = 1$--$20\:$\sfrsdunits (assuming a~\citealp{Salpeter1955} IMF).
}\label{fig:sizemagindi}
\end{figure}

Using the COG method we calculate the half-light radii of the galaxies in our sample and plot them in Fig.~\ref{fig:sizemagindi}.
Reassuringly at the faint-end of our sample ($M_{\rm UV} \simeq -21.5$) we find good agreement with the sizes derived from previous studies.
The sample was split into single and multiple component systems by visual inspection, with the multiple-component galaxies highlighted in Table~\ref{table:sample}.
For the brightest galaxies that tend to appear as multiple component, clumpy, systems, we find considerably larger sizes ($r_{1/2} > 1\,{\rm kpc}$) than for the galaxies with a smoother, single component morphology.
The importance of clumpy or merging-type morphologies in the brightest galaxies at $z \simeq 7$ is clear from Fig.~\ref{fig:sizemagindi}, where these objects are consistently measured to have larger sizes.
The results also show the range of morphologies present in the brightest objects, as for LBGs with $M_{\rm UV} < -22.5$, the derived sizes range from $0.8\,{\rm kpc}$ to $\simeq 3\,{\rm kpc}$.
The measured sizes of the individual clumps in the multiple-component galaxies are consistent with those of the `single component' objects at the luminosity of the clump.

For three LBGs imaged with~\emph{HST}, the size is consistent with a point source within the errors (ID104600, ID35327 and ID271028).
These galaxies are amongst the faintest in the sample, and hence the sizes are more uncertain, however we inspected the SED fits for these objects to identify any potential brown dwarf contamination.
We find that a high-redshift galaxy SED is preferred to that of a brown dwarf in each case.
While the measured sizes are consistent with the galaxies being unresolved at~\emph{HST}/WFC3 resolution, inspection of the imaging available shows extended or elongated emission for these objects.
It is therefore unlikely that they are brown dwarfs, although it cannot be completely excluded with the current data.
An alternative explanation for the compact emission for example in ID104600, which has the smallest $r_{1/2}$ measured in our sample, could be the presence of an Active Galactic Nuclii (AGN).
The faint-end of the quasar luminosity function at $z \simeq 7$ is poorly constrained~\citep{Venemans2013}, however by making realistic assumptions about the form of the LF we expect $< 1$ low-luminosity high-redshift AGN in our sample~\citep{Bowler2014}.
Given the uncertainties in the faint-end slope of the quasar LF at high-redshift however, the presence of one quasar in our sample would not be unexpected (see~\citealp{Willott2010}).

\subsection{Stacked profile}\label{sect:stack}

\begin{figure*}

\includegraphics[width = 0.33\textwidth, trim = 0.5cm 0 0 1.0]{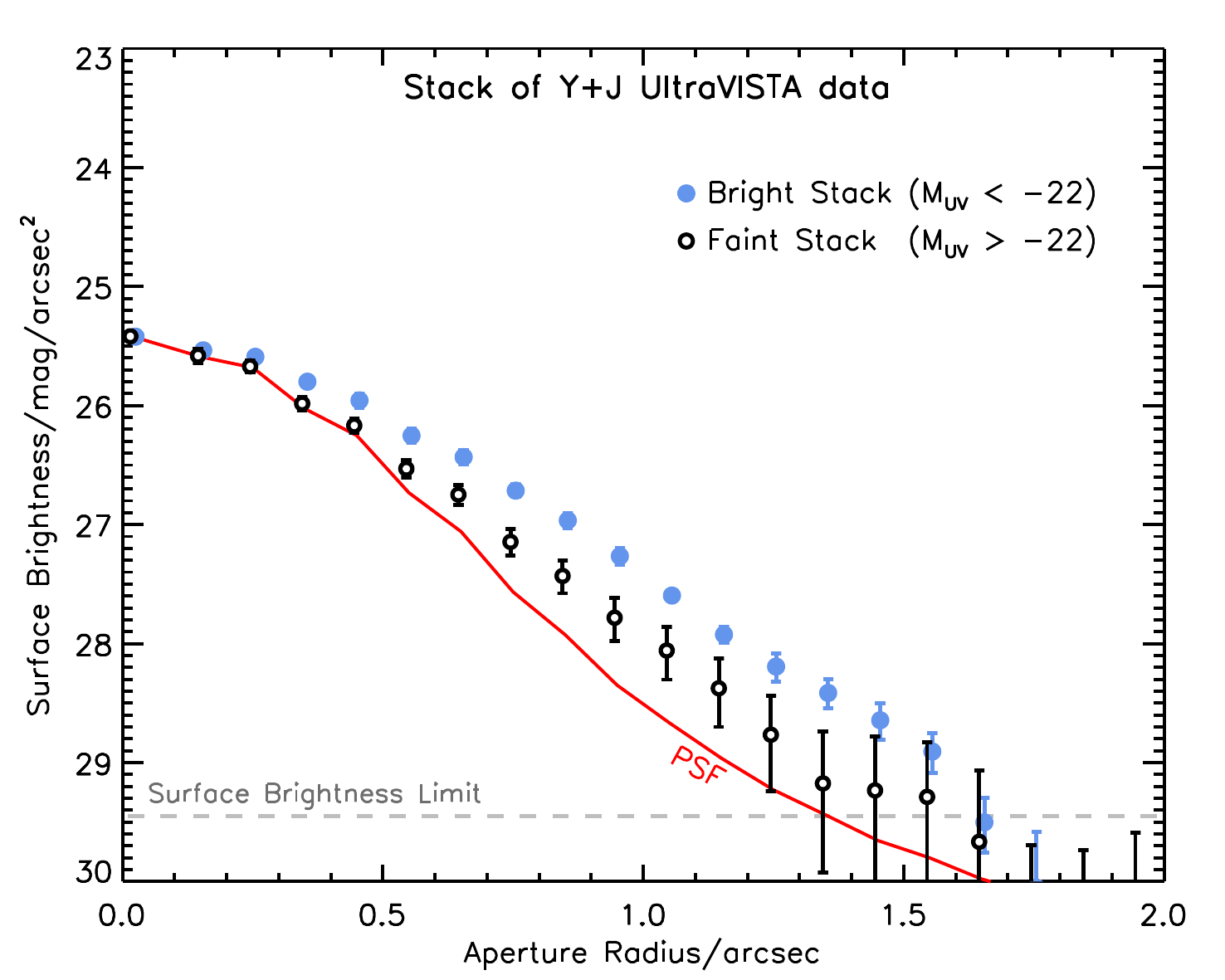}
\includegraphics[width = 0.33\textwidth, trim = 0.5cm 0 0 1.0]{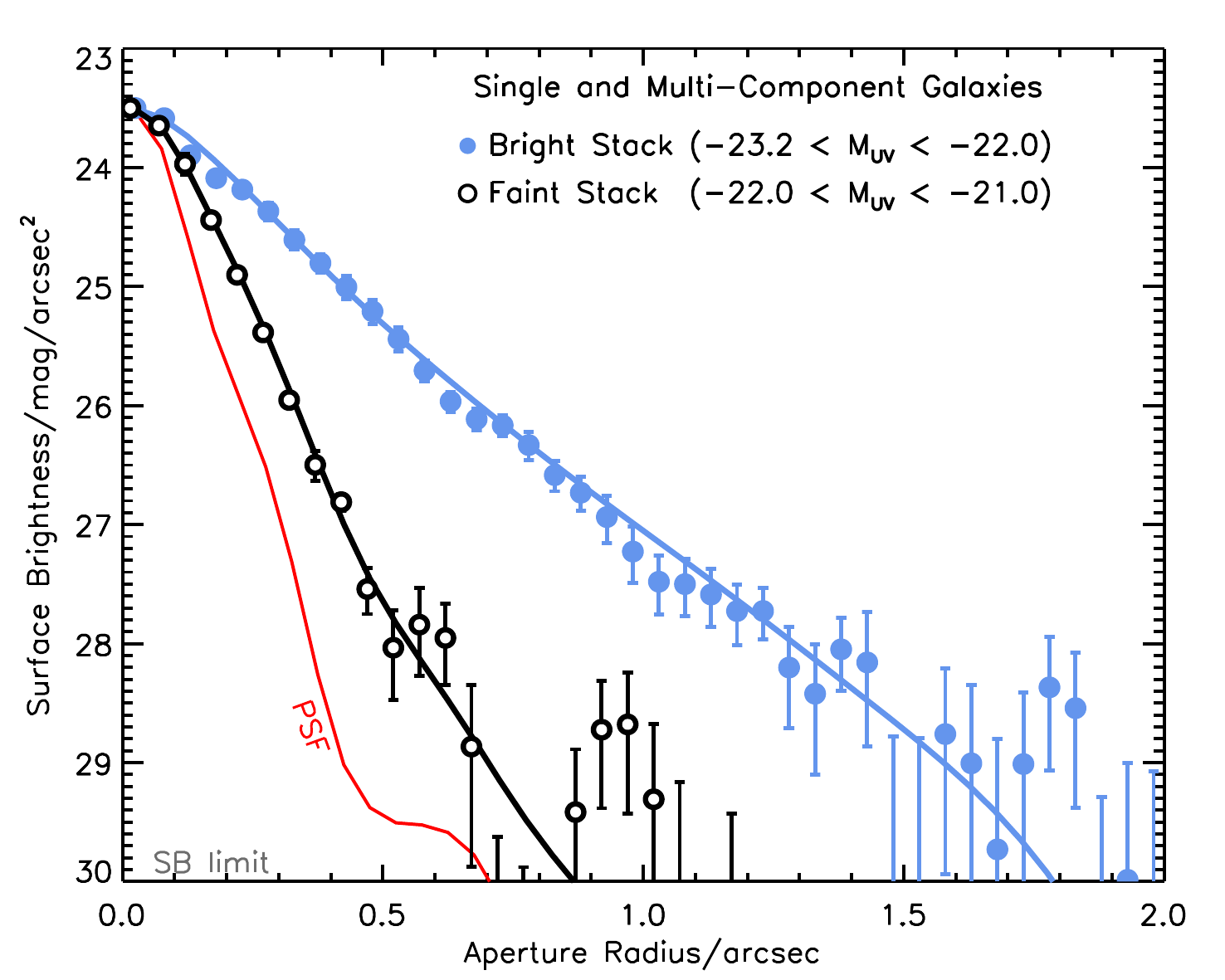}
\includegraphics[width =0.33 \textwidth, trim = 0.5cm 0 0 1.0]{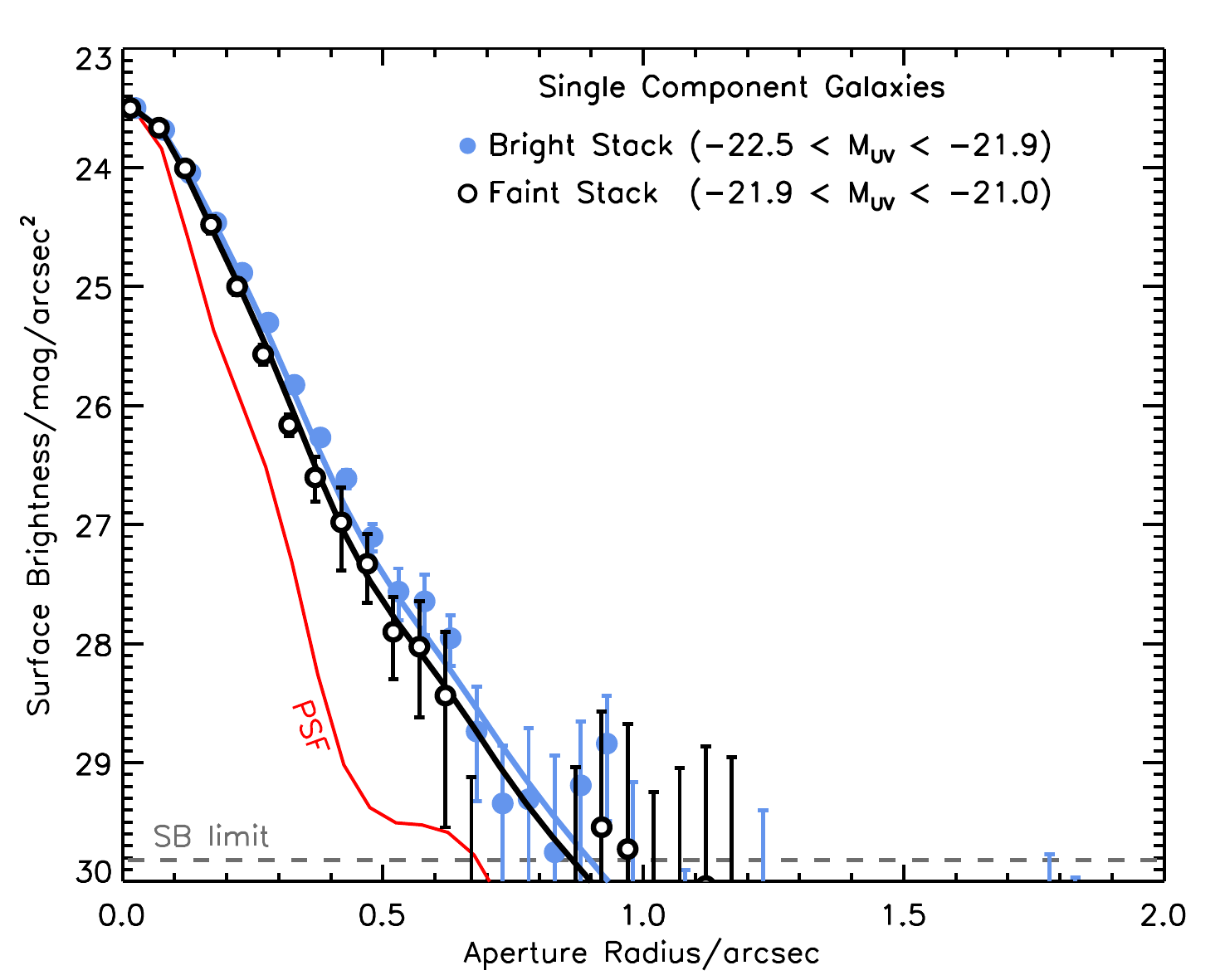}

\caption{The surface-brightness profiles for stacks of our sample of bright $z \simeq 7$ LBGs. 
The left-hand plot shows the results from the ground-based UltraVISTA data, while the central and right-hand plot shows the results from~\emph{HST}/WFC3 imaging.
The PSF for each dataset is shown as the red line, and the surface-brightness limit for each stack is shown as the horizontal dashed line.
The bright sub-set of the data with $M_{\rm UV} \lesssim -22$ is shown as blue circles, and the stack for the fainter objects is shown with open black circles.
The best-fitting S{\'e}rsic model fit for the~\emph{HST}/WFC3 stacks are shown as the solid blue or black line for the brighter and fainter sub-sets respectively.
The peak surface-brightness of the stacks of fainter galaxies have been scaled to match that of the brighter stack in each figure.
}\label{fig:groundstack}
\end{figure*}

To determine the average properties of bright LBGs at $\simeq 7$ we created stacks of the galaxies in our sample.
The range of absolute magnitudes that were included in each stack is presented in Table~\ref{table:stack}.
First, we simply stacked the 10 brighter ($M_{\rm UV} < -22.0$) and 11 fainter ($M_{\rm UV} > -22.0$) LBGs irrespective of morphology (`CR7' was excluded because it does not have imaging in consistent~\emph{HST} filters).
Secondly, we created a bright and a faint stack from the sub-set of galaxies that show only single components in the imaging.
Finally, we created stacks from the ground-based data available to provide a comparison to the new~\emph{HST} results.
The stacks were performed at the barycentre of each object, rather than the peak flux, to attempt to recover the average profile of the full extended system.
Such an approach is also employed for clumpy H$\alpha$ emitters at lower redshift (e.g.~\citealt{Nelson2016}).
To create the stacks, background-subtracted cut-outs of each object were centred to the sub-pixel centroid by sub-sampling the data by a factor of 100, using a bilinear interpolation.
The individual centred stamps were then median stacked, with the errors at each radius calculated from the standard error on the mean.

\begin{table*}
\caption{The measured half-light radius and S{\'e}rsic index for the stacks shown in Fig.~\ref{fig:groundstack}.
The errors were calculated from bootstrap resampling with replacement.
The upper rows show the results from stacks including all galaxies irrespective of visual morphology.
In the lower rows we present the results from stacking only the single component objects.
The first column shows the stack name, followed by the magnitude range of galaxies in the stack.
The half-light radius obtained from a COG analysis (corrected for the PSF in quadrature) is shown in the 3rd column, followed by the circularized half-light radius and S{\'e}rsic Index obtained from {\sc GALFIT} for the stack.
In the final column we show the number of galaxies included in each stack.
}
\begin{tabular}{l c c c c c c}

\hline
Stack Name & Magnitude Range & $r_{1/2}/{\rm kpc} $ COG & $r_{1/2}/{\rm kpc} $ GALFIT & S{\'e}rsic Index & Number \\
\hline

All galaxies (single + multiple comp.) & $     -22.0< M_{\rm UV} <      -21.0$ & $0.78^{+0.18}_{-0.11}$ & $0.57^{+0.14}_{-0.12}$ & $1.3^{+1.2}_{-0.3}$ & 11 \\
All galaxies (single + multiple comp.) & $     -23.2< M_{\rm UV} <      -22.0$ & $2.30^{+0.21}_{-0.76}$ & $2.07^{+0.39}_{-0.81}$ & $1.3^{+2.1}_{-0.3}$ & 10 \\
Single Component Galaxies & $     -21.9< M_{\rm UV} <      -21.0$ & $0.65^{+0.10}_{-0.16}$ & $0.51^{+0.10}_{-0.12}$ & $1.6^{+0.8}_{-0.5}$ & 6\\
Single Component Galaxies & $     -22.5< M_{\rm UV} <      -21.9$ & $0.92^{+0.14}_{-0.19}$ & $0.65^{+0.09}_{-0.17}$ & $1.7^{+1.5}_{-0.2}$& 6\\
\hline
\end{tabular}\label{table:stack}
\end{table*}

The surface-brightness profiles for the six stacks are shown in Fig~\ref{fig:groundstack}.
The stack of the ground-based UltraVISTA data shows that the brightest galaxies in the sample are clearly resolved even under $\simeq 0.8$ arcsec seeing (as expected from the sizes measured in~\citealp{Bowler2014}), with the fainter objects also appearing marginally resolved.
The fact that the galaxies are resolved in the ground-based data is encouraging for future wide-area searches for high-redshift galaxies, for example in the full VISTA VIDEO survey~\citep{Jarvis2013}, where contaminant brown dwarfs significantly exceed the surface densities of LBGs~\citep{Bowler2015}.
The difference in size between the fainter and brighter samples found in the ground-based data is confirmed with the improved spatial resolution provided by~\emph{HST}/WFC3.
When all galaxies are included in the stacks (single and multiple systems), we find the brightest objects are highly extended with flux visible to a radius of $1.5$ arcsec.
The fainter stack however appears compact although it is clearly resolved by~\emph{HST}/WFC3.
The resulting stacked images can be fitted with a S{\'e}rsic profile.
The half-light radii of the stacked profiles obtained using both {\sc GALFIT} and the COG analysis are shown in Table~\ref{table:stack}.
{\sc GALFIT} returns the effective radius along the semi-major axis, and hence we circularize this radius using the axis ratio ($r_{1/2} = r_{e}\sqrt{b/a}$).
Errors on the sizes were obtained by created multiple stacks using bootstrap resampling with replacement.
We present the $68$ percent confidence interval on derived parameters, obtained from the resulting probability distributions.
Reassuringly, the half-light radius obtained from the bright stack of galaxies in the~\emph{HST} data is consistent with that obtained by fitting to the ground-based stack ($r_{1/2} = 1.9\,{\rm kpc}$ with a fixed $n = 1.5$).
The S{\'e}rsic indices are poorly constrained by our data due to the range of input morphologies.
However the distribution of S{\'e}rsic indices peaks around $n = 1.5$ as is commonly assumed at high redshift, with a longer tail to higher values.
As illustrated in Fig.~\ref{fig:groundstack}, we find a large difference in the average profile between galaxies at $M_{\rm UV} \simeq -21.5$ and $M_{\rm UV} \simeq -22.5$, with the average half-light radius increasing by a factor of $\simeq 3$.
If instead we only stack those objects which show little evidence for clumps or interactions (i.e. the single component galaxies), we find that the brighter and fainter stacks are have similar sizes with $r_{1/2} \simeq 0.5$--$0.9\,{\rm kpc}$.
The stacks visually show the strong effect on the derived half-light radius that results from the inclusion of multiple-component systems.
While the fainter stacks in each case (with or without multiple-component galaxies) show similar sizes, the brighter stack has a half-light radius more than twice that of the similarly bright, single component objects.
This indicates that clumpy/merging systems are particularly important at the very bright-end of our sample ($M_{\rm UV} < -22.5$), which we discuss further in the context of the size-luminosity relation in the next section.

\subsubsection{Correcting the size measurements for the PSF}

As shown in Table~\ref{table:stack}, the half-light radii obtained from single S{\'e}rsic profile fitting with {\sc GALFIT} are systematically smaller than those obtained with the COG non-parametric method.
The COG sizes were corrected for the PSF smoothing by subtracting the half-light radius of the PSF in quadrature ($r_{\rm int}^2 = r_{\rm obs}^2 - r_{\rm PSF}^2$), an approach that has been used by several studies at high redshift ~\citep{Oesch2010, Jiang2013b}.
This prescription is analytically valid for the convolution of two Gaussian profiles, however as shown by~\citet{Curtis-Lake2016}, this approximation is not sufficient in the case of a PSF that shows extended wings when using a COG-like size measurement.
The PSF wings act to distribute light from the compact galaxy profile to larger radii, resulting in a larger observed $r_{1/2}$ than that obtained after convolution with a Gaussian function.
The result is that for a given input S{\'e}rsic profile, there is an approximately constant offset between the output sizes after the simple quadrature correction, compared to the input value (see figure 1 of~\citealp{Curtis-Lake2016}).
The COG measured sizes presented in this paper therefore represent upper limits on the likely $r_{1/2}$ measurements for the sample.
The additional correction beyond the simple quadrature subtraction presented above depends on the underlying galaxy light profile and on the aperture diameter used to approximate the total flux of the object.
We investigate the dependencies of the offset for the galaxies in our sample using simple model S{\'e}rsic profiles and the empirically derived PSF.
In agreement with~\citet{Curtis-Lake2016}, we find that if the stack is a simple $n = 1.5$ S{\'e}rsic model, then the sizes measured using the COG (and corrected for the PSF in quadrature) also require an additional correction to smaller sizes of around $\simeq 30$ percent.
Such a correction would result in good agreement between the sizes derived for the stacks from {\sc GALFIT} and the COG analysis.
Due to the unknown underlying morphology for the stacks however (as illustrated by the large errors on the S{\'e}rsic Index), the exact correction is uncertain, and hence we present both size measurements to represent a realistic range in the likely $r_{1/2}$ for the average profile of bright $z \simeq 7$ LBGs.

\subsection{The size-luminosity relation}

\begin{figure}
\includegraphics[width = 0.49\textwidth]{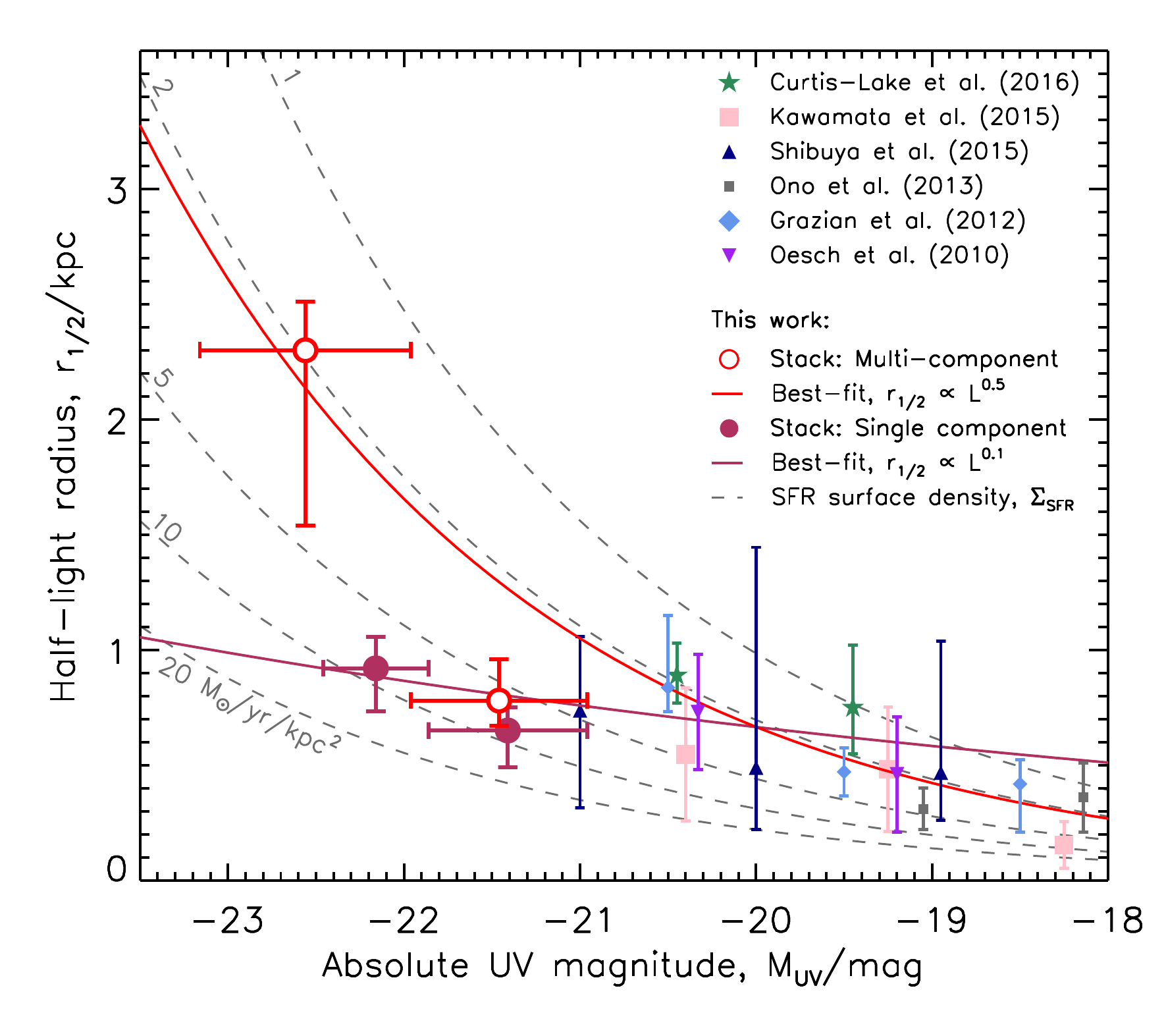}
\caption{The size-luminosity relation at $z \simeq 7$.
The results of this work are shown as the red circles bright-ward of $M_{\rm UV} = -21$.
The half-light radii were derived from the stacked profiles displayed in Fig.~\ref{fig:groundstack} using the COG analysis.
Open circles show the results when all galaxies are included regardless of visual morphology, and the maroon circles show the results when only single component galaxies are considered.
The best-fitting size-luminosity relation to two sets of stacked points are shown as the solid red and maroon lines.
Previous results are shown as in Fig.~\ref{fig:sizemagindi}.
The grey dashed lines show the lines of constant star-formation surface density in the range \sfrsd $ = 1$--$20\,$\sfrsdunits.
}\label{fig:sizemag}
\end{figure}

Using the stacked profiles we are able to measure the average sizes of LBGs at $z \simeq 7$ to bright ($ M_{\rm UV} < -22$) magnitudes for the first time, hence extending the baseline for exploring the size-luminosity relation.
The size-luminosity relation is typically parameterised as a simple power-law, with $r_{1/2} \propto L\,^{\gamma}$, with typical values of $\gamma \sim 0.2$--$0.3$ for star-forming galaxies at low redshift (e.g.~\citealp{Shen2003}, see discussion in~\citealp{Huang2013}).
At $z \simeq 4$--$6$, the slope of the size-luminosity relation has been constrained to $\gamma \sim 0.2$~\citep{Huang2013, Curtis-Lake2016}.
At $z \simeq 7$ however, the slope of the relation is uncertain, due in part to the reduced luminosity range available in samples derived using relatively small area~\emph{HST} surveys.
For example, using the CANDELS data,~\citet{Curtis-Lake2016} found a shallow relation similar to that at lower redshifts ($\gamma = 0.19 \pm 0.38$) whereas~\citet{Grazian2012} found evidence for a significantly steeper size-luminosity relation (with $r_{1/2} \propto L^{1/2}$).
In a comparable study to this work at $ z\simeq 6$,~\citet{Jiang2013b} found a flat size-luminosity relation when including galaxies as bright as $M_{\rm UV} \simeq -22.5$, with a slope of $\gamma = 0.14 \pm 0.03$.
In the sample of bright $z \simeq 7$ galaxies presented in this paper, we have found several galaxies that show multiple component, clumpy morphologies that could be interpreted as merging systems.
If these objects are in a transitional merging state, then for comparison to disk or early-type galaxies at lower redshift (where irregular morphologies are uncommon, e.g.~\citealp{Buitrago2013} and~\citealp{Mortlock2013}) they should be excluded from the size-luminosity relation.
Typically at high redshift, multiple-component galaxies have either been too rare in the sample to dramatically effect the relation~\citep{Oesch2010} or they have been removed from the size-luminosity relation analysis as atypical objects~\citep{Jiang2013b}.
In the calculation of the size-mass relation at $z \simeq 2$--$3$,~\citet{Law2012} use the size of the brightest component as a proxy for the galaxy size in the case of multiple distinct components.
An alternative approach presented in~\citet{Kawamata2015}, is to fit a single S{\'e}rsic profile to galaxies with multiple cores.
We therefore present the results of our size measurement of $z \simeq 7$ galaxies both with and without the inclusion of visually identified multiple component galaxies and compare the results.
In Fig.~\ref{fig:sizemag} we show the sizes derived from using the COG measurement on the stacked galaxy profiles, compared to those from results derived in the UDF~\citep{Ono2012, Oesch2010}, the HFF~\citep{Kawamata2015} and with CANDELS~\citep{Grazian2012, Shibuya2015size, Curtis-Lake2016}.
Given the uncertainties in the PSF correction to the galaxy size, the COG results should be considered upper limits.
For the faintest bin, we find good agreement with the sizes derived from fainter studies, indicating that there is no strong size-luminosity relation apparent faint-ward of $M_{\rm UV} \simeq -22$.
Comparing our results at $M_{\rm UV} = -21.5$ derived with GALFIT with previous studies at $M_{\rm UV} \simeq -20.5$ however, we find a slight tension (at the $2\sigma$ level) with the brightest bin derived by~\citet{Grazian2012} and~\citet{Curtis-Lake2016}.
When including the multiple-component objects in addition to those with smoother profiles, we find significantly larger sizes at the bright end of the sample with a rapid increase in $r_{1/2}$ observed brighter than $M_{\rm UV} \lesssim -22.5$ or equivalently, a ${\rm SFR} > 25\,{\rm M}_{\odot}\,{\rm yr}^{-1}$.

Fitting a power-law to our results combined with those from previous studies at fainter magnitudes provides a rough estimate of the slope of the $z \simeq 7$ size-luminosity relation.
To provide two independent estimates of the slope, we fit our derived $r_{1/2}$ values obtained from the stacks using either {\rm GALFIT} or {\rm SEXTRACTOR}/COG separately, coupled with previous results at fainter magnitudes that used a similar methodology.
For our {\sc GALFIT} results we fit to the points from~\citet{Ono2012},~\citet{Kawamata2015} and~\citet{Shibuya2015size} and for the {\sc SExtractor}/{\sc COG} method we include the points from~\citet{Oesch2010},~\citet{Grazian2012} and~\citet{Curtis-Lake2016}.
The resulting fits are shown in Fig.~\ref{fig:sizemag}.
When considering only the single-component galaxies we find a gradient of $0.14 \pm 0.08$ for the non-parametric method ({$\gamma = 0.22 \pm 0.10$ using {\sc GALFIT}).
This slope is consistent with the previous results at $z = 4$--$6$, which show $\gamma \simeq 0.25$ (\citealp{Curtis-Lake2016};~\citealp{Shibuya2015size};~\citealp{Jiang2013b};~\citealp{Huang2013}).
The slope of our size-luminosity relation is also in good agreement with the results at $z \sim 7$ from~\citet{Shibuya2015size} who found $\gamma = 0.26 \pm 0.06$ (see their figure 10).
The similar gradient obtained from the two size measurement methods indicates that there is a systematic difference in sizes obtained using the two methods over a range of magnitudes, potentially due to the uncertainties in the PSF correction.
Our results are unchanged if we exclude the LAEs CR7 and Himiko, and the LBG with a best-fitting photometric redshift at $z < 6.5$ (without Lyman-$\alpha$ in the fitting).

If we fit to the full sample, including the multiple component systems, we find a steeper gradient of $\gamma = 0.50 \pm 0.07$ for the COG method ($\gamma = 0.49 \pm 0.08$ for {\sc GALFIT}) more consistent with the results of~\citep{Grazian2012}. 
Bright-ward of $M_{\rm UV} \sim -22.5$ we find a sharp increase in the average size of the LBGs in our sample.
It is therefore unsurprising that we find good agreement with the gradient of the size-luminosity relation found by previous studies, as they were typically limited to fainter galaxies found within~\emph{HST} survey data.
Our results are also consistent with the shallow size-luminosity relation found in studies of brighter galaxies at $z =4$--$5$ by~\citet{Huang2013} and at $ z\simeq 6$ by~\citet{Jiang2013b}}, as the samples considered in these studies were limited to $M_{\rm UV} > -22.5$.
Finally, in the~\citet{Huang2013} study, sizes were measured assuming single S{\'e}rsic profiles, which will result in a smaller measured galaxy size in the presence of multiple components, than would be obtained with our non-parametric method.

\section{Discussion}\label{sect:discussion}

The dependence of the measured average sizes and the derived size-luminosity relation on the treatment of multiple component, clumpy galaxies, is clear from Fig.~\ref{fig:sizemag}.
Previous studies of the sizes and morphologies of high-redshift LBGs have tended to focus on the more numerous, but fainter galaxies found within the relatively small area surveys from~\emph{HST}.
Within these surveys, several authors have noted the presence of disturbed and clumpy morphologies (e.g.~\citealp{Oesch2010, Jiang2013b}), however these objects have been sufficiently rare to impact the derived size-luminosity relation.
At the very bright end of our $z \simeq 7$ sample however, we find that objects with distinct components resolvable by~\emph{HST}/WFC3 become increasingly common and dominate at $M_{\rm UV} < -22.5$.
If these objects had been selected in high-resolution~\emph{HST}/WFC3 data, they would have initially been de-blended by {\sc SEXtractor} into several distinct components, rather than classified as a single galaxy as in our ground-based selection.
It is therefore clear that for the bright galaxies in our sample, the imaging resolution and the methodology used to classify a single galaxy can have a significant impact on the derived size-luminosity relation and the galaxy luminosity function.
As we argue in Section~\ref{sect:lf}, the multiple components observed in the brightest galaxies in our sample are very likely to be physically associated, due to the close separation and the enhanced luminosity of the components compared to in the field.
In this case, the selection of the brightest galaxies in ground-based imaging provides an advantage over higher resolution data, as it clearly associates clumps that would have been resolved into separate components with~\emph{HST}.
While the close physical separation of the multiple components we observe implies that they occupy the same dark-matter halo at $z \simeq 7$, the classification of a single galaxy in simulations of structure formation will depend on the algorithm used to group star particles and on the formation mechanism.
For example, if the clumps are formed as part of a gas-rich disk or in the merger of two galaxies, this will impact the underlying structure (such as a diffuse disk-like component, or bridges of stars) that is present in the simulations but is beyond the depth and resolution limit of our data.
In future comparisons to simulations, it is therefore essential to treat clumpy galaxies in a consistent way to provide a meaningful comparison to the observed $z \simeq 7$ rest-frame UV LF and size-luminosity relation, particularly at the bright end. 

\subsection{Clumpy star formation or merging systems?}

We find that the brightest $z \simeq 7$ galaxies in our sample are composed of multiple components, suggestive of clumpy star formation or merging galaxies (Fig.~\ref{fig:hstsb}).
Using deep optical imaging from the ground-based imaging and~\emph{HST}/ACS we can confirm that the individual components are at $z > 6$, however with the currently available data it is not clear whether the observed clumps are formed in-situ or as part of an interaction.
Irregular and clumpy galaxies become increasingly common at higher redshift, with the fraction of clumpy galaxies rising from $z \simeq 1$ to $z \simeq 3$ (e.g. recent work by~\citealp{Guo2015, Shibuya2015clump}).
While the clumps can be embedded in a larger structure or disk (e.g.~\citealp{Law2012}), there are samples of star-forming galaxies such as the clump-cluster objects selected by~\citet{Elmegreen2005,Elmegreen2009} that appear visually similar to the bright galaxies in our sample (see also~\citealp{Wuyts2012}).
The clumps detected at lower redshift are generally thought to have formed by a violent disk instability in gas-rich, turbulent disks, rather than by major mergers (see~\citealp{Shapley2011}).
For the bright, visually disturbed and clumpy galaxies in our sample, we find widely separated components that are resolved by~\emph{HST}.
The `clumps' found in our $z \simeq 7$ LBGs contain the majority of the galaxy flux, whereas on average at $z \simeq 2$--$3$ clumps contribute $ \lesssim 30$ percent of the total galaxy luminosity, with the maximum clump contribution of $\simeq 50$ percent~\citep{Guo2015}.
If instead the clumps were part of a diffuse disk component of $r = 4.0\,{\rm kpc}$ that contained $70$ percent of the galaxy flux, this disk component would be identifiable in the images, which have a surface-brightness limit of $\gtrsim 26.0\, {\rm mag}/{\rm arcsec}^2$ (see Fig.~\ref{fig:hstsb}).
While there are several cases in lower-redshift samples where the clumps also contribute a large fraction of the galaxy luminosity (e.g.~\citealt{Livermore2015}), it is plausible that at $z \simeq 7$, star-formation clumps more typically dominate the luminosity of the galaxy due the increased cold-gas fraction at high redshift (e.g.~\citealp{Mandelker2017}).
Finally, we note that the measurements of size and morphology at $z \simeq 7$ currently available are based on the rest-frame UV emission, which can appear significantly more clumpy than at longer wavelengths (e.g.~\citealp{Wuyts2012}).
The~\emph{James Webb Space Telescope} will provide the first high-resolution view of the rest-frame optical emission of these early galaxies, allowing a unique insight into the spatial distribution of the underlying stellar mass and the star-formation via the rest-frame optical nebular emission lines such as H$\alpha$.

Instead the multiple-component galaxies in our sample could be interpreted as merging systems.
The individual components observed in the brightest galaxies are separated by up to $\sim 5\,{\rm kpc}$, have luminosities from $L \simeq 2.5$--$7\,L^*$ and show sizes that are consistent with fainter galaxies.
Using the Illustris simulation,~\citet{Rodriguez-Gomez2015} showed that the major merger rate at $z \simeq 7$ is around $\sim 1\,{\rm Gyr}^{-1}$ for the most massive galaxies ($M_{\odot} > 10^{10}\,{\rm M}_{\odot}$).
Given that the merger visibility timescale is typically in the range of $400$-$500$ Myrs from simulations~\citep{Lotz2010} and is predicted to rise at high-redshift where galaxies are more gas rich~\citep{Lotz2010a}, it is feasible that we are observing multiple merging systems in the brightest galaxies at $z \sim 7$.
The interpretation of these galaxies as merger-driven star-bursts would provide a natural explanation for the bright magnitudes at which they are observed.
In the local Universe, galaxies in the process of merging are observed to have an increase in the SFR of the system (by a factor of $\sim 1.8$ or $0.6$ mag;~\citealp{Scudder2012}).
If we are observing the merger of two $L^*$ galaxies, then the increase in the SFR induced by the interaction could boost the luminosity of the system and hence increase the observed number density of LBGs at the bright-end of the LF.
Future work comparing to simulations of high-redshift galaxy formation would be instructive to determine whether the observed clumps in our sample likely originate from mergers or instead are formed in-situ at $ z\simeq 7$, and hence determine whether mergers are the main driver for the power-law decline we observe at the bright-end of the rest-frame UV LF.

Finally we note that the most star-forming galaxies are likely to arise in the highest density regions at $z \simeq 7$ (e.g as highlighted by clustering measurements, e.g.~\citealp{Harikane2016, Barone-Nugent2014, McLure2009}), and fainter galaxies should therefore be more highly clustered around brighter objects.
For the intrinsically brightest LBG in our sample, ID169850, we find a companion galaxy with a consistent photometric redshift (ID170216) only $40$ arcsec away ($\sim 200\,{\rm kpc}$ proper distance), in comparison to the next nearest LBG at $z \sim 7$ which is separated by over $12$ arcmin.

\subsection{Star-formation rate surface density}

The galaxy size and rest-frame UV luminosity can be used to calculate the SFR surface density, \sfrsd, at high redshift.
To provide a lower-limit on the \sfrsd~of our sample, we use the SFRs calculated directly from the rest-frame UV luminosity assuming no dust extinction~\citep{Madau1998, Kennicutt1998a}.
This conversion assumes a~\citet{Salpeter1955} IMF\footnote{Dividing the SFRs by a factor of 1.6 converts to a~\citet{Chabrier2003} initial mass function (IMF).}, which we use for comparison to previous results.
Using samples of fainter $z \simeq 7$ LBGs found within the UDF,~\citet{Oesch2010} and~\citet{Ono2012} derived a range of SFR surface density spanning \sfrsd $= 1$--$10\,$ \sfrsdunits.
Similarly,~\citet{Kawamata2015} found a weighted-log-average value of \sfrsd $= 4.1\,$ \sfrsdunits~for a sample of lensed galaxies.
In Fig.~\ref{fig:sizemag} we show the lines of constant \sfrsd~compared to the sizes and luminosities of galaxies in our sample.
We find only a shallow evolution in the size of $z \simeq 7$ LBGs to bright magnitudes, and hence the implied SFR surface density of the galaxies is increased (\sfrsd $\simeq 5$--$20\,$ \sfrsdunits).
Our results are consistent with the higher values derived in samples of brighter $z =6$--$10$ galaxies found within the CANDELS and other~\emph{HST} fields~\citep{Shibuya2015size, Holwerda2015}.

While the brightest objects in our sample show large sizes and hence lower \sfrsd~values, the sizes are driven by the separation of bright clumps, not the surface brightness.
The SFR surface density derived from the individual components of the clumpy/merging galaxies are similar to the single component galaxies in our sample, indicating that the brightest galaxies (and the dominant SF clumps within them) show a higher \sfrsd~on average, than that found in previous studies of fainter galaxies at $z \sim 7$.
The higher implied \sfrsd~for the individual components of the multiple-component galaxies supports our conclusion that the clumps are in a physical association, where the environment (either in an intense star burst or due to a merging process) is impacting the density of star formation. 
Both~\citet{Oesch2010} and~\citet{Ono2012} found an approximately constant SFR surface density of \sfrsd $ \simeq 3\,$\sfrsdunits~independent of galaxy luminosity in the range studied ($L = 0.3$--$1\,L^*_{z = 3}$).
Here the luminosity is expressed as a multiple of the characteristic luminosity at $z = 3$, which corresponds to $M^* = -21.0$~\citep{Steidel1999}.
These previous studies concluded that the dominant driver of the observed size-luminosity relation was an increase in galaxy size, with an approximately constant \sfrsd.
Using the larger dynamic range provided by our sample of luminous LBGs at $z \simeq 7$, we instead find that the sizes of the brightest LBGs (or the dominant SF clumps within multiple-component objects) show little trend with luminosity.
Hence, we find that the inferred SFR surface density is higher for the brightest LBGs at $z \simeq 7$ which have luminosities in the range $\simeq 1$--$7\,L^*_{z = 3}$, and that an evolution in \sfrsd~is the main driver of the observed flat size-luminosity relation.
\citet{Shibuya2015size} similarly conclude that there is an evolution in the average SFR surface density with luminosity or star-formation rate at $z \simeq 7$, with the median values for their sample increasing from  \sfrsd $\simeq 3^{+9}_{-2}\,$ \sfrsdunits~at $0.12$--$0.3\,L^*_{z = 3}$ to \sfrsd $\simeq 14^{+36}_{-10}\,$ \sfrsdunits~in the brightest objects ($1$--$10\,L^*_{z = 3}$).
At $z \simeq 1$--$3$, an observed \sfrsd $\gtrsim 10$ \sfrsdunits~is found predominantly in starburst, merging or infrared selected galaxies (e.g.~\citealp{Daddi2010, Genzel2010}; see figure 8 of~\citealp{Kawamata2015}), and it is plausible that the brightest (in the rest-frame UV) $z \simeq 7 $ LBGs studied here are similarly undergoing an unusually violent episode of star formation as compared to fainter galaxies.

\section{Conclusions}\label{sect:conclusions}

In this work we present new, high-resolution,~\emph{HST}/WFC3 imaging of the brightest star-forming galaxies at $z \simeq 7$.
The galaxy sample was initially selected by~\citet{Bowler2012, Bowler2014} from the ground-based UltraVISTA/COSMOS and UDS/SXDS fields.
By combining our Cycle 22~\emph{HST}/WFC3 imaging with archival data and imaging from the CANDELS survey, we present a complete analysis of the high-resolution~\emph{HST} data available for the galaxies in the original~\citet{Bowler2014} sample.
This consists of imaging for 25 LBGs at $z \simeq 7$ in the magnitude range $ -23.2 < M_{\rm UV} \lesssim -21.0$.
Our main conclusions are:
\begin{enumerate}

\item The brightest $z \simeq 7$ LBGs show a range of visual morphologies including apparently smooth single component profiles and extended and highly clumpy objects.
We find that multiple-component systems account for $> 40$ percent of sample, in agreement with other studies of bright LBGs at $z < 7$.
In the most luminous galaxies ($M_{\rm UV} < -22.5$) we find that multiple-component galaxies, suggestive of merging galaxies or clumpy star-formation, are ubiquitous.
The brightest LBGs studied here are up to $1\,$mag brighter in the continuum than the LAEs `CR7' and `Himiko', and show similarly extended, clumpy morphologies.

\item At the faint-end of our sample, the independent photometry enabled by the new~\emph{HST} data has revealed a new artefact in the VISTA/VIRCAM imaging caused by cross-talk between the CCDs.
The cross-talk artefact accounts for three of the original sample of 30 Lyman-break galaxy candidates selected in the UltraVISTA data in~\citet{Bowler2014}, and must be considered in future selection of faint objects using deep VIRCAM imaging.

\item A deconfusion analysis of the deep~\emph{Spitzer}/IRAC \chone~and \chtwo~data in the fields shows evidence for strong ($EW_{0} $(H$\beta + $\oiii)$ > 600$\AA) rest-frame optical nebular emission lines in the brightest $z \simeq 7$ LBGs.

\item We recalculate the rest-frame UV luminosity function from the~\citet{Bowler2014} sample using the new~\emph{HST}/WFC3 data, accounting for the VISTA/VIRCAM artefact and measuring the total galaxy luminosity using large ($2$--$3$ arcsec diameter) apertures.
We find that a double-power law remains the preferred functional form to fit the bright-end of the $z \simeq 7$ rest-frame UV LF. 
The results support a smooth evolution in the characteristic magnitude of $\Delta M^* \sim 0.5$ from $z \sim 5$ to $z \sim 7$.

\item The half-light radii of the galaxies determined from the~\emph{HST}/WFC3 imaging show a broad range from $r_{1/2} = 0.2$--$ 3.2\,{\rm kpc}$ when calculated with a non-parametric curve-of-growth method.
For the fainter galaxies that appear as single components in our data we find half-light radii of $r_{1/2} \sim 0.5\,{\rm kpc}$, consistent with previous analysis of fainter LBGs.
The brightest galaxies in our sample however, and dominated by multiple-component systems and show significantly larger sizes ($r_{1/2} > 2\,{\rm kpc}$).

\item We use stacked galaxy profiles to constrain the size-luminosity relation at $z \simeq 7$.
When considering only single-component galaxies we find a shallow relation consistent with previous results at lower-redshift, with $r_{1/2} \propto L^{0.1}$.
However if we include multiple-component galaxies, we instead find a significantly steeper slope consistent with $r_{1/2} \propto L^{0.5}$, illustrating the importance of these irregular, clumpy, galaxies at the bright end of our sample.

\item The shallow size-luminosity relation we find for the single-component galaxies and the individual star-forming clumps shows that the brightest galaxies have higher SFR surface densities compared to fainter objects at $z \simeq 7$.
The derived values of \sfrsd $\simeq 5$--$20\,$\sfrsdunits~are similar to those found in starburst galaxies and merger systems at lower redshift, indicating that the size-luminosity relation at $z \simeq 7$ is driven predominantly by a luminosity dependent SFR surface density not a strong evolution in galaxy size.

\end{enumerate}

In future, larger samples of similarly bright (and brighter) galaxies at $z > 6$ will be selected from wide-area surveys from, for example,~\emph{Euclid}~\citep{Laureijs2011} and the Large Synoptic Survey Telescope~\citep{Ivezic2008}.
The results of this work confirm that the number densities of the brightest ($M_{\rm UV} \simeq -22.5$) LBGs remain approximately constant from $z \simeq 5$ to $z \simeq 7$~\citep{Bowler2014,Bowler2015}, and that even within several degrees of imaging, we find galaxies as bright as $m_{\rm AB} = 24$.
These results are encouraging for upcoming surveys, with the most luminous galaxies in our sample being detectable even in the shallower `wide' component of the~\emph{Euclid} survey.
Using our updated DPL (Schechter) fit to the rest-frame UV LF we predict that~\emph{Euclid} will detect $3100^{+1300}_{-1200}$ ($3300^{+1200}_{-600}$) LBGs brighter than $m_{\rm AB} = 26$ at $z \simeq 7$ in the `deep' survey ($40\,{\rm deg}^2$;~\citealp{Laureijs2011}).
The comparable number counts between the DPL and Schechter parameterisation in this case reflects the similarity in these functions around $L^*$.
In contrast for the `wide' component ($15000\,{\rm deg}^2$) we predict $2000^{+5000}_{-1500}$ LBGs brighter than $m_{\rm AB} = 24$ from our DPL function fit, and $\lesssim 50$ for the Schechter fit.
We note that both the DPL and Schechter functions underestimate the observed number of LBGs at very bright magnitudes.
If we simply take the observed number density of galaxies brighter than $m_{\rm AB} = 24$ in our survey ($1$ galaxy over $1.65\,{\rm deg}^2$), we predict $\simeq 10000$ LBGs in the~\emph{Euclid} `wide' survey.
Therefore it is evident that wide area surveys will be able to clearly distinguish between the proposed functional forms for the $z \simeq 7$ rest-frame UV LF.
Cool galactic brown dwarfs have surface densities that outnumber high-redshift galaxies at bright magnitudes ($m_{\rm AB} < 25$;~\citealp{Bowler2015,Ryan2011}), and hence are the dominant contaminant for bright LBG samples selected over very wide areas.
The large observed sizes of the brightest LBGs in our sample however, reassuringly suggest that the $z \simeq 7$ LBGs detected by~\emph{Euclid} will be clearly spatially resolved and be dominated by irregular, clumpy morphologies.

\section*{Acknowledgements}

This work was supported by the Oxford Centre for Astrophysical Surveys which is funded through generous support from the Hintze Family Charitable Foundation.
RAAB and JSD acknowledge the support of the European Research Council via the award of an Advanced Grant (PI J. Dunlop), and the contribution of the EC FP7 SPACE project ASTRODEEP (Ref.No: 312725). 
RJM and DJM acknowledge the support of the European Research Council via the award of a Consolidator Grant (PI R. McLure).
This publication arises from research partly funded by the John Fell Oxford University Press (OUP) Research Fund.
Based on observations made with the NASA/ESA Hubble Space Telescope, obtained [from the Data Archive] at the Space Telescope Science Institute, which is operated by the Association of Universities for Research in Astronomy, Inc., under NASA contract NAS 5-26555. 
These observations are associated with program \#13793.
This work is based on data products from observations made with ESO Telescopes at the La Silla Paranal Observatories under ESO programme ID 179.A\-2005 and on data products produced by TERAPIX and the Cambridge Astronomy survey Unit on behalf of the UltraVISTA consortium. 
This study was based in part on observations obtained with MegaPrime/MegaCam, a joint project of CFHT and CEA/DAPNIA, at the Canada-France-Hawaii Telescope (CFHT) which is operated by the National Research Council (NRC) of Canada, the Institut National des Science de l'Univers of the Centre National de la Recherche Scientifique (CNRS) of France, and the University of Hawaii. 
This work is based in part on data products produced at TERAPIX and the Canadian Astronomy Data Centre as part of the CFHTLS, a collaborative project of NRC and CNRS.  
This work is based in part on observations made with the Spitzer Space Telescope, which is operated by the Jet Propulsion Laboratory, California Institute of Technology under a NASA contract.

\bibliographystyle{mnras}
 \bibliography{library_abbrv.bib}

\appendix

\section[]{Simulations of multiple-component galaxies}

\begin{figure*}

\includegraphics[width = 0.98\textwidth]{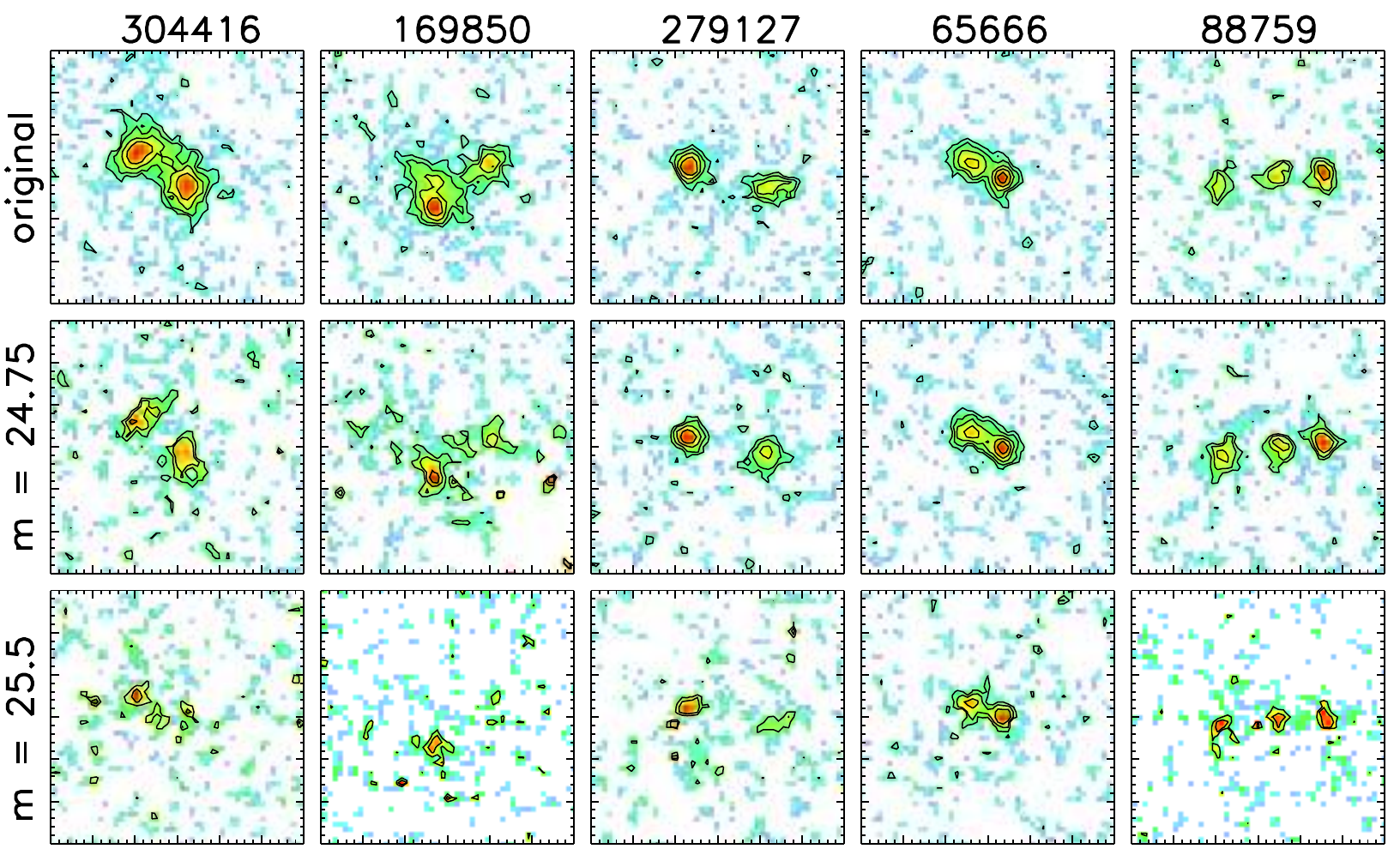}
\caption{Mock observations of multiple-component galaxies in our sample, as they would appear at fainter magnitudes.
From left-to-right the images show the brightest four galaxies in our sample in order of rest-frame UV magnitude, followed by the triple-merger system Himiko (ID 88759), which shows a particularly clumpy morphology.
The stamps are identical to those shown in Fig.~\ref{fig:hstsb} in their scaling and size (3.0 arcsec on a side).
The upper row shows the best-fitting {\sc GALFIT} model to the galaxies, at the original apparent magnitude.
In the lower plots we show the results of dimming the galaxies to progressively fainter total magnitudes of $m_{\rm AB} = 24.75$ and $m_{\rm AB} = 25.5$.
}\label{fig:dim}
\end{figure*}

The results of our~\emph{HST}/WFC3 imaging of bright $z \simeq 7$ LBGs reveals a high prevalence of multiple-component, clumpy systems at the bright-end of the sample.
As the imaging is of constant depth for each object, while the galaxy apparent magnitudes differ by $\simeq 2\,$ mag, we must consider whether this observation represents a genuine trend or is a result of the varying signal-to-noise throughout the sample.
In Fig.~\ref{fig:dim} we show the results of a simple simulation to investigate the sensitivity of our observations to multiple components at fainter magnitudes.
For the four brightest LBGs in our sample we created a noise-free model using {\sc GALFIT}.
The best-fit models are shown in the top row of Fig.~\ref{fig:dim}, where the stamps are scaled in an identical way to the real data shown in Fig.~\ref{fig:hstsb}.
We also included the LAE Himiko (ID 88759) in this analysis for comparison, as it shows a particularly clumpy and extended morphology.
We then progressively reduced the total apparent magnitude (measured in a large 3 arcsec diameter circle apertures) of each galaxy model, and introduced authentic noise by injected the model into blank regions of the~\emph{HST}/WFC3 $J_{140}$ data.

If we first consider the galaxies in our sample brighter than $m_{\rm AB} < 25.0$, our simulations show that multiple-component galaxies similar to those found at the very bright-end of our sample are detectable as multi-component or extended systems at slightly fainter magnitudes.
This is in contrast to the observed profiles of the LBGs in our sample with $ 24.5 < m_{\rm AB} < 25.0$, which show relatively compact, single-component morphologies (e.g. ID238225, ID304384, ID185070) in addition to clumpy morphologies (ID30425, ID28495).
Smooth single component morphologies are not represented in the simulations of the brightest galaxies when dimmed to this fainter magnitude, suggesting that the brightest galaxies do genuinely have a different morphology.
An alternative interpretation could be the effect of orientation, with the very brightest galaxies being the more `edge-on' systems.
Larger samples of extremely bright LBGs are clearly needed to constrain the trend of multiples with luminosity, however with our sample we find that at the very bright-end of our sample ($M_{\rm UV} \lesssim -22.5$) multiple components do appear to be more prevalent than in fainter galaxies.
This is in agreement with the results of~\citet{Shibuya2015clump} who found some evidence for a trend with $M_{\rm UV}$ in the fraction of LBGs that show clumps.

For the faintest galaxies in our sample, with $25.0 < m_{\rm AB} \lesssim 26.0$, our simulations show that for ID304416, ID169850 and ID279127 only the brightest peak in the galaxy profile would be visible if it was at the faint-end of our sample.
This is a result of the flux in these objects being distributed amongst several extended/irregular components rather than compact discrete clumps.
We are therefore only sensitive to multiple-component morphologies at the faint-end of our sample when the profile shows compact emission such as for ID65666 and Himiko/ID88579.

\bsp	
\label{lastpage}
\end{document}